\documentclass[sigconf]{acmart}

\usepackage{amsmath,amsfonts}
\usepackage{amssymb}
\usepackage{wrapfig} % 浮动表格
\usepackage{bbding} %对勾叉号

\usepackage{lipsum}
\usepackage{tikz}
\usetikzlibrary{arrows.meta}
\usetikzlibrary{automata,positioning}

\usepackage{appendix}
\usepackage{float}

\usepackage{fancybox}%use other types of boxes
\usepackage{framed}

\usepackage{lscape}
\usepackage{rotating}
\usepackage[graphicx]{realboxes}
\usepackage{adjustbox}

\usepackage{mathtools}

\usepackage{booktabs} % For formal tables

\usepackage{listings}
\usepackage{blindtext}
\usepackage[ruled,linesnumbered]{algorithm2e}
\usepackage{etoolbox,xstring,mfirstuc,textcase}
\usepackage{booktabs} % For formal tables
\usepackage{listings}

\usepackage{xcolor}
\theoremstyle{plain}                
\theoremstyle{definition}       

\usepackage{indentfirst} 

\usepackage{dashbox}
\usepackage{amsfonts}
\usepackage{booktabs}
\usepackage{siunitx}

\usepackage{multirow}
\usepackage{comment}
\usepackage{indentfirst} 
\usepackage{framed} 

\usepackage[font=small,labelfont=bf,tableposition=top]{caption}
\usepackage{booktabs}
\usepackage{threeparttable}
\usepackage{dashbox}
\usepackage{amsfonts}

\usepackage{listings}
\usepackage{etoolbox}
\usepackage{tkz-euclide}
\tikzset{%
    pics/sema/.style args={#1/#2/#3}{code={%
        \ifstrequal{#2}{0}{%
            \node[circle,minimum width=1mm,draw,fill=#1] {};
        }{%
            \tkzDefPoint(0,0){O}
            \tkzDrawSector[R,fill=#1](O,1mm)(90,90-#2)
            \tkzDrawSector[R,fill=#3](O,1mm)(90-#2,90-360)
    }
    }},
}

\renewcommand\footnotetextcopyrightpermission[1]{} % removes footnote with conference information in first column
\begin{document}
\title{SoK: Diving into DAG-based Blockchain Systems}\thanks{\textcolor{magenta}{$\diamond$} This work has been accepted by ACM Computing Surveys (CSUR) [Nov 2022]. \\ \textcolor{magenta}{$\diamond$} Affiliations are ordered by the text length.}

%=================================================
%author
%=================================================

\begin{comment}

\author{Qin Wang}
\affiliation{%
  \institution{\textit{CSIRO, Data61}}
  \city{Melbourne}
  \country{Australia}
  }
\email{qinwang@swin.edu.au}

\author{Jiangshan Yu}
\affiliation{%
  \institution{\textit{Monash University}}
  \city{Melbourne}
  \country{Australia}
}
\email{jiangshan.yu@monash.edu}

\author{Shiping Chen}
\affiliation{%
  \institution{\textit{Csiro, Data61}}
  \city{Sydney}
  \country{Australia}
}
\email{Shiping.Chen@data61.csiro.au}

\author{Yang Xiang}
\affiliation{%
  \institution{\textit{Swinburne University of Technology}}
  \city{Melbourne}
  \country{Australia}
  }
\email{yxiang@swin.edu.au}

\end{comment}

%\begin{comment}

\author{Qin Wang$^{\S}$, Jiangshan Yu$^{\ddag}$, Shiping Chen$^{\S}$, Yang Xiang$^{\dag}$}
\affiliation{\textit{$\dag$ Swinburne University of Technology, Australia.} \\
 %\textrm{\{qinwang,yxiang\}@swin.edu.au}\\
\textit{$\ddag$ Monash University, Australia}\\ 
 %\textrm{jiangshan.yu@monash.edu} \\
\textit{$\S$ CSIRO Data61, Australia} \\
 %\textrm{Shiping.Chen@data61.csiro.au}
 }

%\end{comment}

%=================================================
%abstract
%=================================================

\begin{abstract}
Blockchain plays an important role in cryptocurrency markets and technology services. However, limitations on high latency and low scalability of classical blockchain systems retard their adoptions and applications. Reconstructed blockchain systems have been proposed to avoid the consumption of competitive transactions caused by linear sequenced blocks. These systems, instead, structure transactions/blocks in the form of Directed Acyclic Graph (DAG) and consequently rebuild upper layer components. The promise of DAG-based blockchain systems is to enable fast confirmation (complete transactions within million seconds) and high scalability (attach transactions in parallel) without significantly compromising security. However, this field still lacks systematic work that summarises DAG techniques. To bridge the gap, this Systematization of Knowledge (SoK) provides a comprehensive analysis of ever-existing and ongoing DAG-based blockchain systems. Based on substantial investigation, we abstract a general model to capture the main features and identify six types of design patterns. Then, we evaluate these systems from the perspectives of structure, consensus, property, security, and performance. We further discuss the trade-off between different factors, open challenges and potentiality of DAG-based solutions, indicating their promising directions for future research.  
\end{abstract}


%=================================================
%body
%=================================================
\maketitle

%=================================================
\section{Introduction}
%=================================================

\begin{comment}
One hindrance to Nakamoto’s Consensus is the reduced
security with frequent block proposals. As the voting power in
the network increases, through node number or otherwise, the
number of possible proposals increases. This resulted in a requirement in Nakamoto’s consensus to have a limited number
of proposals with satisfactory delay between proposals10
.
The increased frequency of block proposals presents a problem for Nakamoto’s consensus, as it is heavily reliant on the
dissemination of blocks prior to any mining. Once the block
creation speed exceeds the propagation time, the probability
of concurrent proposals increases and the honest nodes will
have more conflicting blocks, wasting overall mining power.
At this point, Nakamoto’s consensus will select the longest
chain and as a consequence a number of attacks, such as
double spending, may have a higher success rate without the
required majority voting power.

\end{comment}

\smallskip
\noindent\textbf{Limitations.} Blockchain becomes an emerging technology to realize the distributed ledgers\footnote{In this paper, we regard ``blockchain'' as a general term covering types and models which are based on the same technology. For simplicity, we ignore the difference between ``blockchain'' and ``distributed ledger''. Also, we occasionally use ``DAGs" to represent ``DAG-based Blockchains'' for short. }. The rising interest in blockchain has attracted extensive attention. Thanking to its great potential to tackle critical security and trust challenges in various distributed environments, blockchain technology enjoys rapid development \cite{nakamoto2008peer}\cite{wood2014ethereum} with derived topics evolving into a well-studied field  \cite{bonneau2015sok}\cite{belotti2019vademecum} in both industry \cite{ali2018applications}\cite{xie2019survey} and academia \cite{salman2018security}\cite{conti2018survey}. However, the great impact of blockchain promotes an influx of participants joining the game. Continuously increased traffic results in unavoidably catastrophic congestion due to the performance bottlenecks \cite{croman2016scaling} including \textit{slow confirmation}, \textit{low throughput} and \textit{poor scalability}. Compared to most centralized systems, these factors cannot be easily improved in blockchain systems which require decentralization as the priority. Several studies raise the view of blockchain trilemma \cite{trilemma2019} and the trade-off \cite{kiayias2015speed}, claiming that decentralization, security, and scalability, cannot perfectly co-exist in a blockchain system. For example, one major hindrance to Nakamoto Consensus (NC) \cite{nakamoto2008peer} is the reduced security with an increased block generation rate. The security of NC's \textit{longest chain wins} rule requires that honest nodes should be aware of other's blocks soon after the block's production. If the block creation rate exceeds the propagation time, concurrent blocks increase the possibility of forks happening. To mitigate the bottleneck of performance, multiple approaches have been proposed from different angles.

\smallskip
\noindent\textbf{Potential Solutions.} These approaches include the methods on sharding technique \cite{wang2019sok}\cite{avarikioti2019divide}, layer2 protocols \cite{gudgeon2020sok}, sidechain technique \cite{back2014enabling}, heterogeneous structure \cite{garay2020sok}, hybrid consensus solutions \cite{bano2019sok} and assisted techniques \cite{hafid2020scaling} such as modifying hard-coded parameters \cite{bch17} and cross-chain technique \cite{zamyatin2019sok}. Sharding splits pending transactions into smaller shards and makes them processed in parallel, but it is hard to achieve the consensus across shards due to asynchronization. Cross-chain protocols help to partially solve the problem by establishing the channels between multiple shards (horizontal sharding \cite{zamyatin2019sok}). As a sacrifice, these on-top protocols inevitably result in heavy and redundant systems, weakening efficiency and stability. Layer2 protocols enable participants to perform off-(main)chain transactions through private communication rather than broadcasting to the whole network. Together with such features, the challenge is how to properly and effectively guarantee the validity and consistency between off-chain and on-chain transactions. The technique of sidechain pegs the auxiliary chains to involve more transactions. A heterogeneous structure adds new types of blocks to assign them to different tasks. Hybrid consensus solution combines multiple fundamental consensus mechanisms together (PoX+BFT \cite{bano2019sok}\cite{garay2020sok}) to integrate their benefits. Modifying hard-coded parameters \textit{e.g.,} directly increases the volume of block from 1M to 8M in BCH \cite{bch17}. However, these techniques are still based on the linear-based backbone protocol, limiting the scope of exploitation. Therefore, radical modification -- reconstructing chains from the underlying structure and topology, becomes an emerging solution.

\smallskip
\noindent\textbf{DAG-based Approach.} Linearly structured blockchain systems maintain all the transactions/blocks in one single chain. Concurrent transactions/blocks compete for one valid position each round. This design inevitably leads to slow confirmation due to competitive miners, conflicted transactions, and wasted computations. The performance of the system can only be artificially suppressed (\textit{e.g.,} adjust the confirm time) so that each block is fully attached before the next one's arrival. Very few orphan blocks can be involved in the system. To this end, aiming to enable more transactions simultaneously processed/confirmed motivates the emerge of DAG-based blockchain systems \cite{sompolinsky2013accelerating}\cite{popov2016tangle}\cite{churyumov2016byteball}\cite{sompolinsky2016spectre}. They structure the transactions/blocks in the form of graph topology to underlyingly alter the actual operations. DAG-based systems can improve performance by requiring less communication, computation and storage overhead. However, the sources of existing open-sourced projects and solutions are disparate and disorganized. Various basic concepts in DAGs (\textit{e.g.,} vertex meaning, consensus approach, order sequence) are still confused, and potential challenges have neither been clearly identified. An organized and structured systematic overview is absent for newcomers. Several studies draw their attention to this field, but the works are either rough in summaries \cite{pervez2018comparative}\cite{birmpas2019fairness}\cite{bai2018state}\cite{gorczyca2019distributed}\cite{he2019consensus}\cite{zhou2020solutions}\cite{silvano2020iota}\cite{xiao2020survey}, superficial in analysis \cite{hafid2020scaling}\cite{zander2019dagsim}\cite{lathif2018cidds}\cite{dong2019dagbench}\cite{raderblockchain}\cite{benvcic2018distributed}\cite{saad2019decentralized}\cite{schueffel2017alternative}\cite{webblock18}, or incomplete in evaluations \cite{canon2009evaluation}\cite{li2019direct}\cite{fan2019performance}.

\smallskip
\noindent\textbf{Contributions.} In this SoK, we attempt to consolidate the core knowledge of the structural shift in blockchain systems and review the state-of-the-art DAG-based blockchain systems with comprehensive mechanisms and properties. To drive future research, we also provide multifaceted discussions and comparisons with concurrent scaling techniques of blockchain. Furthermore, we summarise unsolved problems existed in current DAG systems, hoping to highlight the research challenges in this field. We start with a series of simple questions, aiming to provide brief guidance for readers.

\begin{center}
\setlength{\fboxrule}{0.3pt}
\setlength{\fboxsep}{3mm}
\fbox{\shortstack[l]{\textit{Q1. What is the DAG-based blockchain?}  \\
\textit{Q2. How do they structure the ledgers?} \\ 
\textit{Q3. How to run the consensus?}  \\  %  >> Committee selection
\textit{Q4. What are the desired properties?}  \\
\textit{Q5. How secure are they?} \\
\textit{Q6. What is performance improvement?} \\
\textit{Q7. How is this different from other techniques?} \\
\textit{Q8. What are the research challenges?} 
}}
\end{center}

%-------------------------------------------------
\smallskip
\noindent\textbf{Framework.}To understand the DAG-based blockchains, establishing a DAG system step by step is necessary. We provide our analysis framework as follows. This framework helps to clear the mist surrounding the DAG technique applied in blockchain systems and provides a comprehensive systematic overview of such systems. 

\begin{itemize}

   \item[$\diamond$] \textbf{\textit{Overview of DAG.}} Firstly, we review sufficient existing works surrounding DAG-based blockchain systems in \underline{Section \ref{sec-overview}}, covering the topics of protocol design, analysis and discussion, evaluation and simulation, improvement, application, and indirectly relevant studies. 

   \item[$\diamond$] \textbf{\textit{Identified types.}} Secondly, we give an informal mathematical model to generically describe the DAG-based systems in \underline{Section \ref{sec-modeling}}, which contains two elements in different dimensions. Then, we obtain six types of structures by combining each element. According, we collect and categorize in-the-wild projects and studies under our identified types. 
   
   \item[$\diamond$] \textbf{\textit{Consensus.}} Thirdly, we focus on the consensus mechanism of each aforementioned system in \underline{Section \ref{sec-consensus}}. Specifically, we deconstruct the consensus mechanism into several components, assisting in understanding \textit{how these systems operate}. We further discuss the technique features to show their design principles and potential applications to the systems.
   
   \item[$\diamond$] \textbf{\textit{Analysis.}} Then, we analyse the properties in \underline{Section \ref{sec-property}}, securities in \underline{Section \ref{sec-Security}} and performance in \underline{Section \ref{sec-performance}}. These sections enable a deep understanding of the design principles, system features, and unique properties. 

   \item[$\diamond$] \textbf{\textit{Discussion.}} Additionally, we review concurrent techniques surrounding the DAG and clarify a series of open questions in \underline{Section \ref{sec-discussion}, \ref{sec-conclu}}. Detailed discussions help us learn more about current challenges and the potential future of this field. 

   \item[$\diamond$] \textbf{\textit{Summary.}} As a summary, our analysis framework has answered the listed questions in detail. Discussions on elements of the DAG model provide us with a simple but complete classification of current DAG-based systems, answering \textit{Q1} and \textit{Q2}. Deconstructions of consensus mechanisms show us how current DAG-based systems operate with their unique designs, answering \textit{Q3}. Analysis of the properties, securities, and performance enables us to understand \textit{Q4}, \textit{Q5}, and \textit{Q6}. Discussions, comparisons, and challenges answer  \textit{Q7} and \textit{Q8}. Surrounding details about the origination, progress, and application are provided in \underline{Section \ref{sec-discussion}, \ref{sec-conclu}}.

\end{itemize}

From a bird's view, this work provides a roadmap for studying applied-DAG blockchain systems. Based on our scrutinized analysis, we provide several remarks in this part: \textit{a)} DAG-based blockchains are still far away from commercial applications due to their incompatible designs, absence of standards, unreliable security, varied performance, and unfinished implementations. \textit{b)} DAG structure has the potential to improve scalability and performance but inevitably sacrifices certain properties like consistency or finality.  \textit{c)} DAG-based systems vary from one to another. A uniformly formalized model can hardly cover all key points. Instead, a loose and informal model benefits a lot for better understanding. \textit{d)} Although applying the DAG structure to classic distributed systems and blockchain systems is quite challenging, many studies with their proposed systems have greatly progressed and improved. We believe more exciting new protocols/solutions will emerge shortly.

%we deconstruct existed projects by their fundamental components and executed procedures, in the lens of which we based on for further reviews and evaluations. We summarise, as a result, five design patterns to distinguish them. Meanwhile, we also provide evaluations in the lens of individual projects, details (totally 20 projects) are stated \underline{in Appendix}.

%=================================================
\section{Overview of DAG-Based Blockchains}
\label{sec-overview}
%=================================================

This section provides a quick overview of existing studies on DAG blockchain systems. Three aspects are summarized as follows.

\smallskip
\noindent\textbf{Origination and Evolution.}
Directed Acyclic Graph (DAG) represents a finite directed graph with no directed cycles in the mathematic and computer field. Due to the unique topology, it is frequently employed as a basic data structure and applied to various algorithm scenarios such as seeking the shortest path in navigation and data compression in storage. The concept was first introduced to mainstream blockchains by Sompolinsky \textit{et al.} in the work of GHOST \cite{sompolinsky2013accelerating}\cite{sompolinsky2015secure}, aiming to address the concurrency problem by allowing transactions structured in trees. The improved version \cite{lewenberg2015inclusive} was applied to Ethereum \cite{wood2014ethereum} as its core consensus mechanism. GHOST protocol and its variants enable all blocks to quote more than one parent block, and all types of references can be converted into rewards as incentives. Thus, the irreversibility of chains can also be strengthened by blocks that are off the chain. After that, a shift of granularity from block-level to transaction-level was noted by Lerner \textit{et al.} in DAGCoin \cite{lerner2015dagcoin}. Transactions take over the tasks of blocks, directly confirming the pending transactions and maintaining the order of sequence. Consequently, efficiency was leap due to the abandonment of packaging and competing steps. IOTA \cite{popov2016tangle} and ByteBall \cite{churyumov2016byteball}, inheriting the concept of blockless, were proposed with full open-sourced implementation and leads the markets till now. In the following times, several modifications and updates were added to DAGs. Spectre \cite{sompolinsky2016spectre} aims to establish a system that enables concurrent block creation. Hashgraph \cite{baird2016swirlds} proposed a permissoned graph-based blockchain with the consensus inspired by BFT-style protocols \cite{malkhi2018blockchain}. Nano \cite{lemahieu2018nano} proposed a so-called \textit{block-lattice} structure to realize immediate and asynchronous processing by allowing users to maintain their lightweight accounts. Conflux \cite{li2018scaling} brought blocks back to the system for the purpose of a total linear sequence. More cases will be discussed in later sections.

\smallskip
\noindent\textbf{Analysis and Evaluations.} 
Open-sourced and published works provide their basic design patterns. Related studies based on them are rooted in different views. Several works provide analysis by components, including topics on backbone properties \cite{bramas2018stability}\cite{kusmierz2017first}\cite{attias2018tangle}, consensus design \cite{he2019consensus}\cite{tian2020design}, random walk tip selection \cite{kusmierz2019properties}, witness selection \cite{gorczyca2019distributed}, cryptographic hash algorithm \cite{heilman2019cryptanalysis}, timestamps \cite{popov2017timestamps}, signature scheme \cite{buchmann2011security}, memory model \cite{melnykappend}, committee selection \cite{kusmierz2021committee}, pending probability \cite{kusmierz2018probability}, component design \cite{iotacoordicide}, fees \cite{popov2019iota} \textit{etc.} While some studies are based on properties, covering fairness \cite{birmpas2019fairness}, consistency \cite{kiffer2018better}, anonymity \cite{tennant2017improving}\cite{ince2018adding}\cite{sarfraz2019privacy}, performance \cite{fan2019performance}\cite{fan2019performance}, practicability \cite{dong2019dagbench},
robustness \cite{canon2009evaluation}, stability \cite{bramas2018stability}, lightweight client \cite{shafeeq2019curbing}, security \cite{li2019direct}\cite{cullen2020resilience}\cite{cai2019parasite}\cite{staupe2017quasi}\cite{de2018break}\cite{penzkofer2020parasite}\cite{monica2018chainweb}\cite{tanana2019avalanche}, incentive and rewards \cite{szalachowski2019strongchain},  \textit{etc.}  Discussions \cite{pervez2018comparative}\cite{bai2018state}\cite{yuva2018directed}\cite{raderblockchain}\cite{saad2019decentralized} and comparisons \cite{schueffel2017alternative}\cite{benvcic2018distributed} on selected systems (\textit{e.g.,} IOTA, Byteball, Hashgraph, Nano) have identified several features and challenges, but they were technically or theoretically insufficient. In-depth analysis employs different techniques and tools, such as establishing an analysis framework \cite{birmpas2019fairness} and converting complicated environment into simulations \cite{lathif2018cidds}\cite{wang2020security}\cite{zander2019dagsim}\cite{kusmierz2017first}\cite{tarun2018chainweb}\cite{wang2020chainsim}. Meanwhile, several types of formal analyses have also been proposed, including brief models with qualitative arguments \cite{boyen2018graphchain}\cite{sompolinsky2015secure}\cite{sompolinsky2015secure}, formal model analysis \cite{sompolinsky2020phantom}\cite{bentov2017tortoise}\cite{tang2020haootia} and formal model with mathematics induction \cite{fitzi2018parallel}. Beyond that, a few of modified protocols \cite{kan2018improve}\cite{manoppodesigning} and improved schemes \cite{bu2019g}\cite{bu2019metamorphic}\cite{ferraro2018iota} are proposed for further exploitation.

\smallskip
\noindent\textbf{Applications.} 
Systems based on DAG mainly benefit distributed applications (DApp) with high performance and low cost. Existing solutions could be roughly categorized by their combination layers. Directly integrating building blocks with the underlying network \cite{tesei2018iota}\cite{lamtzidis2019novel}\cite{fantowards} helps to enjoy almost equal properties and advantages of DAG, but it requires expertise development skills and costly hardware devices. Establishing applications through official components (\textit{e.g.,} Qubic \cite{qubic20}, MAM \cite{mam} in IOTA) is an alternative selection for developers \cite{lamtzidis2019novel}\cite{zheng2019accelerating}\cite{brogan2018authenticating}. It simplifies the processes of deployment and maintenance but scarifies flexibility and customization to a certain degree. Beyond that, various specific scenarios are considered, which includes the fields on Internet of Things \cite{alsboui2019enabling}\cite{cui2019efficient}, data management \cite{ruiz2018distributed}, vehicular applications\cite{bartolomeu2018iota}\cite{strugar2018m2m}\cite{yang2020ldv}, charging scheduling \cite{guo2020double}, M2M Communications \cite{zivic2019distributed}\cite{assante2019m2m}, manufacturing \cite{raschendorfer2019iota}, smart home infrastructure \cite{fantowards}, precision agriculture \cite{lamtzidis2019novel}, P2P energy trading \cite{murkinblock}, intelligent transport systems \cite{tesei2018iota}, air-quality data-monitoring System \cite{sun2019indoor}, smart transportation services \cite{lamtzidis2019novel}\cite{zichichi2019distributed}, sensor node system \cite{lamtzidis2018iota}, data quality management \cite{wu2018data}, E-health \cite{zheng2019accelerating}\cite{brogan2018authenticating}\cite{zheng2019activity}\cite{shafeeq2019privacy}, smart grid \cite{park2019dag}, feedback control system \cite{cullen2019distributed}, database platform \cite{ding2020dagbase}, voting system \cite{srivastava2018crypto}\cite{srivastava2018phantom}\cite{bahri2019electronic}, named data network \cite{zhang2019dledger}, smart city \cite{ferraro2018distributed} \textit{etc.} It should be noted that proposed schemes cannot simply achieve multifaceted requirements. They need to employ additional techniques such as IPFS file system \cite{benet2014ipfs} for storage, zero-knowledge proof (ZKP) for privacy, access control for authentication  \cite{ruiz2018distributed}, \textit{etc.} Finally, an important fact based on the aforementioned solutions is that all combinations aiming to implement DApps are still staying at the stage of Proof of Concept (PoC), far away from practical realizations and commercial applications.

\smallskip
\noindent\textbf{Guidelines.} For an overview, we summarise related works to provide a full picture of DAG. Among citations in this paper, the topics that are frequently studied include three aspects: the protocol designs (including both projects in industry and schemes in academic); the analysis, comparison, and evaluations; and potential applications to real scenarios. A handful of studies provide improved suggestions or schemes, and several works focus on the techniques or analyses that are (indirectly) related to DAGs (cf. \underline{Table.\ref{tab-guide}}).

\begin{table}[htb!]
 \caption{Guideline of Literature} 
 \label{tab-guide}
  \centering
\resizebox{\linewidth}{!}{  \begin{tabular}{l p{55mm}}
    \toprule
     \textit{Protocol Design} &  \cite{sompolinsky2013accelerating} \cite{lewenberg2015inclusive} \cite{sompolinsky2015secure} \cite{amores2021generalizing} \cite{popov2016tangle}  \cite{iotacoordicide} \cite{churyumov2016byteball} \cite{li2018scaling}    \cite{baird2016swirlds} \cite{Choi2018OPERARA} \cite{choi2018fantom} \cite{Nguyen2019ONLAYOL} \cite{Nguyen2019StakeDagSC} \cite{he2019consensus} \cite{melnykappend}  \cite{wang2021weak}  \cite{wang2019improving} \cite{wang20213ddag} \cite{Nguyen2019StairDagCV}\cite{sompolinsky2018phantom} \cite{sompolinsky2016spectre} \cite{Sompolinsky2017SPECTRES} \cite{lemahieu2018nano} \cite{gkagol2018aleph} \cite{rocket2019scalable} \cite{bagaria2019prism} \cite{danezis2018blockmania} \cite{yang2019prism} \cite{lerner2015dagcoin} \cite{chen2018dexon} \cite{boyen2016blockchain} \cite{yin2019streamnet} \cite{boyen2018graphchain} \cite{Soteria} \cite{zdag} \cite{liu2018vite} \cite{li2020ghast} \cite{li2020decentralized} \cite{itc} \cite{webblock18} \cite{manoppodesigning} \cite{istvan2018streamchain} \cite{popov2018local} \cite{yu2020ohie} \cite{niu2019eunomia} \cite{will2019chainweb} \cite{gupta2019cdag}   \cite{chevalier2019protocol} \cite{cao2019high} \cite{zhou2020hotdag} \cite{zhang2020c} \cite{abram2020democratising} \cite{swaroopa2020scalable} \cite{tian2020design} \cite{Idit2019dagrider} \cite{nguyen2021lachesis} \cite{giridharan2022bullshark} \cite{muller2022tangle} \cite{wang2021securing}; \\
     
    \midrule
    \textit{Analysis/Discussion} &  \cite{kiayias2017trees} \cite{kovalchuk2018number} \cite{popov2019equilibria} \cite{kusmierz2018extracting} \cite{bramas2018stability} \cite{heilman2019cryptanalysis} \cite{silvano2020iota}  \cite{melnykappend} \cite{attias2018tangle} \cite{buchmann2011security} \cite{popov2017timestamps}  \cite{popov2019iota} \cite{kusmierz2018probability} \cite{kusmierz2019properties} \cite{kiffer2018better} \cite{tennant2017improving} \cite{staupe2017quasi} \cite{li2019analysis} \cite{li2019direct} \cite{kan2018improve} \cite{fan2019performance} \cite{pervez2018comparative} \cite{bai2018state} \cite{kannengiesser2019mind} \cite{raderblockchain} \cite{benvcic2018distributed} \cite{saad2019decentralized} \cite{qin2020security} \cite{fitzi2018parallel} \cite{monica2018chainweb} \cite{tanana2019avalanche} \cite{divya2018iota}; 
     \\

    \midrule
    \textit{Evaluation/Simulation}  &   \cite{popov2019equilibria} \cite{kusmierz2017first} \cite{kusmierz2018probability} \cite{birmpas2019fairness} \cite{park2019performance} \cite{gorczyca2019distributed} \cite{wang2020security} \cite{dong2019dagbench} \cite{lathif2018cidds} \cite{zander2019dagsim} \cite{tarun2018chainweb} \cite{wang2020chainsim} \cite{wang2020prism} \cite{demotest}; \\ 
    
    \midrule
    \textit{Attack/Defense}  & \cite{cullen2020resilience} \cite{cai2019parasite} \cite{penzkofer2020parasite} \cite{staupe2017quasi} \cite{tarun2018chainweb} \cite{de2018break} \cite{psdetection} \cite{qin2020security};  \\ 
     
    \midrule
    \textit{Improvement} & \cite{shafeeq2019curbing}  \cite{kusmierz2021committee} \cite{ferraro2018iota} \cite{vigneriachieving} \cite{sarfraz2019privacy} \cite{tennant2017improving} \cite{ince2018adding} \cite{bu2019g} \cite{bu2019metamorphic} \cite{gkagol2019aleph} \cite{cullen2020congestion}; 
      \\ 
      
    \midrule
    \textit{Application/Cases} &  \cite{yuva2018directed} \cite{bartolomeu2018iota} \cite{yang2020ldv} \cite{raschendorfer2019iota} \cite{ruiz2018distributed} \cite{wu2018data} \cite{lamtzidis2018iota} \cite{brogan2018authenticating} \cite{zheng2019accelerating} \cite{lamtzidis2019novel} \cite{zichichi2019distributed} \cite{sun2019indoor} \cite{murkinblock} \cite{xiao2019large} \cite{cui2019efficient} \cite{zhou2019dlattice} \cite{xiang2019jointgraph} \cite{park2019dag} \cite{tesei2018iota} \cite{fantowards} \cite{alsboui2019enabling} \cite{strugar2018m2m} \cite{cullen2019distributed} \cite{srivastava2018crypto} \cite{srivastava2018phantom}  \cite{zivic2019distributed} \cite{assante2019m2m} \cite{bahri2019electronic} \cite{zheng2019activity} \cite{ferraro2018distributed} \cite{shafeeq2019privacy} \cite{zhang2019dledger} \cite{guo2020double};   \\
    
    \midrule
    \textit{Related Discussion}   & \cite{wang2019sok} \cite{bano2019sok} \cite{zamyatin2019sok}  \cite{bonneau2015sok} \cite{gudgeon2020sok} \cite{gramoli2017blockchain} \cite{szalachowski2019strongchain} \cite{makhdoom2019blockchain} \cite{wang2019survey} \cite{xiao2020survey} \cite{vukolic2017rethinking} \cite{kaur2020scalability} \cite{malkhi2022maximal}. \\ 

    \bottomrule
  \end{tabular}}
  
\end{table}

%=================================================
\section{Modeling}
\label{sec-modeling}
%=================================================

This section defines an informal DAG model and network model. Then, we classify current systems into six types accordingly. 

%-------------------------------------------------
\subsection{From Classic Blockchain To DAG}
%-------------------------------------------------

%Blockchain is a linear chain
Blockchain records activities in the form of transactions. Transactions are organized into a hierarchical structure as a block, and blocks are arranged in an irreversibly ordered sequence. New blocks can only be attached to the main chain which are strictly sequenced by predecessors and successors. Although consensus mechanisms vary from system to system, classic blockchains base on the same linear-based chain structure. Transactions cannot be tampered or compromised due to the global view shared by all participants, integrity and traceability are consequently provided. Other types of blockchain modeling could be referred to \cite{sompolinsky2015secure}\cite{lewenberg2015inclusive}\cite{atzei2018formal}\cite{garay2015bitcoin}.

%chain-based cause throughput bottleneck / Blockchain to DAG
The performance bottleneck of the linear structure is mainly caused by consensus mechanisms. Specifically, a group of nodes compete for the right of block packaging though many methods such as rotating leaders (BFT-style), solving puzzles (PoW), holding stakes (PoX), \textit{etc.} Only winners are able to determine the validity and confirmation of transactions, while left transactions are pended in the pool or discarded off the chain. Adding blocks at the same time causes more conflicts. In contrast, DAG-based structure aims to enable multiple transactions confirmed in one round. Each unit (transactions/blocks/events) of the ledger could refer to more than one parent units, and also could be referenced by numerous subsequent units. This structural design supports concurrent operations. Multiple nodes can simultaneously add units to the ledger, thereby significantly improving the throughput.

%-------------------------------------------------
\subsection{DAG-based Model}
%-------------------------------------------------

A directed acyclic graph $\mathcal{G}$ consists of a point set $\mathcal{V}$ and an edge set $\mathcal{E}$. Firstly, each element in the point set corresponds to a \textit{unit}. A unit can be instantiated as a transaction $\mathsf{Tx}\in\mathcal{T}$, a block $\mathsf{B}\in\mathcal{B}$, or an event $\mathsf{E}\in\mathcal{E}$ in protocols, where $\mathcal{T}, \mathcal{B},\mathcal{E}$ represents the sets of elements. Secondly, the element in the edge set is a pair $(u, v)$, which represents the partial order relationship between two points $u$ and $v$. The relationship, in most cases, indicates that one of the \textit{unit} references another unit. For instance, $ u\gets v $ means $v$ confirms/verifies/witnesses/sees $u$, where $\{u,v\} \subseteq \mathcal{V}$. We define the DAG-based model with two properties as follows:

\begin{definition}[DAG-based blockchain model]\label{eq-model}
The model $\mathcal{G}$ is defined as,
\begin{equation*}
\begin{array}{ll}
\begin{aligned}
   \mathcal{G} & = (\mathcal{E},\mathcal{V})^{\dag}{^{\ddag}}, \,\, \textit{where}\\
  \mathcal{V} & = \{u\, | u \in \{\mathcal{T} \cup \mathcal{B} \cup \mathcal{E}\} \}, \,\,
  \mathcal{E}  = \{(u,v)\,|\, u\gets v \land \{u,v\} \subseteq \mathcal{V} \}. \\
   & \dag:  \,\, \forall u\gets v  \nRightarrow v \gets u; \\
  & \ddag: \, \textit{\textbf{Assume that}} \,\, u_i \in \{  u_1,u_2,...,u_l \} \subseteq \mathcal{V}: \,\,\, \forall i,j,k \in [1,l],\, \\
  & \quad where \,\, k>j>i,\,\,  \{j\}  \subset\{\varnothing, ...,\{i+1,...,k-1\}\} \,\,  \\
   &  \quad \textbf{if}\,\, u_i \gets u_j, u_j \gets u_k,\,\, \textbf{then,}\,\,  u_k \gets u_i \,\,\textit{does not exist}.\\
\end{aligned}
\end{array}
\end{equation*}
\noindent\textit{Here, the operation ``$a|b$'' means $a$ satisfies the condition of $b$. ``$\gets$'' represents an action that happens in the network. The property means (i) \textit{unidirectional$^\dag$}: the references point in the same direction;  (ii) \textit{acyclic$^\ddag$}: no loop exists in graph. It enables the units in the network appended-only and orderable. (more details refer to \underline{\textrm{Appendix A}})}
\end{definition}

%-------------------------------------------------
\subsection{Identified Types}
%-------------------------------------------------

The abstracted model describes the protocol in a theoretical and mathematical form. \textbf{We first specify the exact representations of units $\mathcal{V}$ and edges $\mathcal{E}$ in the model}. These elements define the structure and topology of a DAG-based system.  We start our analysis framework from here.

\begin{itemize}
\item[-] \textbf{\textit{Unit Representation. }} This shows the underlying element of a system, either transactions, events, or blocks. We have two types of options:  \textit{ $1^{od}$}  and  \textit{ $2^{od}$}. The former represents the request that is immediately handled whenever it is received, without having to wait for more requests from peers. The forms of this type include transactions and triggering-events. The latter represents the request that needs further handling. In most cases, such requests are pre-computed or packaged by powerful parties (\textit{e.g.}, miner, validator) and then be disseminated. The forms of this type contain blocks.

\item[-] \textbf{\textit{Graph Topology. }}
A single edge may be insufficient to represent all relationships within a DAG-based system. The system is usually meshed by a set of edges. We identify three types of typologies based on their formed graphs, namely \textit{Divergence}, \textit{Parallel} and \textit{Convergence} (abbreviated as $\widehat{D},\widehat{P},\widehat{C}$, respectively). Specifically, \textit{divergence} means the units sparsely spread in unpredictable directions without predetermined orders. \textit{Parallel} means the units are maintained in the form of multiple (parallel) chains. \textit{Convergence} means the units are organized in a determined sequence or tend to converge in a determined sequence.

\end{itemize}

The \textit{unit representation} indicates the element of the system. It also reflects the ledger execution model, indicating how a transaction is completed in DAG-based systems. Generally, two types of models are included: \textit{UTXO-based} model and \textit{account-based (\textit{acct})} model. The former one means all operations are completed through atomic transactions. Users can calculate the balance by tracing the history of previous transactions. For the latter one, each user holds an account and the transaction is merely configured as one of the fields in its structure (\textit{e.g.}, the changes of \textit{balance}). Users calculate the balance directly in their accounts.

The \textit{topology} mainly focuses on an eventual trend of these units in the network. The units that initially spread to multiple directions but finally converge to the main chain, can still be classified into \textit{convergence}. For example, the blocks in GHOST \cite{sompolinsky2015secure} attach to their parents in a random way, but they are eventually structured into the main chain. This metric indirectly shows the finality and consistency of a blockchain system.

\begin{figure*}[!hbt]
\centering
\includegraphics[width=0.9\textwidth]{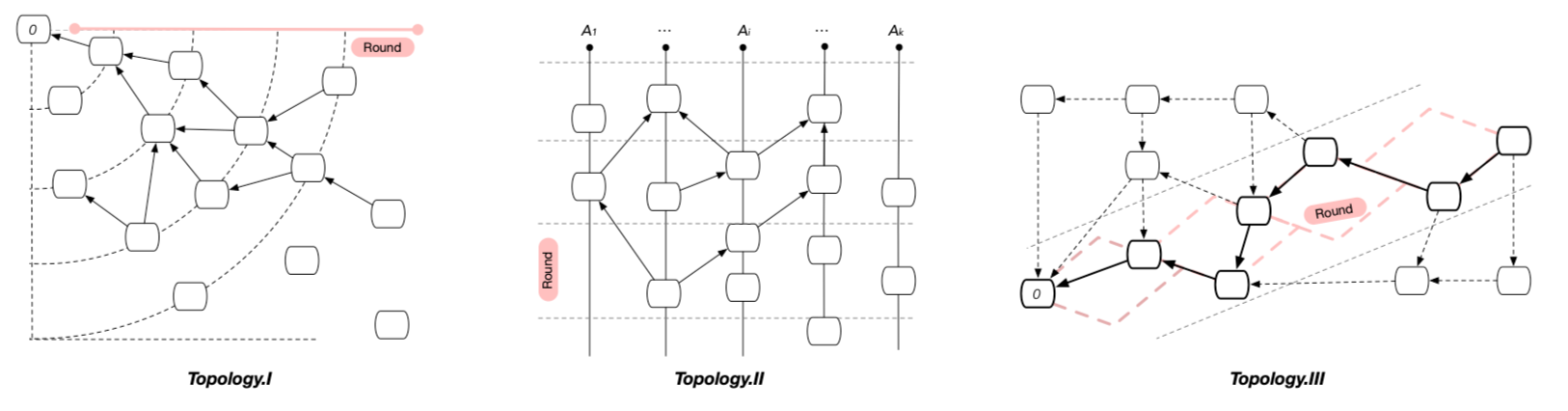}
\caption{Topology Patterns}
\label{fig-type}
\end{figure*}

\smallskip
\textbf{Then, we identify six types based on the combinations of the options in each metric.} We collect 30+ DAG-based blockchains systems (from 2011 to date) and classify these systems into the aforementioned types. Detailed classifications are shown in \underline{Table.\ref{tal:tps}}. 

\begin{itemize}
    \item[$\diamond$] \textit{$(\widehat{D},1^{od})$}. For better understanding, we describe the situations with terms frequently used in blockchain. This type means the data structure is blockless and the topology is a natural graph. Transactions are in equal and fair entities. Exemplified systems include IOTA \cite{popov2016tangle}, Graphchain \cite{boyen2018graphchain} and Avalanche \cite{rocket2019scalable}. We denote this type as \textit{Type I}.

    \item[$\diamond$] \textit{$(\widehat{D},2^{od})$}. Transactions needed to be organized in blocks for packaging, and the topology is a natural graph. Exemplified systems contains Spectre \cite{sompolinsky2016spectre},  Phantom \cite{sompolinsky2020phantom} and Meshcash \cite{bentov2017tortoise}. This type is denoted as \textit{Type II}.

    \item[$\diamond$] \textit{$(\widehat{P},1^{od})$}. The data structure is blockless, and transactions are maintained by individual nodes which form multiple parallel chains. Exemplified systems cover Nano \cite{lemahieu2018nano},  Hashgraph \cite{baird2016swirlds}, DLattice \cite{zhou2019dlattice}, Jointgraph \cite{xiang2019jointgraph}, Chainweb \cite{will2019chainweb}, Aleph \cite{gkagol2019aleph}, Vite \cite{liu2018vite}, Caper \cite{amiri2019caper} and Lachesis-class protocols \cite{Choi2018OPERARA}\cite{choi2018fantom} \cite{Nguyen2019ONLAYOL}\cite{Nguyen2019StakeDagSC}\cite{Nguyen2019StairDagCV}. This type is denoted as \textit{Type III}.

    \item[$\diamond$] \textit{$(\widehat{P},2^{od})$}. Transactions are structured in blocks to form parallel chains. The systems include Prism \cite{bagaria2019prism},  OHIE \cite{yu2020ohie}, Shpnix \cite{wang2021weak}, Blockmania \cite{danezis2018blockmania}, Blockclique \cite{forestier2018blockclique},  Eunomia \cite{niu2019eunomia}, Dexon \cite{chen2018dexon} and PARSEC \cite{chevalier2019protocol}. We denote this type as \textit{Type IV}.

    \item[$\diamond$] \textit{$(\widehat{C},1^{od})$}. The data structure is blockless, and transactions converge into the main chain. Examples are Byteball \cite{churyumov2016byteball}, Haootia \cite{tang2020haootia} and JHdag \cite{he2019consensus}. The type is denoted as \textit{Type V}.

    \item[$\diamond$] \textit{$(\widehat{C},2^{od})$}. Transactions are structured in blocks with the main chain. The systems include GHOST \cite{sompolinsky2015secure}, Inclusive \cite{lewenberg2015inclusive}, 3D-DAG \cite{wang20213ddag}, Conflux \cite{li2018scaling}\cite{li2020ghast}\cite{li2020decentralized}, CDAG \cite{gupta2019cdag} and StreamNet \cite{yin2019streamnet}. We denote this type as \textit{Type VI}.
    
\end{itemize}

\noindent\textit{Compared with Previous Classification.} 
Previous classifications are coarse-grained that make newcomers confused, such as the blocklattice mentioned in \cite{chen2018dexon}, blockDAG in \cite{forestier2018blockclique}, GHOSTDAG in \cite{sompolinsky2020phantom}, Lattice-based DAG in \cite{natoli2019deconstructing}, \textit{etc}. We observe that the gap is caused by the lack of classification metrics or dimensions. We, instead, classify existing systems in fine-grained forms by two dimensions (the topology made by edges, the units that indicate structures). With this in hand, we find that so-called TxDAG/blockDAG are named by their unit representations, while the terms GHOSTDAG/TreeDAG are based on their topology. Our classification method, naturally, leads to the fact that each term in previous classifications maps more than one type in ours. For example, Lattice-based structure (equiv. parallel-chains) covers \textit{Type III/IV} defined in our method.

%-------------------------------------------------
\subsection{Key Parameters of the Structure} 

We capture several qualitative metrics as the basic terminologies to describe the systems. In contrast, we omit the quantitative parameters (such as confirmation time, transaction fees, propagation time). These parameters can hardly be estimated due to the absence of implementations.

\begin{itemize}

   \item[-] \textbf{\textit{In/Out degree}}, denoted as $(|In|,|Out|)$, describes the number of connections of each unit. Equivalently, the in-degree  $|In|$ represents the number of a certain unit's successors. The out-degree $|Out|$ shows the number of a unit's ancestors. 

   \item[-] \textbf{\textit{Transaction model}} describes how to complete a transaction. Two main model types have been identified. \textit{UTXO} \cite{nakamoto2008peer} stands for the \textit{unspent outputs}. The transaction in \textit{UTXO model} is atomic and unsplittable. Every operation has to be completed through these transactions.  \textit{Account model} \cite{wood2014ethereum} maintains the \textit{balance} field in its data structure.  Transactions are completed via the changes in the user's balance. The user modifies its data in the local view and then synchronizes it to the network. Additionally, in several cases (\textit{e.g.}, \cite{lemahieu2018nano}), a complete token transfer, relying on the so-called \textit{pair model}, consists of two closely coupled transactions: a ``send'' transaction signed by the sender and a ``receive'' transaction by the receiver. These two transactions are required to occur within a negligible time interval.

   \item[-] \textbf{\textit{Confidence}} is a cumulative number that is used to show the confidence of a unit being verified by subsequent units both directly and indirectly. It also reflects the probability of a unit that will be selected in the next round. The parameter can be instantiated as different forms in the systems, such as the cumulative weight in IOTA \cite{popov2016tangle} and the witness in Hashgraph \cite{baird2016swirlds}, the score in Phantom \cite{sompolinsky2018phantom}. 

   \item[-] \textbf{\textit{Identifiers}} are used to identify the unit in the network. Each system has its customized parameters. These parameters include: \textit{height} ($h$) represents the length of a path, where the path starts from genesis transaction to a specified transaction; \textit{depth} ($d$) means the longest reverse-oriented path, starting from the tip\footnote{A \textit{tip} transaction indicates the latest transaction that is involved in branches.} transaction to a certain transaction; \textit{chain index} ($i$) identifies the chain with an index number when multiple chains coexist in the network; \textit{logic clock} ($t$) plays the role of capturing chronological and causal relationships.
   
\end{itemize}

Here, we provide details of \textit{types}, \textit{In/Out degree} and \textit{transaction model} in \underline{Table.\ref{tab-consensus}} to describe the system structure. \textit{Confidence} and \textit{identifiers} will be mentioned when discussing mechanisms and properties in latter sections.

\begin{table*}[htb!]
 \caption{Identified Types with Corresponding Systems} 
  \centering
   \resizebox{0.95\linewidth}{!}{
  \begin{tabular}{lccc}
    \toprule
      &  \textit{$\widehat{D}$ivergence}   &\textit{$\widehat{P}$arallel} &\textit{$\widehat{C}$onvergence} \\
    \midrule
    
     \multirow{4}{*}{ \textit{ $1^{od}$} }  & IOTA\cite{popov2016tangle},  &  Nano\cite{lemahieu2018nano},  Hashgraph\cite{baird2016swirlds}, DLattice\cite{zhou2019dlattice}, Jointgraph\cite{xiang2019jointgraph},  & Byteball\cite{churyumov2016byteball},    \\
     
     & Graphchain\cite{boyen2018graphchain},  &   Chainweb\cite{will2019chainweb}, Aleph\cite{gkagol2019aleph}, Vite\cite{liu2018vite}, Caper\cite{amiri2019caper},   &   Haootia\cite{tang2020haootia}, \\
     
     & Avalanche\cite{rocket2019scalable} & Lachesis\cite{Choi2018OPERARA}\cite{choi2018fantom}\cite{Nguyen2019ONLAYOL}\cite{Nguyen2019StakeDagSC}\cite{Nguyen2019StairDagCV}  & JHdag\cite{he2019consensus} \\

     & (\textit{Type I})   & (\textit{Type III})  & (\textit{Type V})  \\
     
   \cmidrule{2-4}
   
    \multirow{4}{*}{ \textit{ $2^{od}$} }  &  Spectre\cite{sompolinsky2016spectre},  &  Prism\cite{bagaria2019prism},  OHIE\cite{yu2020ohie},  Shpnix\cite{wang2021weak}, Blockmania\cite{danezis2018blockmania},  & GHOST\cite{sompolinsky2015secure}, Inclusive\cite{lewenberg2015inclusive}, 3D-DAG\cite{wang20213ddag} \\
    
    &  Phantom\cite{sompolinsky2020phantom}, & Blockclique\cite{forestier2018blockclique}, Eunomia\cite{niu2019eunomia}, &   Conflux\cite{li2018scaling}, CDAG\cite{gupta2019cdag}, \\

   &  Meshcash\cite{bentov2017tortoise}  & DEXON\cite{chen2018dexon}, PARSEC\cite{chevalier2019protocol}  &  StreamNet\cite{yin2019streamnet} \\

   & (\textit{Type II})   & (\textit{Type IV})  & (\textit{Type VI})  \\
    
    \bottomrule
  \end{tabular}
   }
  \label{tal:tps}
\end{table*}

%=================================================
\section{Consensus of DAG Systems}
\label{sec-consensus}
%=================================================

This section aims to explain consensus in a simple way through the deconstruction. We present several key aspects of discussion in consensus mechanisms and provide the details of each system. Then, we further conclude technical features and give our discussions.

%======================================
%======================================
\subsection{Deconstructed Components.} 

Achieving the consensus considers three aspects: a) \textit{who conducts the consensus?} b) \textit{how do they operate?} and c) \textit{what is the technique?} We answer questions by presenting consensus mechanisms of each system (summaries in \underline{Table.\ref{tab-consensus}}). Specifically, we deconstruct the consensus into several decoupled components for a better understanding of different systems and capture their commons. The nodes that can conduct the consensus procedures make up a consensus group, called the committee. \textit{Openness} indicates whether this committee is open to every node. \textit{Membership selection} defines the rules of becoming a committee member.  \textit{Unit allocation}, \textit{unit positioning}, \textit{extension rule} and \textit{conflict solving} determine the methods to reach the agreement. \textit{Technical feature} highlights the distinguished techniques in each system. Further, we provide discussions from perspectives of key techniques and consensus algorithms that are frequently utilized in systems. 

\begin{itemize}

     \item[-] \textbf{\textit{Openness}} indicates whether an arbitrary node can run the consensus algorithm without permission. Two types of selections are included, namely \textit{permissionless} and \textit{permissioned}. A permissionless DAG system means every node can join/leave the committee without restrictions. The size of the committee is dynamic. A permissioned DAG system means the new-coming nodes are required to obtain permission when joining the committee. The permitted conditions are unusually predefined by founder teams or core members. The size of the committee is fixed. 

     \item[-] \textbf{\textit{Membership selection}} defines the rules that are used to select the nodes into the committee. For a permissionless system, every node can participate in the competition for producing units. Nodes are ranked for priority according to their owned resources. The resources includes computing work (Proof of Work, \textit{PoW}), stake (Delegated Proof of Stake, \textit{PoS}/DPoS), vote (voting for favorable nodes, \textit{Elect}). The node that holds/obtains the most resources has the highest probability to win the competition. For a permissioned system, nodes have to prove the fact that they can meet the predefined rules. The committee membership is designated by their developing teams (\textit{Assign}).

    \item[-] \textbf{\textit{Unit allocation}}  is a prelude for consensus. Units in this step are required to be allocated with an identifier (\textit{Role}) or a destination (\textit{App}, \textit{Chain}). The identifier refers to the unit that need to play a specific role with responsibilities for tasks. The destination indicates that a newly appended unit is uniquely allocated to a specific place, instead of broadcasting to the network. This design arranges the units into isolated zones in advance to reduce potential conflicts. The option \textit{App} refers to a unit that is allocated to a specific application, whilst \textit{Chain} represents a unit that is allocated (randomly or by certain rules) to a certain chain among all chains. 
    
    \item[-] \textbf{\textit{Unit Positioning}} is a way to locate the unit in the network. This step is essential to sort the units into a total linear order. In classic blockchain systems, locating a block can reference the parameter of \textit{height} ($h$). However, locating the unit in DAG systems is hard. For the systems based on \textit{Topology $\widehat{D}$}, unstructured units make it almost impossible to precisely locate the unit in the expanding network. Only when the units are sorted into a linear sequence and buried deep enough, a unique location in the form of height ($h$) is obtained. In contrast, for the systems based on \textit{Topology $\widehat{P}$}, chains with unique indexes ($i$) are structured in parallel. A unit can be positioned via the orthogonal parameters $(i, h)$. Similarly, for the systems based on \textit{Topology $\widehat{C}$}, any unit can be roughly located by key blocks for the first step on the main chain, and then be precisely searched inside these blocks by sub-parameters like hash, timestamp.

    \item[-] \textbf{\textit{Extension rule}} empathizes how to extend the chains/graphs and break ties. Tie-breaking occurs when equivalent forks (subgraphs) compete for one winner, which is essential to maintain consistency by deleting the overlapped branches and reducing the data overburden. \textit{Extension rule}s conclude the main consensus mechanisms. Feasible methods includes Nakamoto consensus ($NC$) by the longest chain, variant Nakamoto consensus  (\textit{variant-NC}) by the heaviest weighted sub-tree, asynchronous Byzantine agreement (\textit{async-BA}), classic PBFT protocols (\textit{PBFT}) and its variation (\textit{smpl-PBFT}), tip selection algorithm (\textit{TSA}), some special algorithms like greedy algorithm (\textit{GA}), recursive traverse algorithm (\textit{RTA}) and \textit{sampling} algorithm. For the systems without explicit rules, we use \textit{natural} to denote the case.

    \item[-] \textbf{\textit{Conflict solving}} presents a set of parameters that determine the priority of conflicting units. The parameters include three types: a) the confidence of each unit, instantiated by the forms of \textit{weight}, \textit{confidence}, \textit{fee}, \textit{score}, \textit{fitness}; b)  the random beacon like \textit{Hash}; and c)  a natural sequence, containing \textit{logic clock}, \textit{appearance} and \textit{rank}. Sometimes, the decision is made by powerful authorities such as the leader and coordinator. We denoted them as the \textit{trusted role}s (TRs). Additionally, conflict solving differs from tie-breaking in their scope of adoption. Tie-breaking happens at the level of (main) chains, whereas conflict solving impacts on every single unit and mostly happens in the sorting algorithm with the aim to order units in linearization. In several systems, both tie-breaking and conflict solving are integrated into the extension rule. The consensus mechanism, by this way, reaches the agreement in one-step. 
    
    \item[-] \textbf{\textit{Technical feature}} provides the techniques that are different with other systems. We highlight one main feature of each system to distinguish them from peers. 

\end{itemize}

%Committee are used to differentiate the consensus nodes from other nodes. Committee management grants a group of nodes ability to run consensus. Extension rule determines the methods on chain growth. Unit sorting serves for organize the sequence of units, reducing forks and reaching agreement on a certain agreement. Generally, an individual DAG system may only cover several of these procedures.

%======================================
%======================================
\subsection{Consensus Algorithms in Each Type.} 
This subsection provides our reviews of the consensus mechanisms of current DAGs. We mainly analyze them in the view of our identified types, which helps to find some insights.

%-----------------------
%-----------------------
\smallskip
\noindent\textbf{Consensus on Type I.}  Type I systems are blockless. The topology of transactions is a natural expanding graph. This type of systems include IOTA \cite{tang2020haootia}, Graphchain \cite{boyen2018graphchain} and Avalanche  \cite{rocket2019scalable}.

%-----------------------
\textit{IOTA} \cite{popov2016tangle} is a permissionless network where each node can freely participate and leave it. IOTA adopts the UTXO model as the data structure. This design makes IOTA establish the system through transactions. The transactions issued by nodes constitute the site set of the \textit{tangle}, which is the ledger for an up-to-date history of transactions. All nodes in an IOTA network store a copy of the tangle and reach a consensus on its contents. Specifically, the extension of the tangle follows the rule\footnote{It should be noted that, at the current stage, IOTA relies on a central coordinator \cite{coordnator} issuing milestones to periodically confirm the transactions for stability. They claim the coordinator will be removed in a short future time.} where one tip (pending/unapproved transaction) is required to approve two ancestor transactions. Thus, users who issue a transaction will contribute to the security of the system. However, as tips are continuously generated and attached to the tangle, formed sub-graphs inevitably spread in different directions. To prevent the network from splitting into isolated cliques, \textit{tip selection} algorithms are essential for stability. The algorithms help to maintain a uni-direction graph by controlling the way to select tips. Three types of mechanisms are provided in \cite{popov2016tangle}: \textit{uniform random}, \textit{unweighted random walk} and \textit{weighted random walk}. All these mechanisms are based on statistical probability to simulate real scenarios. The most advanced mechanism is a weighted random walk algorithm, which is an application of Markov Chain Monte Carlo (MCMC) algorithms. MCMC could be transformed into other strategies through $\alpha$, a configurable parameter used to control the effectiveness. When $\alpha$ converges towards $0$, tip selection becomes uniformly random; while towards $1$, tip selection becomes deterministic. Additionally, we introduce two types of modified tip selections as improvements:

\begin{itemize}
    \item \textit{G-IOTA} \cite{bu2019g} modifies the rule (the weighted random walk tip selection rule) that a tip can approve three ancestor transactions at one time, instead of two. The extra one selects a left-behind tip in the tangle, used to increase the fairness in terms of transaction confidence for all honest transactions. The algorithm allows left-behind tips to regain the opportunity to be approved by incoming tips. G-IOTA further discusses the incentive mechanism to punish conflicting transactions and introduces a mutual supervision mechanism to reduce the benefits of speculative and lazy behaviors.
    
   \item  \textit{E-IOTA} \cite{bu2019metamorphic} propose a parameterized algorithm to provide the randomness for tip selection. The algorithm sets two specific parameters $ p_1, p_2$ where $0<p_1<p_2<1$. When appended to the network, a tip generates a random number $r$ where $r\in(0,1]$  to decide which types of mechanisms to conduct: $r\in(0,p_1)$ -- uniformly random selection; $r\in[p_1,p_2)$ -- low $\alpha$ weighted selection; or $r\in[p_2,1)$ -- high $\alpha$ weighted selection. Thus, E-IOTA controls the distribution of different types of mechanisms by changing the parameters $ p_1 , p_2$. This assists to establish a self-adjusted tangle. The algorithm suits both IOTA and G-IOTA.
   
\end{itemize}

\begin{figure}[H]
\centering
\includegraphics[width=0.35\textwidth]{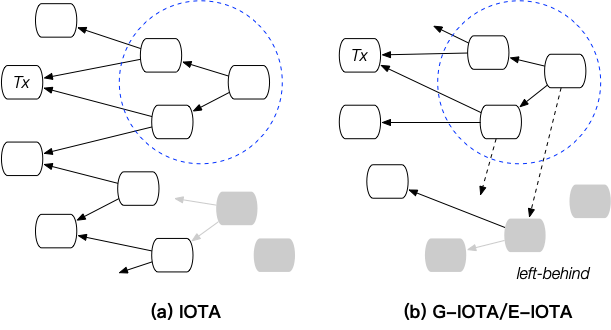}
\caption{IOTA with its Improvements}
\label{fig-ziota}
\end{figure}

%-----------------------
\textit{Graphchain} \cite{boyen2018graphchain} is a permissionless network that has a similar design to IOTA. Graphchain is formed by tasking each transaction to confirm its ancestors. Specifically, a transaction must verify several (at least two) ancestor transactions and each transaction carries a PoW of elective difficulty. PoW cumulatively and transitively affirms the prior valid history. In contrast, Graphchain differs from IOTA in its incentive mechanism. IOTA removes incentives from participated nodes where all transactions are feeless. The system progresses under the tip selection rules and is free from the influence of incentives. Graphchain, instead, introduces an incentive mechanism to maintain the graph.  Transactions are required to post transaction fees as the offering for collection. That means each transaction must refer to a group of ancestor transactions to deplete their deposit (fees) when verifying the validity. Meanwhile, ancestor transactions must have enough fees for collection. Fees are depleted by incoming transactions from the oldest ancestors in paths to reach a prescribed total amount.  High-fee transactions attract powerful miners working in parallel for rapid confirmation, while low-fee transactions will finally be picked up by a small miner. Based on this design, Graphchain makes the current transactions quickly become enshrined in the ancestry of all future transactions.

\begin{figure}[H]
\centering
\includegraphics[width=0.27\textwidth]{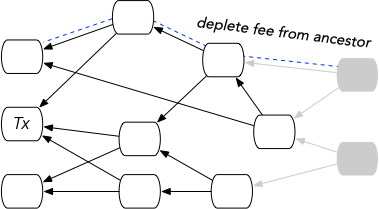}
\caption{Graphchain}
\label{fig-zgraphchain}
\end{figure}

%-----------------------
\textit{Avalanche} \cite{rocket2019scalable}\cite{rocket2018snowflake} is a permissionless system based on a new type of approach to reach the consensus. Deviating from BFT-style and Nakamoto mechanisms, Avalanche constructs its underlying protocol called \textit{Slush}, a CFT-tolerant mechanism, by capturing the concepts from the gossip algorithm and epidemic networks. More specifically, Avalanche randomly samples a small group of nodes to obtain their bias in a bivalent state. Regardless of the size of the network, the size of the sampled group ranges in a small intervals (such as from 10 to 20) to facilitate the execution time at each round. Then, the system arguments Slush to the extended algorithms with a series of parameters  (like the single counter in Snowflake and the confidence in Snowball) for finality and security. These parameters assist to yield a threshold result for their historical colors via countering the number of \textit{queries} (as votes in BFT protocols). Last, the augmented algorithm is applied to the whole network which is naturally formed in topological DAG. Thus, the final state, represented by a color in \textit{blue} (b) or \textit{red} (r), is achieved by repeated sampling of the network and customized guidelines of bias. Avalanche also provides a demo to show the process in visualization \cite{demotest}.

\begin{figure}[H]
\centering
\includegraphics[width=0.33\textwidth]{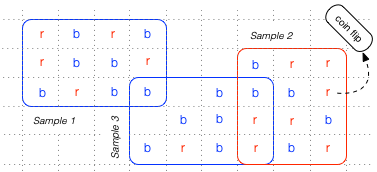}
\caption{Avalanche}
\label{fig-zAvalanche}
\end{figure}

%-----------------------
%-----------------------
\noindent\textbf{Consensus on Type II.}  Type II systems are based on blocks. Blocks are structured as a natural expanding graph. This part includes systems of Spectre \cite{sompolinsky2016spectre}, Phantom \cite{sompolinsky2020phantom}, and Meshcash \cite{bentov2017tortoise}. 

%-----------------------
\textit{Spectre} \cite{sompolinsky2016spectre}\cite{Sompolinsky2017SPECTRES} is a permissionless network. The key technique behind Spectre is a recursive weighted-voting algorithm based on the precedence of blocks in the underlying topology. The voting procedure is completed by blocks, instead of miners. Newly attached blocks are required to submit votes for every pair of blocks, denoted as $(x,y)$, according to their locations. The vote, $(-1,0,1)$, over such a pair of blocks, is inherently a preference ordering of the selected blocks, such as $x$ arrives before $y$ (denoted as $x<y$) and vice versa ($y<x$). Consequently, the final decision of the pairwise ordering is measured in the absolute values of aggregate votes (\textit{a.k.a,} weights). A consistent set of transactions is extracted according to the majority of collected votes. The ordering is not necessarily linearizable. This design relaxes the assumption of the ordering where any two transactions must be agreed upon by all non-corrupt nodes. Instead, Spectre decides its ordering only by honest nodes. Furthermore, the block renewal rate in Spectre has to slow down, to ensure the voting procedure has been completed by honest nodes.

\begin{figure}[H]
\centering
\includegraphics[width=0.26\textwidth]{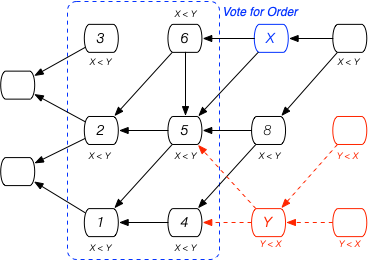}
\caption{Spectre}
\label{fig-zspectre}
\end{figure}

%SPECTRE’s underlying model falls into the category of partial synchronous networks: its security depends on the existence of some bound on the delivery time of messages between honest participants, but the protocol itself does not contain any parameter that depends on this bound. Hence, while other protocols that do encode such parameters must operate with extreme safety margins, SPECTRE converges according to the actual network delay.

%Key to SPECTRE’s achievements is the fact that it satisfies weaker properties than classic consensus requires. In the conventional paradigm, the order between any two transactions must be decided and agreed upon by all non-corrupt nodes. In contrast, SPECTRE only satisfies this with respect to transactions performed by honest users. We observe that in the context of money, two conflicting payments that are published concurrently could only have been created by a dishonest user, hence we can afford to delay the acceptance of such transactions without harming the usability of the system. Our framework formalizes this weaker set of requirements for a cryptocurrency’s distributed ledger. We then provide a formal proof that SPECTRE satisfies these requirements.

%-----------------------
\textit{Phantom} \cite{sompolinsky2020phantom} is a PoW-based protocol for the permissionless network. Each block in Phantom contains multiple hash references to predecessors and referenced by multiple successors. The protocol, firstly, identifies a set of well-connected blocks to exclude blocks (with high probability) created by dishonest nodes. Then, Phantom utilizes a recursive $k$-cluster algorithm (equally, a greedy approximation algorithm) to achieve the partial ordering of the identified set of blocks (cluster) to a full topological order. The greedy algorithm incentives selected blocks inside the cluster while penalizing outside blocks via iterative rounds. The parameter $k$ is used to adjust the level of tolerance of concurrent blocks. After the sorting of blocks, transactions in blocks are ordered according to the order of their appearance. Consistent transactions will be accepted and confirmed as the final state. The iteration over transactions enables a total order of the network. Similar to Spectre, Phantom also relies on honest nodes to agree upon this robust ordering of blocks and transactions. But Phantom differs from Spectre in that it enforces a strict linear ordering over blocks and transactions in the network.

\begin{figure}[H]
\centering
\includegraphics[width=0.33\textwidth]{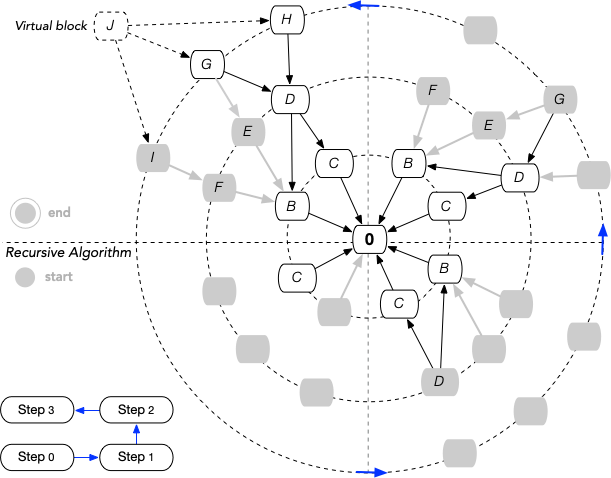}
\caption{Phantom}
\label{fig-zphantom}
\end{figure}

%It provides a total ordering among the blocks added to the ledger. It makes Phantom suitable for smart contracts, but the performance degrades severely in the presence of conflicting blocks.

%-----------------------
\textit{Meshcash} \cite{bentov2017tortoise} is a layered DAG system that allows the blocks to coexist at the same time. Firstly, to generate a block, the mining strategy follows classic Nakamoto PoW computations but with the modifications where each block: a) points to every block in the previous \textit{layer} (a specified field as \textit{round} in other systems); b) points to every block with 0 in-degree; and c) increases the counter of layer $i$ when seeing more than threshold blocks in the previous layer. To achieve the consensus, two types of protocols are integrated: a slow PoW-based protocol \textit{tortoise} to guarantee long-term consensus and irreversibility of blocks, and an interchangeable fast consensus protocol \textit{hare} to get quick consensus. A block belonging to which type of the protocol depends on its appearance in the network, measured by its layer. For blocks in old layers (earlier than $i-s$), the system follows the rules in \textit{tortoise} protocol. The protocol is inherently a leaderless BFT consensus, which decides the consistency (be part of the canonical history) of a block according to the weighted votes of its subsequent blocks. This protocol guarantees a total ordering of blocks. If the same transaction is contained in multiple blocks, the transaction appearing in an earlier block is preferred. For blocks in recent layers (from $i-s$ to latest updates), Meshcash shifts to the \textit{hare} protocol. The protocol is an interchangeable protocol with the aim of quickly settling down the blocks. Meshcash provides both a simple but limited-security method and a complex but attack-resistance one by utilizing the off-chain asynchronous byzantine agreement protocol (ABA). It enlarges the gap between honest and bad blocks and ensures that the honest parties tend to vote in the same direction once completing the protocol.

\begin{figure}[H]
\centering
\includegraphics[width=0.33\textwidth]{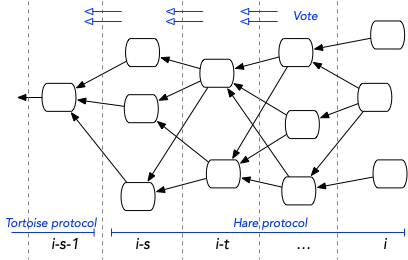}
\caption{Meshcash}
\label{fig-zmeshcash}
\end{figure}

%-----------------------
%-----------------------
\noindent\textbf{Consensus on Type III.}  Type III systems are blockless data structures. Transactions are maintained by individual nodes and finally form multiple parallel chains. This part includes systems of Nano \cite{lemahieu2018nano}, Hashgraph \cite{baird2016swirlds}\cite{baird2018hedera}, DLattice \cite{zhou2019dlattice}, Jointgraph \cite{xiang2019jointgraph}, Chainweb \cite{will2019chainweb}, Aleph \cite{gkagol2019aleph}, Vite \cite{liu2018vite}, Caper \cite{amiri2019caper} and Lachesis-class protocols \cite{Choi2018OPERARA}\cite{choi2018fantom}\cite{Nguyen2019ONLAYOL}\cite{Nguyen2019StakeDagSC}\cite{Nguyen2019StairDagCV}. 

%-----------------------
\textit{Nano} (RaiBlocks)  \cite{lemahieu2018nano} is a permissionless network with two types of entities involved, namely \textit{account holder} and \textit{representative} (short for \textit{Rpst}). Account holders can select a representative to vote on their behalf in case of offline leaves. Elected representatives serve for solving conflicted transactions.  Transactions in Nano are atomic, and an account record all the transaction history related to himself. A complete transfer in Nano consists of two parts -- a \textit{send} transaction and a \textit{receive} transaction. First, the sender creates a \textit{send} transaction by referring to the latest block in his account. At this time, corresponding amounts have been deducted from his account. Then, the receiver creates a \textit{receive} transaction by also referring to the \textit{send} transaction and the latest block in her account. When forks occur, a representative creates a vote referencing the conflicted transaction. Then, it starts to observe incoming votes from other representatives. This process lasts for four voting periods and spends one minute in total. Finally, it confirms the winning transaction with the highest cumulative votes (weights, $w$).

%However, this design requires nodes to be always online, since transaction is considered to be accepted only when both the \textit{send} transaction and \textit{receive} transaction are completed.

\begin{figure}[H]
\centering
\includegraphics[width=0.27\textwidth]{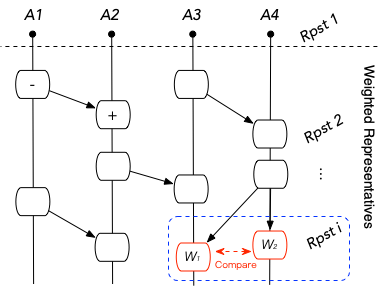}
\caption{Nano}
\label{fig-znano}
\end{figure}

%-----------------------
\textit{Hashgraph} \cite{baird2016swirlds}\cite{baird2018hedera} is a permissioned network. Each participated node maintains a separate chain, and nodes mutually interact via the gossip protocol. The node locally creates an \textit{event} to record the history of received information. An event mainly contains three fields, including \textit{timestamp} for synchronization, \textit{transaction} for trading and \textit{hash} for cross-references. Paralleled chains interact with each other based on the design of cross-references. These references point to the latest event on his chain, and also to the events from synchronized neighbor chains. Events carrying the complete history of ledgers' views are transmitted through the gossip protocol and nodes will eventually obtain a full history. The hashgraph achieves the consensus by a virtual Byzantine agreement consensus. One event to be finally deemed as valid has to go through a three-stage procedure, namely \textit{see}, \textit{strongly see}, and \textit{decide}. Each procedure needs to collect the votes more than a threshold --- 2/3 \textit{famous witness}es, who are the proposers elected by committee members in each round. This design simulates a conventional BFT consensus and integrates the concepts into parallel chains. Events, in this way, can be structured in a global total order. Meanwhile, they are able to reach finality when consensus is completed in current rounds.

\begin{figure}[H]
\centering
\includegraphics[width=0.27\textwidth]{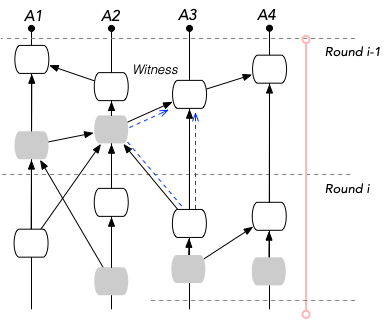}
\caption{Hashgraph}
\label{fig-zhashgraph}
\end{figure}

%-----------------------
\textit{DLattice} \cite{zhou2019dlattice} follows the structure of Nano and Hashgraph where parallel chains maintained by individual nodes make up a DAG system. The system consists of a Genesis Header and multiple nodes. Genesis Header organizes the involved nodes in the form of Merkle Patricia Tree (MPT), whilst the nodes mainly complete the tasks such as token transferring. Similarly to Nano, the action of a token transferring is split into the \textit{send} transaction and the \textit{receive} transaction between the sender and the receiver. If a node observes any forks of transactions, a consensus program is launched to solve the conflicts. DLattice uses a so-called \textit{DPoS-BA-DAG} (PANDA) protocol to reach consensus among users. DPOS provides the way of committee formation and BA shows how to achieve consensus in their DAG. Specifically, the nodes who satisfy the customized PoS condition are able to locally and secretly generate their identities based on the voting power by the Verifiable Random Function (VRF) technique \cite{gilad2017algorand}. After broadcasting the messages, other nodes can verify whether an identity is selected as the consensus identity. The consensus is divided into two phases \textit{Vote} and \textit{Commit}. Committee members select a transaction to vote at the Vote phase and start the commit according to collected votes at the Commit phase. If the count of committed votes exceeds the threshold, the consensus reaches the current agreement.

\begin{figure}[H]
\centering
\includegraphics[width=0.27\textwidth]{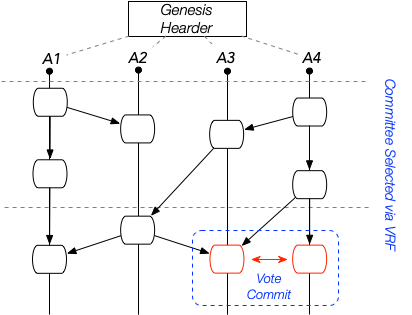}
\caption{DLedger}
\label{fig-zDLedger}
\end{figure}

%-----------------------
\textit{Jointgraph} \cite{xiang2019jointgraph} is a simplified protocol based on Hashgraph. The consensus in Hashgraph needs at least two voting rounds for events, while Jointgraph, instead, simplifies the voting process into one round by introducing a powerful \textit{supervisor} node. This node mainly monitors the misbehaving nodes by replacing them with honest ones, and periodically takes snapshots of system states to release the memory. Specifically, each event  (carrying transactions) in Jointgraph is broadcast to peer nodes through the gossip protocol. Upon receiving the event from another member, the node will verify its validity (signatures and hash). If all passed, the node votes the event. The finality of this event can be confirmed if it receives more than 2/3 of all the nodes, \textit{one of them must be voted by the supervisory node}. If any conflicts happen, the supervisory node plays the role of a judge to make the final decision. After a certain time of running, two special types of events are issued by the supervisor node, namely the snapshot event and the storage event. These two events are used to take snapshots and store them. Once finishing the permanent storage, the previous memories could be released.

\begin{figure}[H]
\centering
\includegraphics[width=0.27\textwidth]{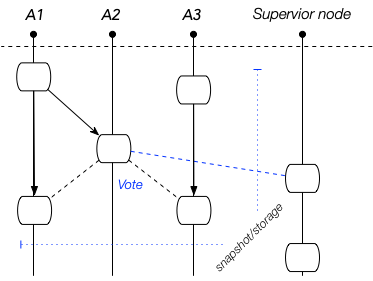}
\caption{Jointgraph}
\label{fig-zJointgraph}
\end{figure}

%-----------------------
\textit{Chainweb} \cite{will2019chainweb} is a permissionless system attempting to scale Nakamoto consensus by maintaining multiple parallel chains. The system has two features. On the one hand, Chainweb uses the cross-reference of hashes to connect parallel chains. Individual chains in the system are based on a PoW consensus that incorporates each Merkle root from others to increase the hash rate. A block needs to reference the header of its ancestor block, and additionally, reference the headers of peers at the same block height. Thus, each block references all peer chains to form a weaved web according to the base graph (Petersen graph, the graph topology of Chainweb). On the other hand, Chainweb employs simple payment verification (SPV) proof to complete token transfers. The procedure follows a similar approach to Nano \cite{lemahieu2018nano}. Each token to be transferred cross-chain has to move under the SPV proofs. The method forces to destroy tokens from one chain and create equal amounts on another chain. This requires a strict assumption that all chains can grow at a synchronous pace, and fast-growing chains need to be periodically stalled and cleared. As a result, SPV proofs of token transfer can only be validated until the blocks are recorded in other chains.

\begin{figure}[H]
\centering
\includegraphics[width=0.26\textwidth]{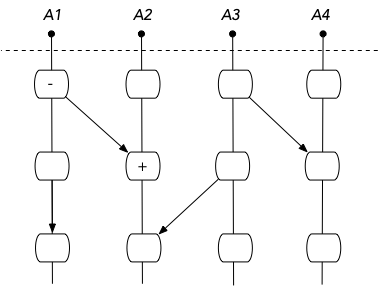}
\caption{Chainweb}
\label{fig-zChainweb}
\end{figure}

\textit{Aleph} \cite{gkagol2019aleph} is a permissioned BFT-style distributed system. The essential idea of Aleph is learned from blockchain systems that adopt the pattern of \textit{Type III}. Specifically, Aleph modifies the classic BFT consensus by removing the role of the leader. Instead, it enables each node to equally and concurrently issue messages, also denoted as \textit{units}. This design makes the units able to be asynchronously and efficiently transmitted in the network. Similar to the aforementioned systems, The units issued by the same node naturally form a chain. The units issued by the different nodes are organized in parallel chains and each of them is independent where they can freely create, disseminate, and vote. The key is to build a collectively total ordering among these units. Aleph utilizes leaderless BFT consensus to achieve the consensus. Firstly, each unit is configured with a round number, and the units with the round number ranging from $0,1, ...,r-1$ to $r$ are collected in a batch. The system sorts the batches in linearization according to their round numbers. Then, the units within batches break ties by their hashes. Next, with a uniform set of units, the system conducts the voting processes for each unit in the set, to verify whether all nodes can see this unit. If collecting more than $2f$ votes, the unit is deemed as confirmed and finishes the consensus procedure.

\begin{figure}[H]
\centering
\includegraphics[width=0.31\textwidth]{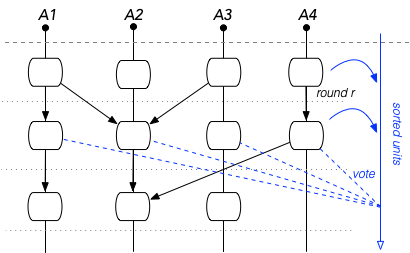}
\caption{Aleph}
\label{fig-zAleph}
\end{figure}

%-----------------------
\textit{Vite} \cite{liu2018vite} follows the basic (lattice) structure of Nano, but introduces a global \textit{snapshot chain} to the achieve the total order sequence. Each account in Vite individually creates transactions in parallel. A complete transaction is split into a trading pair based on an asynchronous \textit{request-response} model, similar to the \textit{send-receive} model in Nano. Transactions can be instantly written into the ledgers without being blocked by the procedure of confirmation. The confirmation is completed by a \textit{snapshot chain}, a storage structure to maintain consistency. Snapshot blocks play the role to store the states (balance, Merkle root, the hash of the last block) of Vite ledgers. The snapshot chain adopts the Nakamoto consensus to structure the blocks linearly. The confidence of confirmation is measured by an accumulated value. The value will increase as the chain grows. If forks happen, the \textit{snapshot chain} breaks the tie by selecting the \textit{longest-chain}. Thus, the transaction is deemed valid in a probabilistic way.

\begin{figure}[H]
\centering
\includegraphics[width=0.27\textwidth]{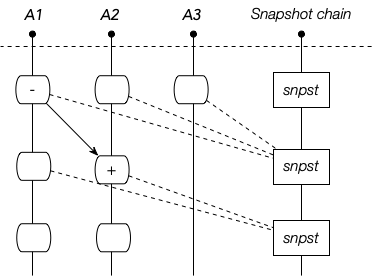}
\caption{Vite}
\label{fig-zvite}
\end{figure}

%%%%Concurrent with and independent of this work, there are two online non-refereed technical reports [5], [15] proposing a similar approach to composing multiple parallel chains. While their high-level designs share similarities with ours, in their designs, one of the k parallel chains is designated as a special chain, and blocks on the other chains are related to blocks on that special chain [5], [15].  More importantly, these two concurrent and independent works [5], [15] do not have any implementation details or experimental evaluation, while we have a prototype implementation as well as large-scale evaluation on Amazon EC2. In particular, our large-scale experiments confirm that propagating many parallel blocks does not negatively impact block propagation delay, which lays the empirical foundation for parallel chain designs.

%-----------------------
\textit{Caper} \cite{amiri2019caper} is a permissoned system with aims to support distributed applications. The system considers both the confidentiality of internal states generated in each application and the interoperability of external states that come from the cross-application transactions. Ledgers in Caper are formed as parallel chains. Deviating from the aforementioned systems, the units are created by different types of applications, rather than nodes. Each application maintains its view of the ledger, including both the internal state and all external states. To achieve the consensus, Caper introduces three consensus protocols by \textit{a separate set of orderers}, \textit{hierarchical consensus}, and \textit{one-level consensus}. These protocols are commonly based on BFT-style consensus mechanisms (CFT, BFT). The mechanism is pluggable and each application can adopt any fitted one according to the specific scenarios. Here, we take the most complex one, hierarchical consensus, as an example. The hierarchical consensus contains two main procedures: a) achieving the local view consensus in each application, and b) reaching a global view consensus across multiple applications. Every application operates its local consensus to internally decide on a state. These states from multiple applications are regarded as the input of the global consensus protocol. The global consensus, then, employs an asynchronous Byzantine fault-tolerant protocol to reach the agreement, where all states are structured in a total linear ordered sequence.

\begin{figure}[H]
\centering
\includegraphics[width=0.33\textwidth]{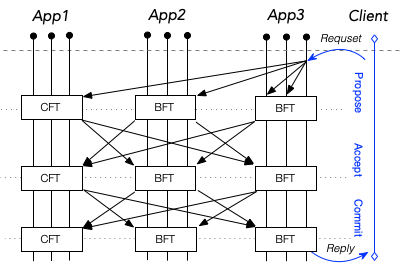}
\caption{Caper}
\label{fig-zCaper}
\end{figure}

%-----------------------
 \textit{Lachesis-class} \cite{Choi2018OPERARA}, denoted by $\mathcal{L}$, is a new family of consensus protocols, designed for distributed networks with features of being asynchronous, leaderless, and resistant against Byzantine fault. This family of protocols integrates a series of algorithms such as peer selection, synchronization, block creation, and consensus. We focus on how to achieve consensus in Lachesis. Generally, the Lachesis has two main steps that corresponded to its hybrid structure.  This structure includes two layers.  The first layer maintains the DAG structure (called the \textit{Opera chain}) to enable the parallel processing, while the second layer keeps the \textit{Mainchain} to achieve a total order of events. The bridge between these two layers is to select a set of key events, called \textit{Atropos}. A general algorithm of the Opera chain takes multiple events (in the DAG network) as inputs and outputs a selected set of Atropos events. Then, the algorithm of the Mainchain inputs these events and outputs a topological ordering of events. Each node in the network will share this sequence. Thus, all nodes have the same view of the network and the protocol completes the consensus. Lachesis instantiates a family of protocols, including  \textit{Fantom} \cite{choi2018fantom}, \textit{Onlay}  \cite{Nguyen2019ONLAYOL}, \textit{StakeDAG} \cite{Nguyen2019StakeDagSC}, and \textit{StairDAG} \cite{Nguyen2019StairDagCV}. The difference between these frameworks is their customized ways to implement the sub-algorithms. More details are shown as follows.

 \begin{itemize}
     \item  \textit{Fantom} \cite{choi2018fantom}  adopts the leaderless BFT mechanisms to select the key set. In the Opera chain, firstly, several events become the \textit{root} events once collecting more than the threshold (2/3) references of previous roots. Next, a root event becomes the \textit{Atropos} event if more than the threshold (2/3) participated nodes agree on this root event. Last, these Atropos events form the \textit{Mainchain}, a core subgraph of the Opera chain with aims to order the events in the graph. Each Atropos events are created by adding several parameters as their identities including \textit{frame}, \textit{consensus time} and \textit{Lamport timestamp}. In the \textit{Mainchain}, the sorting algorithm is determined based on the allocated identities of each Atropos event, and it follows the priorities of a) the frame value of each Atropos; b) the consensus time when multiple Atropos events coexist at the same frame; c) the Lamport timestamp when more than one Atropos having any of the same consensus time on the same frame; d) the hash value when above conditions stay same.

     \item  \textit{Onlay} \cite{Nguyen2019ONLAYOL} employs an online layering algorithm to achieve leaderless BFT mechanisms. This layering provides a better structure of DAG to select the key set. In Onlay, the protocol introduces the H-OPERA chain, built on top of the OPERA chain. To compute the H-OPERA chain, the protocol applies a layering algorithm, such as LongestPathLayer (LPL), to the OPERA chain. The algorithm is inherently a list-scheduling algorithm that produces a hierarchical graph with the smallest possible height. A group of Atropos events is, as a consequence, selected to be the key set. As for the \textit{Mainchain}, the protocol orders the Atropos vertices based on priorities of a) their layer, b) Lamport timestamp, and c) hash information of the event blocks.

     \item  \textit{StakeDAG} \cite{Nguyen2019StakeDagSC} adopts the PoS mechanism to select the key set, instead of BFT-style protocols. Three steps are included in the Opera chain: a) initialize accounts with stakes, b) compute the validation score of an event, and c) assign weights to new roots. Correspondingly, confirming whether some block is a root in StakeDag is different from that in Fantom: StakeDag requires more than 2/3 of the validating power (total stakes) while Fantom requires more than 2/3 of the total number of nodes.  The key set is obtained by such a procedure. The following sorting algorithm in the \textit{Mainchain} is similar to which in Onlay.

     \item  \textit{StairDAG} \cite{Nguyen2019StairDagCV} builds on top of StakeDAG with the difference to distinguish participants into  \textit{validators}  (high stake) and \textit{users}  (low stake) by their stake. Validators can expose more validating power to complete the onchain validation for faster consensus, while users can participant in the system as observers or monitors to retrieve DAG for post-validation. Both of them create and validate event blocks and maintain a DAG network. Similar to calculating the stakes in StakeDAG, each event in StairDAG is associated with a validation score, which is measured by the total weights of the roots reachable from it. The key set is obtained when events have been validated by more than two-thirds of the total validating power in the OPERA chain. After that, the sorting algorithm is launched in the same way as StakeDAG.

 \end{itemize}

\begin{figure}[H]
\centering
\includegraphics[width=0.44\textwidth]{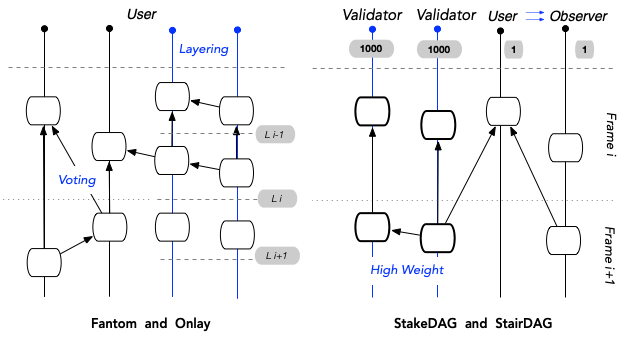}
\caption{Lachesis-class Protocols}
\label{fig-zLachesis}
\end{figure}

%-----------------------
%-----------------------
\noindent\textbf{Consensus on Type IV.}  Type IV systems are block-based structures. Blocks are maintained by individual nodes. This part includes Prism \cite{bagaria2019prism},  OHIE \cite{yu2020ohie}, Sphinx \cite{wang2021weak}, Blockmania \cite{danezis2018blockmania},  Blockclique \cite{forestier2018blockclique}, Eunomia \cite{niu2019eunomia}, Dexon \cite{chen2018dexon}, and PARSEC \cite{chevalier2019protocol}.

%%%%%%%%%%%%%%%%%%%
%%%%%%%%%%%%%%%%%%%
%%%%%%%%%%%%%%%%%%%

%-----------------------
\textit{Prism} \cite{bagaria2019prism} is designed in a similar parallel chain approach. Prism decouples the functionalities of Nakamoto consensus into transaction proposing, validation and confirmation, and utilizes three types of blocks to take these functionalities. Blocks are hence divided into \textit{transaction block},  \textit{voter block}, and \textit{proposer block}. Transaction blocks only generate and carry transactions, acting as the fruit in FruitChain \cite{pass2017fruitchains}. Voter blocks are used to vote for the proposer blocks and specify a leader block according to their heights. Proposer blocks will pack these transaction blocks and extends the chain under the \textit{longest-chain rule}. The leader (winning miner) confirms the integrity and validity of transactions to form the final ledger. These three types are relatively independent but as well as interacted with each other. Miners will simultaneously work on one transaction block, one proposer block, and multiple voter blocks. A block to be allocated with which type of functionalities depends on a random hash value when the block is successfully mined. In Prism, transactions are confirmed before being ordered, and unrelated ones are simultaneously processed. Meanwhile, a leader block will not be stopped by waiting for its voters becoming irreversible since reverting a majority of voter chains in a short time interval spends much more computing power than reverting one voter chain as in the Nakamoto consensus. Thus, Prism, armed with its deconstruction approach, separately scales each functionality to its physical limits for maximum optimization.

\begin{figure}[H]
\centering
\includegraphics[width=0.34\textwidth]{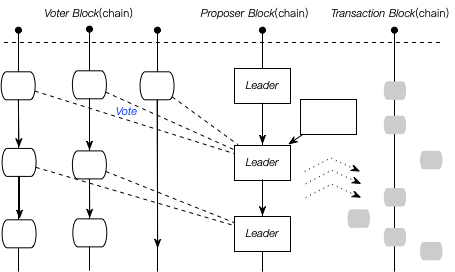}
\caption{Prism}
\label{fig-zprism}
\end{figure}

%-----------------------
\textit{OHIE} \cite{yu2020ohie} shares similarities in composing multiple parallel chains. OHIE is a permissionless blockchain system where chains are equivalent and symmetric. The system operates the standard Nakamoto consensus on each individual chain and allocates each block into these chains according to their hash values. Then, OHIE deterministically sorts the blocks to obtain a global order across parallel chains. To achieve that, the system defines a special tuple embedded in each block, denoted as (\texttt{Rank},\texttt{NextRank}). The field of $\texttt{Rank}$ represents the current index of a block located on each individual chain, while the field of $\texttt{NextRank}$ specifies a reference to the next block. Here, \texttt{NextRank} is used to balance the length of parallel chains in case of a huge gap between them. It is usually pointed to the block with a high $\texttt{Rank}$. 
Blocks with ranks higher than the bar are deemed as totally confirmed, otherwise are partially confirmed. In fact, the tuple together with the chain index inherently locates a block in the network, and their positions assist to sort blocks in linearization. Generated blocks across parallel chains are sorted by $\texttt{Rank}$s with a tie-breaking of $\textit{chain index}$. Specifically, the block with a smaller $\texttt{Rank}$ number will be arranged in front of a block with a higher $\texttt{Rank}$. This is the first priority. Then, if several blocks with the same $\texttt{Rank}$ number, the block holding a smaller $\textit{chain index}$ will take the advance. Thus, a total linear ordering achieves by adjusting the special $\texttt{Rank}$ fields.

\begin{figure}[H]
\centering
\includegraphics[width=0.32\textwidth]{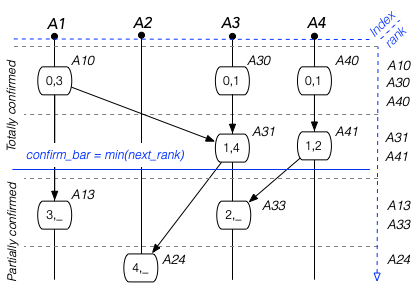}
\caption{OHIE}
\label{fig-zohie}
\end{figure}

%%%% When mining a block in OHIE, the miner needs to include the parent block hash of the current block in each chain. Once a valid proof-of-work puzzle is solved, the block hash determines which chain the new block belongs to. The parent blocks for each chain are selected following Nakamoto consensus. Each individual chain has a low block generation rate to match the security requirement in Nakamoto consensus. The parallel chain remains a low block generation rate for security and all the chains achieve a high throughput collaboratively. However, such design increases the cost in metadata extremely. In order to achieve desirable performance, OHIE runs 640 parallel chains and generates 64 blocks per second. In security analysis of OHIE, a block will be confirmed in OHIE only if it is confirmed in the individual chain it belongs to. So its confirmation time is worse than the Nakamoto consensus.

%-----------------------
\textit{Sphinx} \cite{wang2021weak} introduces a weak consensus mechanism to relax the requirement of strict consistency in most distributed systems. Strict consistency indicates that only one block can be confirmed as valid in each round, where most competitive blocks are afterward abandoned. In contrast, weak consensus aims to weaken such requirements by allowing a loose consistency: only the relative positions between two blocks are guaranteed, no matter how many blocks are inserted between them. For instance, in the following figure, the proposed mechanism only ensures the sequence of (B1-> B2 ->B3) in peer nodes. It is also correct when inserting other blocks between these blocks like (B2 -> A3 ->B3) in A or (B2 -> D1 ->B3) in D. Weak consensus is a variant of PBFT protocol, changing the leader-based protocol to a leaderless protocol. All the committee nodes maintain their chains and independently process received transactions. From a system perspective, the transactions are executed in parallel and the performance and scalability are thus improved. Based on this concept, the authors apply the weak consensus to a blockchain system called \textit{Sphinx}. The implementation contains other underlying components covering the P2P network, account, incentive model, \textit{etc}.

\begin{figure}[H]
\centering
\includegraphics[width=0.4\textwidth]{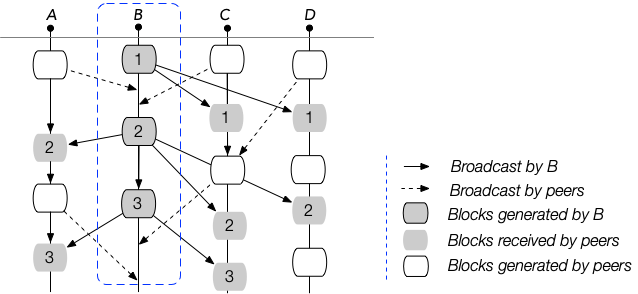}
\caption{Sphinx}
\label{fig-zsphinx}
\end{figure}

\begin{figure}[H]
\centering
\includegraphics[width=0.35\textwidth]{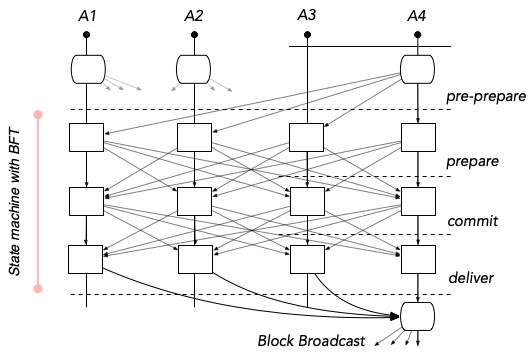}
\caption{Blockmania}
\label{fig-zBlockmania}
\end{figure}

%-----------------------
\textit{Blockmania} \cite{danezis2018blockmania} is a BFT-style protocol which mainly bases on PBFT \cite{CastroL99}. Blockmania simplifies the PBFT protocol by a) deciding on a position rather than a sequence number. The position of a block is denoted as $ B_{n, k}$, where $n$ is the index of nodes and $k$ is the height of blocks emitted by the node; b) becoming leaderless which is egalitarian for involved nodes; and c) removing several complex procedures like ``checkpointing''.  Nodes in Blockmania separately create blocks by themselves and jointly form the DAG network. Here, each node can only propose one block for the given position without any equivocation. Then, the modified PBFT consensus is executed by these nodes in parallel and a decision is required to execute a classic three-stage (\textit{pre-prepare}, \textit{prepare}, \textit{commit}).  Besides, an incentive mechanism, as another essential factor, is introduced to sort transactions in the total order. To solve potential conflicts, each transaction included in a block will be associated with both the height of the block $k$ and deposited fee $\phi$. Clients who expect a priority of confirmation can send transactions in earlier rounds or spend higher fees. Upon reaching a decision from sufficient blocks (2/3 threshold), transactions are sorted across parallel nodes according to the tuples $(k,-\phi)$. Alternatively, fees-based tiebreakers may also be replaced by a traditional way --- by the value of their hashes, as used in Hashcash \cite{back2002hashcash}. Note that, the formed DAG system is interpreted as a state machine that contains the blocks. Each state machine carrying the interpreted information  (view number, a count for prepare and commit messages, stored in square blocks in Fig.\ref{fig-zBlockmania}) will be sent and received through blocks. In real network communication, only the blocks are broadcast to peers.

%The goal of the Blockmania protocol is to ultimately ensure that all honest nodes arrive at the same ordering of transactions (safety) despite a subset of nodes being byzantine; and for the protocol to finalize an ordering of transactions from5 honest nodes, despite the actions of byzantine nodes (liveness)

%-----------------------
\textit{Blockclique} \cite{forestier2018blockclique} shares a similar design of the parallel-chain model, though it names with \textit{multi-thread}. The system is a permissoned protocol that adopts the \textit{proof of ownership of a resource} (PoR, including PoS, PoW, \textit{etc.}) to select its committee members. The selected members take responsibility for creating blocks. Then, Blockclique allocates the received transactions separately into different chains according to their hashes. The transaction sharding avoids incompatibility when blocks are produced in parallel. A block in Blockclique can only pack the transactions with input addresses assigned to this thread, whereas the outputs can point to any thread. Thus, the links between blocks make different threads connected. Meanwhile, Blockclique embeds a scalar \textit{fitness} value for each block to measure the required resources. The fitness is calculated by the total number of involved addresses during the block creation. To achieve the consensus, Blockclique identify two types of incompatibility cases, namely \textit{transaction incompatibility} and \textit{grandparent incompatibility}. Transaction incompatibility represents the conflict transactions that reference the same parent in one thread, while \textit{grandparent incompatibility} means the grandparent references of two blocks are reversed across different threads. Then, avoiding incompatible cases, the system recursively searches for the maximal clique of compatible blocks according to total fitness values contained in blocks. If two cliques hold the same fitness values, the clique with the smallest sum of block hashes is preferred.

\begin{figure}[H]
\centering
\includegraphics[width=0.27\textwidth]{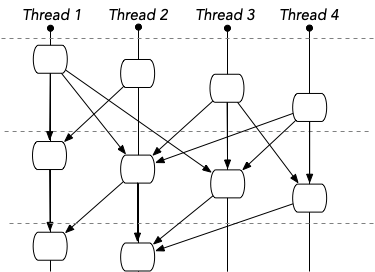}
\caption{Blockclique}
\label{fig-zBlockclique}
\end{figure}

%-----------------------
\textit{Eunomia} \cite{niu2019eunomia} is a permissionless system with multiple parallel chains. The system achieves the consensus by the following steps. Firstly, to avoid the conflicts of transactions, Eunomia utilizes a fine-grained UTXO sharding to avoid double-spending. A transaction can only take the UTXOs with the same sharding index as input and generate the output UTXOs with random indexes ranging from $0$ to $m-1$. Then, to package these cross-reference transactions, miners adopt the \textit{m-for-1} PoW mechanism \cite{garay2015bitcoin}\cite{fitzi2018parallel} where they can simultaneously create blocks on $m$ chains, but a block only extends one chain for each time. Generated blocks across $m$ chains are sorted by their logical clocks with a tie-breaking of chain index. Specifically, each chain in Eunomia is identified by the chain index $i$, and the blocks in individual chains are measured in a \textit{virtual logical clock} $v$, as a counter for increments.  Since each chain maintains a local state of the clock, nodes need to involve a hash reference of the updated blocks to synchronize clocks across different chains. The block with the largest $v$ can be recognized in priority. Here, $v$ indicates the epochs to provide separated time slots as mentioned in other protocols \cite{li2020decentralized}\cite{gupta2019cdag}. Thus, the ordering algorithm based on such positions $(i,v)$ sorts the blocks in the priorities of a) the virtual logical clock $v$ and b) the chain index $i$.

\begin{figure}[H]
\centering
\includegraphics[width=0.29\textwidth]{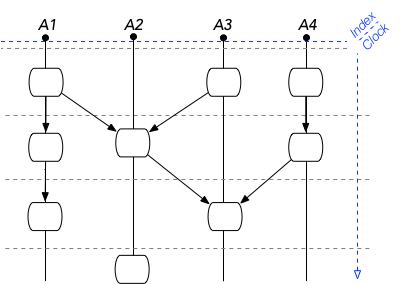}
\caption{Eunomia}
\label{fig-zEunomia}
\end{figure}

%-----------------------
\textit{Dexon} \cite{chen2018dexon} is a permissioned system. The system is maintained by two special sets of nodes: the \textit{CRS set} to generate public randomness and the \textit{notary set} used to propose blocks.  Dexon confirms each parallel chain through the reference field called \texttt{ack}, and determines the global order through these references. Specifically, the consensus of Dexon consists of four steps: a) To avoid the waste of space utility of blocks, Dexon makes each transaction only be packed into one single chain according to the residue of its hash value. This design acts as a \textit{load balance} to schedule the incoming transactions to the individual chains. b) Upon receiving transactions, every single chain individually generates blocks and achieves consensus by employing the technique of a modified Algorand \cite{gilad2017algorand} (mainly VRF) scheme. Nodes in CRS  set periodically generate updated common random numbers to maintain the variations in each epoch. c) The involved chains form the blocklattice, and each node executes the \textit{total ordering algorithm} with this blocklattice as input. The output is a globally-ordered compaction chain with all blocks sorted in linearization. d) The compaction chain applies the \textit{timestamping algorithm} to compute the consensus time for each block. Then, the consensus timestamp, the height of blocks, and the threshold signature together determine a unique block in the compaction chain. Thus, the validity of the block can be further verified by the following blocks.

%Then, it proposes a sorting mechanism to determine the total order of blocks across parallel chains. The sorting scheme is event-driven, which is triggered only when chains receives new blocks. To avoid the waste of space utility of blocks, Dexon utilizes the residue of its hash value, making a transaction only be packed into one single chain. However, the removed redundancy mechanism sacrifices the security of system. If the specified chain is crashed, the history of transactions are disappeared. 

\begin{figure}[H]
\centering
\includegraphics[width=0.29\textwidth]{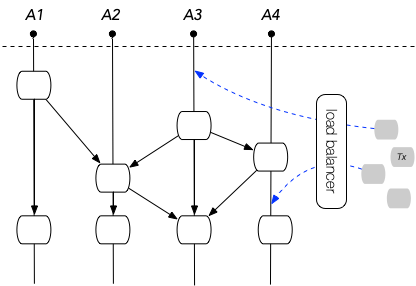}
\caption{Dexon}
\label{fig-zDexon}
\end{figure}

%-----------------------
\textit{PARSEC} \cite{chevalier2019protocol} is a permissioned protocol with the creation of an asynchronous BFT mechanism that is resilient to 1/3 byzantine nodes. The protocol follows the main concepts from Hashgraph \cite{baird2016swirlds}, where involved nodes spread their events via the gossip algorithm and the consensus of these events is completed by virtual voting. PARSEC deviates from Hashgraph in that PARSEC additionally defines a new type of data structure called \textit{stable block}. The block inputs a selected set of events after the virtual voting in the graph and sorts them in linearization. Then, it finds the next valid block according to the votes from events. The stable blocks record an order of events that all nodes agree upon, which can also be regarded as periodical checkpoints to confirm the states of the network. The consensus of PARSEC requires two steps: \textit{obtain a set of valid events} from its graph network and \textit{sort the selected events and find the next block}. For the first step, we narrow the focus down to its graph network. In PARSEC, the nodes execute the virtual voting algorithm to decide the qualification of events. The algorithm has two similar procedures as Hashgraph, namely \textit{see} and \textit{strongly see}, to collect the votes (in the form of binary values). The event that receives more than the supermajority  (2/3 of total nodes) is considered to be qualified in the next step. Here,  the virtual voting elects a set of events agreed upon by all the nodes as a temporary agreement. Next, for the second step, we focus on the operation surrounding blocks. The key point is to achieve a consensus on both selected events and the next block without any conflicts. The algorithm considers the following three aspects: a) multiple events in the same individual chain are sorted according to their appearance, measured by the embedded counter;  b) conflicting blocks seen by multiple events are elected by their collected votes; and c) the tiebreaker is based on the lexicographical sequence.

\begin{figure}[H]
\centering
\includegraphics[width=0.26\textwidth]{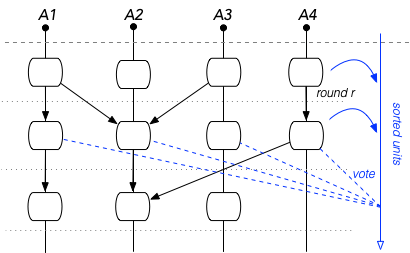}
\caption{PARSEC}
\label{fig-zPARSEC}
\end{figure}

%-----------------------
%-----------------------
\noindent\textbf{Consensus on Type V.}  Type V systems are blockless. Appended transactions gradually converge into the main chain. This part includes the systems of Byteball \cite{churyumov2016byteball}, Haootia \cite{tang2020haootia} and \textit{JHdag} \cite{he2019consensus}.

\textit{Byteball} \cite{churyumov2016byteball} is a permissionless network. Byteball organizes units (transactions) in the graph topology but later forms a main-chain with the help of trustful and reputable witness nodes. These nodes distinguish common nodes by periodically generating witness units. Each unit is marked with an index called Main Chain Index (MCI) that links to a witnessed unit. Conflicting states are resolved by MCI, where the unit with the lower MCI is deemed valid and the higher one fails. If both nonserials hold the same MCI, a tiebreaker rule is applied that the unit with the lower hash value (as represented in base64 encoding) is deemed valid. In Byteball, a total of 12 witnesses are selected to protect against the occasional failures. Witnesses can be replaced by common nodes, who change with better candidates in his list. But the changes happen only gradually since the majority of users are required to achieve an agreement on a new candidate.

\begin{figure}[H]
\centering
\includegraphics[width=0.24\textwidth]{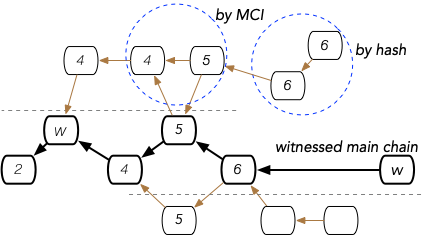}
\caption{Byteball}
\label{fig-zbyteball}
\end{figure}

%-----------------------
\textit{Haootia} \cite{tang2020haootia} is a permissioned network that consists of three types of roles, \textit{transaction proposers}, \textit{miners}, and \textit{committee members}. Transaction proposers can only send/synchronize transactions; miners can dynamically become the committee members by solving the hash puzzles; committee members are able to participate in its (intra-committee) PBFT consensus process to decide the blocks. In Haootia, a two-layer consensus framework is induced. The first layer embraces generated transactions and organizes them in a naive graph topology. The second layer is a PoW-based backbone chain with key blocks to decide the total order of transactions. PBFT consensus runs on the backbone chains to achieve linearization by directly sorting key blocks, where key blocks confirm the trees of attached transactions (denoted as \textit{increment tree}). Each transaction in the tree may reference multiple ancestors. Only one reference with the smallest lexicographic order could be saved and other parental references are erased. Then, through concatenating the reversed breath-first traverse (RBF-traverse) sequence of the increment trees of key nodes, Haootia achieves an append-only total ordering.

\begin{figure}[H]
\centering
\includegraphics[width=0.3\textwidth]{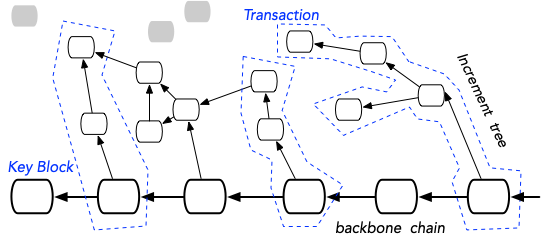}
\caption{Haootia}
\label{fig-zhaootia}
\end{figure}

%-----------------------
\textit{JHdag} \cite{he2019consensus} is a permissionless system that applies Nakamoto consensus to the DAG-based protocol. Instead of packing a batch of transactions, each block only contains one transaction to limit the block size and simplify the cryptographic puzzles of PoW. Fast confirmation of blocks is achieved without having to wait for peers (so that we categorize this system in $1^{od}$). In JHdag, two types of blocks exist: \textit{common block} for carrying transactions and \textit{milestone block} for making decisions. Becoming which types of blocks depends on their hashes of PoW, where a common block needs 10 consecutive bits of 0 and a milestone block requires 15 consecutive bits. Then, to create a block, each miner specifies three pointers: a) points to the miner's previous block to form a peer chain representing the state of that miner; b) points to the previous milestone to form the main chain under the longest chain rule, and c) points to another miner's common block to enhance the connectivity for peer chains. Each milestone block can verify multiple common blocks surrounding themselves, and such a clique is denoted as a \textit{level set}. The level set essentially acts as the role of a (Nakamoto) block in Bitcoin, since all blocks in the system are directly or indirectly confirmed by the longest milestone chain. Therefore, the consensus of JHdag inherits similar principles and properties as Bitcoin.

\begin{figure}[H]
\centering
\includegraphics[width=0.35\textwidth]{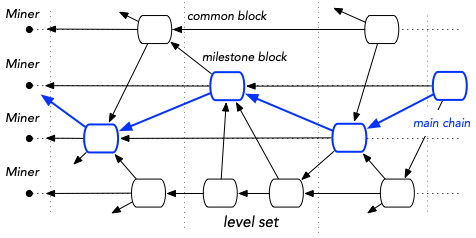}
\caption{JHdag}
\label{fig-zJHdag}
\end{figure}

%-----------------------
%-----------------------
\noindent\textbf{Consensus on Type VI.}  Type VI systems are based on blocks. These blocks gradually converge into the main chain. This part includes protocols and systems of GHOST \cite{sompolinsky2015secure}, Inclusive \cite{lewenberg2015inclusive}, 3D-DAG\cite{wang20213ddag}, Conflux \cite{li2020decentralized}, CDAG \cite{gupta2019cdag} and StreamNet \cite{yin2019streamnet}.

%-----------------------
\textit{GHOST}, Greedy Heaviest Observed Sub-Tree \cite{sompolinsky2013accelerating}\cite{sompolinsky2015secure}, is a permissionless network with two types of entities, \textit{common node} and \textit{miner}. Each common node in the network can compete for the rights of packaging blocks, and the winner becomes a miner by successfully resolving a hash puzzle as Bitcoin does. GHOST introduces the \textit{weight} to measure the number of blocks in the subtree attached to each block. The difference focuses on the extension rule: the chain grows by recursively selecting the heaviest (weight) sub-tree, instead of the longest chain. Here, a sub-tree is formed by blocks rooted in the same ancestors. In this scenario, DAG topology is converged to one main chain among several growing sub-trees.

\textit{Inclusive} protocol \cite{lewenberg2015inclusive} is an variant of GHOST, which allows one block to reference multiple ancestors. Only one of the referenced blocks with the same height can be elected as the \textit{father}, earning most of the rewards, while others are denoted as \textit{uncle}, sharing the rest of the rewards. The ancestors are considered within the order of seven generations of the chain. This design makes off-chain transactions involved in the protocol and incentive miners continuously contribute to the network. Ethereum (Casper) adopts this protocol in its implementation.

\begin{figure}[H]
\centering
\includegraphics[width=0.29\textwidth]{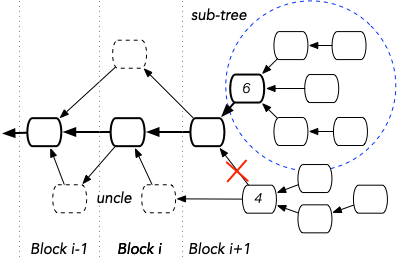}
\caption{GHOST and Inclusive Protocols}
\label{fig-zghost}
\end{figure}

%-----------------------
\textit{3D-DAG} \cite{wang20213ddag}\cite{wang2019improving} establishes a hybrid architecture that combines multiple sidechains for parallel processing and a mainchain for linearization ordering. The sidechains are based on the natural DAG topology where each transaction can reference more than one previous transaction. In this stage, the transactions are validated by not sorted. The ``validated'' means the token has been permanently spent, but unrelated transactions cannot be compared. The ``sorted'' means all transactions are linear sequences. On-top applications like the smart contract can thus be supported. The design of the sidechain partially follows the model of IOTA \cite{popov2016tangle} but modifies with additional functions. One of the significant differences is that every transaction has to vote for a valid member in the main-chain. The voting procedure utilizes the VRF technique to hide the identity of the elected leader. After that, the leader in the mainchain launches the procedure of sorting transactions and packing blocks. Once the block is broadcast and agreed upon by enough committee members, the ledger achieves finality.

\begin{figure}[H]
\centering
\includegraphics[width=0.35\textwidth]{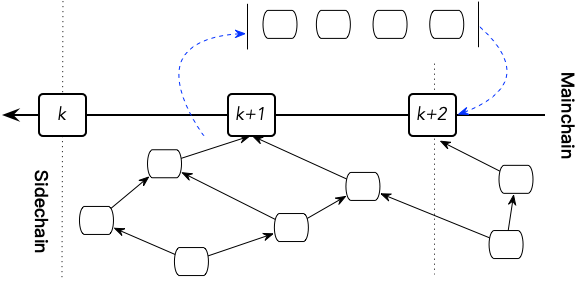}
\caption{3D-DAG}
\label{fig-z3ddag}
\end{figure}

%-----------------------
\textit{Conflux} \cite{li2018scaling}\cite{li2020decentralized} is a permissionless system attempting to scale NC-based blockchain for fast confirmation and high throughput. The system achieves the goal by proposing its improved protocol of GHOST, named \textit{GHAST} (Greedy Heaviest Adaptive Sub-Tree, \cite{li2020decentralized}).  The design introduces an adaptive weight mechanism, which enables Conflux to switch between two strategies: an optimistic strategy (similar to GHOST) for fast confirmation and a conservative strategy to resist the liveness attack. Specifically, each block in Conflux has one \textit{parent} edge to form a pivot chain and multiple \textit{reference} edges to operate concurrent blocks. A newly generated block will be adaptively assigned with weights according to its past sub-graph (\textit{a.k.a}, Tree-Graph). If all ancestors in the pivot chain are secured (with a low probability of being reversed), the system assigns 1 to the new block; otherwise, the system assigns a block with weight $h$ within $1/h$ chance ($h$ is a protocol parameter configured as 600 \cite{li2020decentralized}).  Blocks with their weights are organized in several sub-trees, and the pivot chain grows by selecting the heaviest sub-tree among them. Every block on the pivot chain is marked with one epoch. Then, Conflux derives a total order of blocks. The priority of sorting blocks follows the metrics: a) their epochs, b) topological order, and c) PoW quality or block hash. Last, the priority of sorting transactions follows a) the orders of enclosed blocks, and b) the time of the appearance in the same block.

\begin{figure}[H]
\centering
\includegraphics[width=0.32\textwidth]{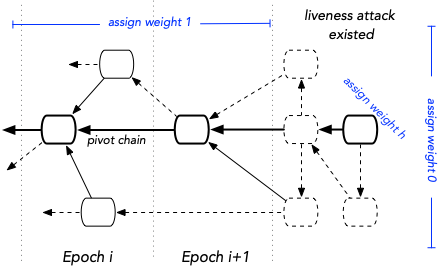}
\caption{Conflux with its consensus GHAST}
\label{fig-zghast}
\end{figure}

%-----------------------
\textit{CDAG} (Converging Directed Acyclic Graph) \cite{gupta2019cdag} is a permissioned network consisting of two types of entities, namely  \textit{Colosseum} and \textit{block proposer}. Colosseum is designed for permissioned blockchains where the identities of participants are known. Members of Colosseum compete to become the block proposer in a fair and randomized two-player game for each time slot (contain $log_2 N$ where $N$ is the number of participants asynchronous rounds). These proposers partition the transactions into non-intersecting buckets and then select a random bucket to generate blocks and disseminate them into the network. The generated blocks are required to be consistent with a \textit{Converging Block} (C-Block), which is a collection of blocks proposed in a time slot and works as a single point of reference for the next set of blocks. C-Block in each slot enables the system to progress as a chain and maintains a total ordering among blocks. Temporary forks that may frequently occur in the formed chain are solved by using a variant of the heaviest chain selection approach \cite{sompolinsky2015secure}.

\begin{figure}[H]
\centering
\includegraphics[width=0.28\textwidth]{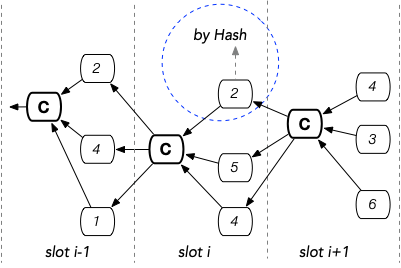}
\caption{CDAG}
\label{fig-zcdag}
\end{figure}

%C-Block (Converging block) is the converging point in CDAG. It is introduced as an extra layer between two layers of blocks providing a point of convergence to the blocks of the previous slot and a single point of reference for all the blocks of the current slot.

%-----------------------
\textit{StreamNet} \cite{yin2019streamnet} adopts the concept from Conflux \cite{li2018scaling} and IOTA \cite{popov2016tangle} to form a permissionless network with a main chain. The system combines the rule of tip selection from IOTA (two ancestors are approved by one transaction) and the structure from Conflux (a pivot chain is required to sort blocks). Specifically, StreamNet consists of multiple \textit{machines} where each machine locally creates the block and instantly broadcasts the blocks through a gossip protocol. To achieve the consensus, a new attached block will approve two ancestor blocks: one is deterministic, and the other is random. The first selected ancestor is deemed as the \textit{parent} block to construct a pivot chain for the purpose of total ordering. The algorithm recursively advances to the previous child block with the highest \textit{score} child of the sub-tree. In case of the same score between two child blocks, the one who has the largest hash value wins. The second ancestor is randomly selected through the technique of Markov Chain Monte Carlo (MCMC) to scale out. The random walk starts from the genesis to the latest attached blocks. Next, a total ordering algorithm is conducted recursively to order the blocks in previous epochs. The algorithm breaks ties by comparing the hash of each block. Upon every machine reaching a unified view of DAG, the globally total ordering is completed.

\begin{figure}[H]
\centering
\includegraphics[width=0.28\textwidth]{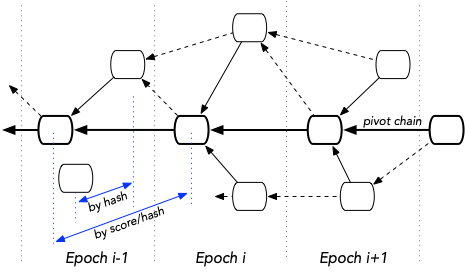}
\caption{StreamNet}
\label{fig-zStreamNet}
\end{figure}

\begin{comment}
\textcolor{blue}{HotDAG \cite{zhou2020hotdag}} %%%%%%%
\textcolor{blue}{C-DAG \cite{zhang2020c}} %%%%%%%
\textcolor{blue}{MAdag\cite{abram2020democratising} } %%%%%%%
\cite{lamtzidis2019novel}
\end{comment}

%======================================
%======================================
\subsection{Analysis} 

\underline{Table.\ref{tab-consensus}} provides the summary of aforementioned systems. This part provides discussions focusing on techniques and algorithms.

%-----------------------
\smallskip
\noindent\textbf{Techniques.} We abstract technical features in each system, and some of them are frequently adopted. Details are shown as follows.

\begin{itemize}
\item[$\diamond$]   \textbf{\textit{Cross-referencing}} enables chains ``twisted'' together for improved throughput, high scalability, or low confirmation times.  This technique enables the systems to involve orphaned units and increase the in-degree and out-degree of a single unit. As shown in the \underline{Table.\ref{tab-consensus}}, we use $(x,y)$ to show that each unit can reference multiple ancestors, as well as be referenced by multiple successors, where $x$ and $y$  are flexible positive integers larger than $1$.  Several systems explicitly define the rules of in/out-degree, such as $(x,2)$ in IOTA, Hahsgraph \cite{baird2016swirlds} and StreamNet \cite{yin2019streamnet}.  Furthermore, the references might be allocated with different roles, which generally includes the \textit{parent} edge and \textit{reference} edges.  This design is adopted by Conflux \cite{li2020decentralized} and StreamNet \cite{yin2019streamnet} to form the main chain, or by Chainweb \cite{will2019chainweb} and OHIE \cite{yu2020ohie} to form parallel chains.

\item[$\diamond$]  \textbf{\textit{Trusted authority}} represents the powerful authority who makes the final decision. The authority can be instantiated as the \textit{leader} in PBFT \cite{CastroL99}, the \textit{coordinator} in IOTA \cite{popov2016tangle}, the \textit{snapshot chain} in Vite \cite{liu2018vite}, the \textit{supervisory nodes} in Jointgraph \cite{xiang2019jointgraph}, the \textit{representatives} in Nano \cite{lemahieu2018nano}, the \textit{proposer block} in Prism \cite{bagaria2019prism}, and the \textit{witness nodes} in Byteball \cite{churyumov2016byteball}. Trusted authorities can either directly affect the consensus or indirectly solve the conflicts. This technique speeds up the confirmation of units but sacrifices the decentralization, bringing additional risks. The adversary may launch several types of attacks against these centralized parties.

\item[$\diamond$]  \textbf{\textit{Pairwise vote}} is a $2$-for-$1$ voting selection, rather than $n$-for-$1$ in normal voting algorithms. Each unit according to its knowledge decides one element has a high priority than another. The procedure only guarantees partial consistency. This technique is adopted by Spectre \cite{sompolinsky2016spectre} as the consensus, as well as by several systems to solve the conflicts when sorting the units. Additionally, the pairwise vote differs from the \textit{coin-flip}  \cite{tanana2019avalanche} in their randomness. Pairwise vote relies on the willingness and knowledge of the node, whereas coin-flip is based on a mathematically random selection with even distribution between two values.

\item[$\diamond$]  \textbf{\textit{Transaction sharding}} is to allocate the transactions into different chains,  in advance, to prevent potential duplication and conflict in sorting procedures. The technique is adopted by the models of parallel chains, including Blockclique \cite{forestier2018blockclique}, Eunomia \cite{niu2019eunomia}, and Dexon \cite{chen2018dexon}.  In general cases, the allocation follows the rules according to the residue of the transaction hash. Transaction sharding relies on an implicit assumption that each transaction will be only processed by one miner/validator in the network.

\item[$\diamond$]  \textbf{\textit{The PoW mechanism}} in several DAG systems, \textit{e.g.}, IOTA \cite{popov2016tangle}, Nano \cite{lemahieu2018nano}, is simply used as an anti-spam tool that can be computed within seconds. PoW for subsequent blocks is pre-computed once the transaction is sent. The design enables transactions instantaneous to an end-user.

\end{itemize}

%-----------------------
\smallskip
\noindent\textbf{Algorithms.} We further discuss the consensus with the following types. This provides a general perspective when designing and analyzing DAG-based blockchain systems. 

\begin{itemize}
\item[$\diamond$]  \textbf{\textit{Tip selection algorithm}} is a set of rules specifying how an ancestor transaction selects its subsequent transactions, or how a new transaction attaches to the ancestors. The algorithm is widely adopted by the systems in topology $\widehat{D}$, such as IOTA \cite{popov2016tangle}, Graphchain \cite{boyen2018graphchain}, and StreamNet \cite{yin2019streamnet}. The advantage of applying TSA in DAG systems is to replace complex consensus mechanisms with the simplest rules. This design enables higher throughput and better scalability, and tolerable resistance against forks. But as sacrifices, they reduce security and strict consistency to some degree.

\item[$\diamond$] \textbf{\textit{Recursive algorithm}} represents a family of algorithms that makes each round's outputs progressively approach stable values. Recursion obtains the result by using functions that call themselves several times. This design is widely adopted by consensus mechanisms to enable the disordered units convergence into a total sequenced chain. The systems can extend the graph in a determined way. This family of algorithms include the \textit{recursive transverse algorithm} in Spectre \cite{sompolinsky2016spectre}/Blockclique \cite{forestier2018blockclique}, \textit{greedy algorithm} in Phantom \cite{sompolinsky2020phantom}, and \textit{sampling} algorithm in Avalanche \cite{rocket2019scalable}.

\item[$\diamond$] \textbf{\textit{BFT-style consensus}} in DAG systems can be classified into three types. The first is to directly apply the classic BFT consensus protocols (PBFT, BFT, CFT) into the system. It requires selecting a committee in advance to operate the consensus. The committee members are selected according to their owned resources (PoW, PoS).  The approach is adopted by the systems of Blockmania \cite{danezis2018blockmania}, Dexon \cite{chen2018dexon}, and Haootia \cite{tang2020haootia}. The second type is modifying the protocols to fit parallel chains. It is acknowledged as \textit{async-BA},  and sometimes known as \textit{leaderless BFT} protocols. The key design of this type is to remove the single-leader related phrases from the protocols. Each chain can freely produce and broadcast blocks and vote for anyone it favors. Any block that collects enough (2/3 of the total) votes is deemed as confirmed. Since blocks in each chain get confirmed and confirm peers in asynchronization, it is difficult to achieve a global linear order. Additional techniques are required, such as \textit{gossip by gossip} in Hashgraph \cite{baird2016swirlds}, or the \textit{sorting algorithm} in PARSEC \cite{chevalier2019protocol}, Fantom \cite{choi2018fantom}. The third type is integrating the former two types to form a \textit{hierarchical} protocol such as Caper \cite{amiri2019caper}. Classic BFT protocols are first applied to each separated zone, and async-BA is adopted by an upper-layer protocol to achieve the final consensus across zones.

\item[$\diamond$] \textbf{\textit{Nakamoto consensus and its variants}} are the most prevailing consensus approaches in current blockchain systems. Classic NC selects the longest sub-chain whilst the variant NC selects the highest weighted sub-chain. NC and variant-NC are adopted by the systems that can form a pivot (main) chain, such as GHOST \cite{sompolinsky2015secure}, Inclusive \cite{lewenberg2015inclusive}, Conflux \cite{li2020decentralized}, Vite \cite{liu2018vite}, CDAG \cite{gupta2019cdag}, and Prism \cite{bagaria2019prism}. This design relies on complex rules to extend the chains and then sort the rest of the units in each round. NC/variant-NC can also be applied to multiple chains that simultaneously grow, such as Eunomia \cite{niu2019eunomia}, JHdag \cite{he2019consensus}, and OHIE \cite{yu2020ohie}. But singly NC/variant-NC, in this case, cannot guarantee the linear order of units so additional techniques are necessary.

\item[$\diamond$] \textbf{\textit{Sorting algorithm}} aims to sort the units in total linear order. This is essential to guarantee global consistency.  Sorting algorithm relies on several types parameters, including cumulative confidence like \textit{weight, score, fee}, random beacon like \textit{hash}, or natural sequence like \textit{lexicographic order} and \textit{appearance}.  Then, the algorithm sorts the units by setting a priority among these parameters. For example, Conflux \cite{li2020decentralized} organizes the blocks under the priorities of a) their epochs, b) topological order, and c) PoW quality or block hash. This procedure is used as a complementary mechanism to sort filtered units after NC or BFT protocols. Half of DAG systems apply the sorting algorithm to their protocols for strict consistency and better security. A linear order of units would support upper-layer constructions such as the state-transited smart contract.

\end{itemize}

%============================Table
%============================Table
%============================Table
%============================Table
%============================Table
%\begin{landscape}
\begin{table*}[!ht]
%\begin{sidewaystable*}[!]

\newcommand{\tabincell}[2]{\begin{tabular}{@{}#1@{}}#2\end{tabular}}
\caption{The Structures, Consensus, Properties of DAG-based Blockchain Systems\tnote{1} }
\label{tab-consensus}
\begin{center}
\begin{threeparttable}

\resizebox{\textwidth}{!}{
\begin{tabular}{l p{2mm}p{7mm}p{2mm}p{5mm}p{6mm}p{5mm}p{7mm}p{6mm}p{5mm}p{15mm}p{10mm}p{19mm}p{2mm}p{2mm}p{2mm}}
\toprule
& \multicolumn{1}{c}{} 
& \multicolumn{4}{c}{\textbf{Structure}} 
& \multicolumn{7}{c}{\textbf{Consensus}}
& \multicolumn{3}{c}{\textbf{Property}} 
% \multicolumn{2}{c}{\textbf{Network Configuration}} 
\\
\cmidrule(lr){3-6} 
\cmidrule(lr){7-13}
\cmidrule(lr){14-16}

 & 
\rotatebox{60}{\textit{Types}} & 
\rotatebox{60}{\textit{Unit Representation}} & 
\rotatebox{60}{\textit{Topology}} & 
\rotatebox{60}{\textit{In/Out Degree}} & 
\rotatebox{60}{\textit{Transaction Model}} &
\rotatebox{60}{\textit{Openness}} & 
\rotatebox{60}{\textit{Membership Selection}} & 
\rotatebox{60}{\textit{Unit Allocation}} &
\rotatebox{60}{\textit{Unit Positioning}}  &
\rotatebox{60}{\textit{Extension Rule}} & 
\rotatebox{60}{\textit{Conflict Solving}} & 
\rotatebox{60}{\textit{Technical Feature}}  & 
\rotatebox{60}{\textit{Consistency}} &
\rotatebox{60}{\textit{Ordering}} &
\rotatebox{60}{\textit{Finality}} 
\\   
\midrule %******************************** 
IOTA \cite{popov2016tangle}$^\dag$$^\ddag$     & \textit{I}  & Bundle   & $\widehat{D}$ &  (x,2)  & UTXO  & .less & - & -  &  - & TSA(MCMC)&  Weight & Blockless & \tikz\pic{sema=black/180/white};  & \tikz\pic{sema=black/180/white};  & \tikz\pic{sema=black/180/white}; \\

IOTA \cite{coordnator}$^\star$     & \textit{I}  & Bundle   & $\widehat{D}$ &  (x,2)  & UTXO  & .ed & Assign  & -   &- & TSA(MCMC) & TR &   Coordinator & \tikz\pic{sema=black/180/black}; &  \tikz\pic{sema=black/180/white}; & \tikz\pic{sema=black/180/black};    \\

Graphchain \cite{boyen2018graphchain}$^\dag$$^\ddag$   & \textit{I}  & Tx  & $\widehat{D}$ &  (x,y)   &   UTXO  & .less  &  - & - &  - & TSA &  Fee & Incentive  &  \tikz\pic{sema=black/180/white}; &  \tikz\pic{sema=black/180/white}; &\tikz\pic{sema=black/180/white}; \\

Avalanche \cite{rocket2019scalable}$^\dag$$^\ddag$   & \textit{I}  & Tx  & $\widehat{D}$&  (x,y)   & Acct &  .less &  -  & - & - & Sampling &    Query &   Coin-flip  & \tikz\pic{sema=black/180/white};  & \tikz\pic{sema=black/180/white};  & \tikz\pic{sema=black/180/black};    \\

Spectre \cite{sompolinsky2016spectre}$^\dag$$^\ddag$   &  \textit{II}  & Block  & $\widehat{D}$ &  (x,y)  &  Acct & .less &  -  & - & - & RTA  & Vote & Pairwise vote  &  \tikz\pic{sema=black/180/white};  &  \tikz\pic{sema=black/180/white};  &  \tikz\pic{sema=black/180/white};  \\

Phantom \cite{sompolinsky2020phantom}$^\dag$$^\ddag$    & \textit{II}  & Block  & $\widehat{D}$  &   (x,y)   &  Acct &.less  &  -  & - &  & GA   & Score &  $k$-cluster   &  \tikz\pic{sema=black/180/black};  &  \tikz\pic{sema=black/180/black}; &  \tikz\pic{sema=black/180/white};    \\

Meshcash \cite{bentov2017tortoise}$^\dag$    & \textit{II}&  Block & $\widehat{D}$ &    (x,y)  &  Acct & .less &  -  & - & - &   async-BA & -  & Layered ptcl  & \tikz\pic{sema=black/180/white};    & \tikz\pic{sema=black/180/white};   &   \tikz\pic{sema=black/180/black};       \\

Nano \cite{lemahieu2018nano}$^\dag$$^\ddag$  & \textit{III} & Tx  & $\widehat{P}$  & (1,1) &  Pair & .less & Elect   & - & $(i,h)$  &  Natural$^\natural$  &  TRs & Trading pair  & \tikz\pic{sema=black/180/white};   & \tikz\pic{sema=black/180/white};   & \tikz\pic{sema=black/180/white};    \\

Hashgraph \cite{baird2016swirlds}$^\ddag$$^\flat$  & \textit{III} & Event & $\widehat{P}$  & (x,2) &  Acct  &  .ed  & Elect  & - & $(i,h)$ & async-BA  & Witness & Gsp by Gsp  &  \tikz\pic{sema=black/180/black};     &   \tikz\pic{sema=black/180/white};    &  \tikz\pic{sema=black/180/black};      \\

DLattice \cite{zhou2019dlattice}$^\dag$$^\ddag$  & \textit{III}  & Tx  & $\widehat{P}$  &  (1,1)  &  Pair & .less &  DPoS  & - & $(i,h)$ & async-BA   & - &  Trading pair   & \tikz\pic{sema=black/180/white};   &  \tikz\pic{sema=black/180/white};   &  \tikz\pic{sema=black/180/black};  \\

Jointgraph \cite{xiang2019jointgraph}$^\dag$$^\ddag$   & \textit{III}  &  Event &  $\widehat{P}$ &  (x,2)  &  Acct & .ed  &  Assign  & - &  $(i,h)$  & async-BA  & TR  & Supervisory & \tikz\pic{sema=black/180/white};  &  \tikz\pic{sema=black/180/white};  &   \tikz\pic{sema=black/180/black};    \\

Chainweb \cite{will2019chainweb}$^\dag$  & \textit{III}  & Tx  & $\widehat{P}$  &  (x,y)  &  Acct & .less &  PoW  & - &   $(i,h)$  & Natural & Length &  Cross-Ref &  \tikz\pic{sema=black/180/white};    &\tikz\pic{sema=black/180/white};     & \tikz\pic{sema=black/180/white};    \\

Aleph \cite{gkagol2019aleph}$^\dag$  & \textit{III}  &  Unit & $\widehat{P}$  &  (x,y)  &  Acct & .ed &  Elect  & - &  $(i,h)$ & async-BA &  Hash  &  Sorting ptcl  &    \tikz\pic{sema=black/180/black};  &    \tikz\pic{sema=black/180/black};   &    \tikz\pic{sema=black/180/black};        \\

Vite \cite{liu2018vite}$^\ddag$ & \textit{III}  & Tx  & $\widehat{P}$  &   (1,1)  &  Pair & .less  & Assign  & - &  $(i,h)$ & NC &  TR  &  Snpst chain &    \tikz\pic{sema=black/180/black};     &  \tikz\pic{sema=black/180/black};   &    \tikz\pic{sema=black/180/black};     \\

Caper \cite{amiri2019caper}$^\dag$$^\ddag$  & \textit{III}  & Tx  & $\widehat{P}$  &  (x,y)  &  Acct &  .ed & Assign & App & $(i,h)$ & async-BA  & - & Hierarchical &   \tikz\pic{sema=black/180/black};  &  \tikz\pic{sema=black/180/black};   &     \tikz\pic{sema=black/180/black};   \\ %$^\S$

$\mathcal{L}$-Fantom \cite{choi2018fantom}$^\dag$ & \textit{III}  &  Event & $\widehat{P}$  &  (x,y)   &  Acct &  .less &  Elect  & - & $(i,h)$  & async-BA & Apprnce & Sorting ptcl &  \tikz\pic{sema=black/180/black};  &  \tikz\pic{sema=black/180/black};   &    \tikz\pic{sema=black/180/black};  \\

$\mathcal{L}$-Onlay \cite{Nguyen2019ONLAYOL}$^\dag$ & \textit{III}  &  Event & $\widehat{P}$  &  (x,y)   &  Acct &   .less&  Elect  & - & $(i,h)$  & async-BA & Apprnce & Layered DAG   & \tikz\pic{sema=black/180/black};  &  \tikz\pic{sema=black/180/black};  &      \tikz\pic{sema=black/180/black};  \\

$\mathcal{L}$-StakeDag \cite{Nguyen2019StakeDagSC}$^\dag$ & \textit{III}  & Event  & $\widehat{P}$  &  (x,y)  &  Acct & .less  &  PoS & - & $(i,h)$ & Natural & Weight &  Validator & \tikz\pic{sema=black/180/black};   &  \tikz\pic{sema=black/180/black};    & \tikz\pic{sema=black/180/black};  \\

$\mathcal{L}$-StairDag \cite{Nguyen2019StairDagCV}$^\dag$ & \textit{III}  & Event  & $\widehat{P}$  & (x,y)  &  Acct &  .less &  DPoS & - &  $(i,h)$& Natural & Weight &  $w$-Validator  & \tikz\pic{sema=black/180/black};   & \tikz\pic{sema=black/180/black};   & \tikz\pic{sema=black/180/black};    \\

Prism \cite{bagaria2019prism}$^\dag$$^\ddag$  & \textit{IV}  & Block  & $\widehat{P}$  &  (x,y)  &  Acct & .less  &  Elect  & Role & $(i,h)$ &  NC & TR & Decoupling &  \tikz\pic{sema=black/180/black};   & \tikz\pic{sema=black/180/white};    &  \tikz\pic{sema=black/180/black};   \\

OHIE \cite{yu2020ohie}$^\dag$$^\ddag$ & \textit{IV}  & Block  &  $\widehat{P}$ & (x,y)  & Acct  &   .less  & PoW &  -  & $(i,h)$ & NC   & Rank  & Sorting ptcl & \tikz\pic{sema=black/180/black};   & \tikz\pic{sema=black/180/black};    &  \tikz\pic{sema=black/180/white};   \\

Sphinx \cite{wang2021weak}$^\dag$$^\ddag$ & \textit{IV}  & Block  &  $\widehat{P}$ & (x,y)  & Acct  &   .ed  & Elect &  -  & $ (i,h) $ & async-BA   & -  & position & \tikz\pic{sema=black/180/white};   & \tikz\pic{sema=black/180/white};    &  \tikz\pic{sema=black/180/white};   \\

Blockmania \cite{danezis2018blockmania}$^\dag$$^\ddag$  & \textit{IV}  & Block  & $\widehat{P}$  & (x,y)   &  Acct &  .less &  PoS  & - &  $(i,h)$ & smpl-PBFT & Fee & weak rules   &   \tikz\pic{sema=black/180/white};   &   \tikz\pic{sema=black/180/white}; & \tikz\pic{sema=black/180/black};     \\

Blockclique \cite{forestier2018blockclique}$^\dag$$^\ddag$  & \textit{IV}  & Block  & $\widehat{P}$  &  (x,y)  & Acct & .less & PoR &  Chain & $(i,h)$ &  RTA & Fitness   & Tx sharding & \tikz\pic{sema=black/180/white};    &   \tikz\pic{sema=black/180/white};   & \tikz\pic{sema=black/180/white};   \\

Eunomia \cite{niu2019eunomia}$^\dag$  & \textit{IV}  & Block  & $\widehat{P}$  &  (x,y)  &  UTXO & .less &  PoW  & Chain & $(i,h)$  & NC  & Clock & $m$-for-1 PoW &  \tikz\pic{sema=black/180/black};   &   \tikz\pic{sema=black/180/black}; & \tikz\pic{sema=black/180/black};    \\

Dexon \cite{chen2018dexon}$^\dag$  & \textit{IV}  & Block  & $\widehat{P}$  &   (x,y)  &  Acct & .ed &  Assign  & Chain &  $(i,h)$ & BA & Apprnce & VRF & \tikz\pic{sema=black/180/black};   & \tikz\pic{sema=black/180/black}; &    \tikz\pic{sema=black/180/black};      \\

PARSEC \cite{chevalier2019protocol}$^\dag$  & \textit{IV}  &  Event & $\widehat{P}$  &  (x,y)  &  Acct &  .ed & Assign   & - & $(i,h)$ & async-BA &  LO & Stable block & \tikz\pic{sema=black/180/black};   & \tikz\pic{sema=black/180/black}; &    \tikz\pic{sema=black/180/black};   \\

Byteball \cite{churyumov2016byteball}$^\dag$$^\ddag$  & \textit{V}  &  Unit & $\widehat{C}$  &  (x,y)  &  UTXO & .less &   Elect & - & - & Natural & TRs & Witness node  & \tikz\pic{sema=black/180/black};   & \tikz\pic{sema=black/180/white};   &   \tikz\pic{sema=black/180/black};       \\

Haootia \cite{tang2020haootia}$^\dag$$^\ddag$  & \textit{V}  &  Unit & $\widehat{C}$  &  (x,y)  &  Acct & .less &  PoW  & - & $(h,i)$  & PBFT & LO &  Hybrid ptcl & \tikz\pic{sema=black/180/black};  &  \tikz\pic{sema=black/180/black}; & \tikz\pic{sema=black/180/black};      \\

JHdag  \cite{he2019consensus}$^\dag$  & \textit{V}  &  Tx & $\widehat{C}$  &  (x,y)  &  Acct & .less &  PoW  & - &  $(h,i)$   & NC &  Hash & Flexible PoW     &   \tikz\pic{sema=black/180/black};   &    \tikz\pic{sema=black/180/black}; & \tikz\pic{sema=black/180/white};       \\

GHOST \cite{sompolinsky2015secure}$^\dag$$^\ddag$  & \textit{VI}  & Block  & $\widehat{C}$ & (x,1)  &  UTXO &  .less  & PoW  &  -   & $(h,i)$   &  variant-NC & Apprnce & Earliest DAG  &  \tikz\pic{sema=black/180/black};   &   \tikz\pic{sema=black/180/black};   & \tikz\pic{sema=black/180/white};   \\

Inclusive \cite{lewenberg2015inclusive}$^\dag$$^\ddag$    & \textit{VI}  & Block  & $\widehat{C}$ & (x,1)  & UTXO & .less   &  PoW   & - & $(h,i)$   &  variant-NC & Apprnce & Involve Uncles  &   \tikz\pic{sema=black/180/black};  &    \tikz\pic{sema=black/180/black};  & \tikz\pic{sema=black/180/white};   \\

3D-DAG \cite{wang20213ddag}$^\dag$$^\ddag$   & \textit{VI}  &  Block & $\widehat{C}$ &  (x,1)  &  UTXO &  .ed & Elect  & - & $(h,i)$  & BA & Apprnce &    hybrid ptcl  &  \tikz\pic{sema=black/180/black};   &  \tikz\pic{sema=black/180/black};     & \tikz\pic{sema=black/180/black};   \\

Conflux \cite{li2020ghast}$^\dag$$^\ddag$   & \textit{VI}  &  Block & $\widehat{C}$ &  (x,y)  &  UTXO &  .less & PoW  & - & $(h,i)$  & variant-NC & Apprnce &    Adaptive ptcl  &  \tikz\pic{sema=black/180/black};   &  \tikz\pic{sema=black/180/black};     & \tikz\pic{sema=black/180/white};   \\

CDAG \cite{gupta2019cdag}$^\dag$$^\ddag$ & \textit{VI}  &  Block & $\widehat{C}$ &  (x,y)    &  Acct &  .less   &  PoW  & - & $(h,i)$  & variant-NC & Hash & C-Block CDAG &   \tikz\pic{sema=black/180/black};   &   \tikz\pic{sema=black/180/black};     & \tikz\pic{sema=black/180/white};  \\

StreamNet \cite{yin2019streamnet}$^\dag$   & \textit{VI}  & Block  & $\widehat{C}$ &   (x,2)  &  Acct &  .less &  -  & - & $(h,i)$   & TSA(MCMC) & Weight & Pivot chain &     \tikz\pic{sema=black/180/black};   &     \tikz\pic{sema=black/180/black}; & \tikz\pic{sema=black/180/white};     \\

\bottomrule %************************************
\end{tabular} }

\begin{tablenotes}
       \footnotesize
       \item[] \tikz\pic{sema=black/180/black}; = totally provided (global consistency; linear sorting; deterministic finality).
         \tikz\pic{sema=black/180/white}; = partially provided (partial consistency; topological sorting; probabilistic finality).
       
      \item[] $-$ = does not provide property; $\star$ = Temporary Settings. $\natural$: This represents the tip attachment procedure without any predefined rules.
      \item[] $\dag$ = Academic documents (Preprint included) available; $\ddag$ = Implementation available;  $\flat$ = Patent Protection; default = Whitepaper/Concept.
     
     % \item[] $\S$: App: Units are identified in different applications; Role: allocate the unit with the role (functionalities); Chain: allocate the unit into an individual chain.
             
      \item[] (x,y) represents the number of in-degree and out-degree edges, satisfying $1<x\leq m, 1<y\leq m$ where $m$ is the maximum number of generated units at one round.
      \item[] $h$ = unit index/height/clock in the (sorted) chain; $(i,h)$ = the unit with index/height/clock $h$ in the $i$-th chain; $(h,i)$ = the unit $i$ inside the $h$-th block/set.
     
     \item[] \textbf{Abbreviation}: Tx = transaction;  Acct = account; ptcl= protocol;  strc = structure; smpl = simple; Gsp = Gossip; Apprnce = Appearance; $w$-Validator = Weighted validator.
      \item[] TR = Trusted Role;  TRs = a small group of TR; LO =  lexicographic order;  MCMC = Markvo Chain Monte-Carlo \cite{popov2019equilibria}; Adaptive ptcl = Adaptive weighted protocols. 
      \item[] TSA/RTA/GA = Tip Selection / Recursive Traverse / Greedy Algorithm; async-BA = asynchronous Byzantine Agreement; variant-NC = NC with the weightest subgraph;
     \end{tablenotes}
   \end{threeparttable}
\end{center}

%\end{sidewaystable*}
\end{table*}

%=================================================
\section{Property Analysis}
\label{sec-property}
%=================================================

Properties reflect the inherent protocol designs. We capture three properties related to the consensus procedure, namely \textit{consistency}, \textit{ordering} and \textit{finality}. Other properties (that are potentially related, but not frequently mentioned) are omitted in this paper. In the following subsections, we respectively provide the reasons, definitions, and analyses of these properties.

%============================
\subsection{Properties Between BA and NC.} 
%============================
Byzantine Agreement (BA-style) \cite{CastroL99}\cite{cachin2017blockchain} and Nakamoto consensus (NC) \cite{nakamoto2008peer} are the most prevailing protocols adopted by current blockchain systems \cite{bano2019sok}\cite{vukolic2017rethinking}. BA-style protocols have been proposed to achieve consensus in the presence of malicious nodes, where the tolerance is maximal $2/3$ of the total nodes \cite{CastroL99}. NC protocols, in recent decades, stand out in one critical aspect thanks to its remarkable simplicity. NC introduces a new method to extend the chain -- the longest subchain wins. This design breaks the assumption in BA-style protocols that only the closed committee can conduct the consensus. Instead, it enables all participants to get involved in the consensus process by the ways like PoW~\cite{nakamoto2008peer}, PoS~\cite{kiayias2017ouroboros}, \textit{etc} (PoX). These two types of protocols are radically different in the mechanism designs, but they follow similar rules in properties. Here, we start our observation.

Classic BA-style protocols follow three well-known properties:  ($a1$) \textit{agreement}: for a given consensus instance, if a correct node decides the block $B$, then all correct nodes decide $B$;  ($a2$) \textit{validity}: if all correct nodes propose a valid transaction before starting a consensus instance, then the block decided in this instance is not empty; and ($a3$) \textit{termination}: all correct nodes eventually have the decisions.  Nakamoto consensus slightly modifies the definitions.  Two key properties of NC  protocols \cite{garay2015bitcoin}\cite{kiayias2015speed} are:  ($b1$)  \textit{persistence}: if a transaction is seen in a block deep enough in the chain, it will stay in the position; and ($b2$) \textit{liveness}: if a transaction is given as input to all honest nodes, it will be eventually inserted in a block that is deep enough in the chain.

%BFT and NC are different.
Properties between $(a1, a2, a3)$ and $(b1, b2)$ are different. The difference roots in their system formulations. The BA-style protocols rely on the interactive model of communication, where transmitted messages need to be responded from peers. The decision, which cannot change anymore, is completed once enough participants agree on it. In contrast, Nakamoto consensus is based on the non-interactive model and the states get updated in every phase. The probability of being reversed depends on its depth. The formulation of NC protocols shifts from the send-receive replication to state machine replication \cite{garay2015bitcoin}\cite{pass2017analysis}. These modifications make the above-mentioned properties presented in different forms.

%BFT and NC are same.
However, in fact, the properties from BA and NC essentially pursue the same goals. Firstly, the \textit{liveness} property of NC $(b2)$ is phrased as chain growth and chain quality, and a block is inserted in the chain with a negligible probability of being reversed. As one block is sufficiently buried in the chain, the property $a3$ and $b2$ could be regarded as approximately identical. Several studies even provide the boundary of reversibility \cite{pass2017analysis}\cite{ren2019analysis}.  Secondly, the \textit{consistency} property ($a1$ and $b1$) ensures that a system can obtain the global view of states. All blocks see the same blocks at a specific height.  Decisions in BA-style consensus are deterministic since peers are allowed to negotiate at each round. Decisions in NC protocols are probabilistic.  The confidence increases as the chain develop under the \textit{longest chain wins} rule.  Thirdly, the \textit{validity} property ($a2$) is guaranteed by the verification of hashes in each block header, as a default setting in systems.

Their similarities enable us to inherit the common principles of both BA-style and NC properties and also left gaps to add specialized properties concerning DAG. Specifically, we provide three properties, namely \textit{consistency}, \textit{ordering} and \textit{finality}.  The  \textit{consistency} is similar to the properties of $a1$ and $b1$, while the  \textit{finality} matches $a3$ and $b2$. The property \textit{ordering} is added as the distinguished property when adapting to DAG-based blockchain systems. This property is neglected in BA-style consensus and NC protocols due to their implicit structure: all units in their systems must be sorted in linearization. Details are shown in the next subsection.

%============================
\subsection{Properties}
%============================

%-----------------------Consistency
%-----------------------Consistency    
\noindent\textbf{Consistency.}
Scalability and performance in blockchain systems are closely related to the requirements of consistency. Smart contract-supported platforms rely on strict consistency. The state transition must be aligned in a linear sequence since the state-disorder may fail operations. In comparison, some scenarios may only require a weak assumption -- partial consistency, where the sortation happens in associated transactions, rather than all transactions. It fits the scenarios that do not provide strict integrity, such as the proof of existence in certificate systems, the token transferring inside organizations. Here, we define two types of consistency as follows.

\begin{itemize}
    \item[-] \textbf{\textit{Strict consistency.}} Each honest node has the same decisions for a given position on the ledger so that the units can be precisely positioned in the system. For example, BA-style consensus and Nakamoto consensus rely on strict consistency by default. 
    
    \item[-] \textbf{\textit{Partial consistency.}} 
    Only associated nodes have the same view of decisions, whereas the decisions of disjoint nodes (no intersect backward or forward) are unknown to each other. The number of associated transactions ranges from two (equiv. pairwise order) to multiple.
\end{itemize}

\noindent\textit{Comparison with the global view.} The view of a ledger/node represents its corresponding local state. The strict consistency does not necessarily require an arbitrary node has a global view since the assumption of \textit{global} is much stronger than that of strict consistency.  Assuming that all nodes maintain the same ledger (like Bitcoin\cite{nakamoto2008peer}/GHOST\cite{sompolinsky2015secure}/\textit{etc.}), the property of strict consistency can be approximately regarded as a global view because all nodes synchronize the same (latest) states that have been agreed and synchronized to their local chains. In contrast, if every node only maintains its own ledger (like Nano\cite{lemahieu2018nano}/OHIE\cite{yu2020ohie}/Prism\cite{bagaria2019prism}/\textit{etc.} in \textit{Type III/IV}), the property of global views can hardly be achieved due to the isolation of their units. But achieving strict consistency is feasible with the help of special operations such as the sorting algorithm.

%-----------------------Ordering        
%-----------------------Ordering
\smallskip
\noindent\textbf{Ordering.} Ordering indicates how the units in the systems are organized. Classic blockchain systems arrange the units in linearization as default. In contrast, DAG-based systems break this implicit assumption by introducing the concept of graph-based structure. Units attached to the network may form any shapes of topologies, either the divergent graph or parallel-line graph. The difference between DAG systems is whether they sort the disordered units into a linear sequence with an additional step. We use the term \textit{topological sorting} to describe the systems that stay the status of disordered units. We define two types of ordering to distinguish the desired property as follows.

\begin{itemize}
    \item[-] \textbf{\textit{Topological ordering.}} The topological ordering in a DAG system is an ordering of the vertices satisfying the properties of \textit{unidirectional} and \textit{cyclic} as defined in Equation \ref{eq-model}. 
   
    \item[-] \textbf{\textit{Linear ordering.}} Besides satisfying the properties of \textit{unidirectional} and \textit{cyclic} in topological ordering, linear ordering additionally requires the vertices (units) in the network being ordered in a \textit{total linear sequence}. BA-style consensus and NC protocols rely on the linear sorting by default. 
\end{itemize} 
 
\noindent\textit{Approaches to achieving the linear ordering.} The systems, where multiple nodes maintain the same ledger, sort units recursively. This means the sorting algorithm runs instantly once a unit is attached to the main chain (Phantom\cite{sompolinsky2020phantom}/Aleph\cite{gkagol2019aleph}/GHOST\cite{sompolinsky2015secure}). Every unit will be recognized with a height or an index after the serialization. In contrast, the systems, where each node only maintains its ledger, rely on a periodical way to sort the units. The systems collect the qualified units (\textit{e.g.} with sufficient votes) during each time interval. Then, these units are sorted according to their confidence (appearance time, weight, hash, lexicographical order, \textit{etc.}) into a linear sequence. The sorting algorithms also might be operated by trusted authorities (Vite\cite{liu2018vite}/Jointgraph\cite{xiang2019jointgraph}). But in most cases, they are embedded in the default systems.

\smallskip
\noindent\textit{Features of the topological ordering.} Topological ordering includes all the typologies except for linearly structured ones. The systems in \textit{Type V/VI} are inherently designed with the main chain for unit linearization. In contrast, topological ordering appears in systems belonging to \textit{Type I-IV}. We have identified two types of topological ordering from the perspective of whether an arbitrary unit can be traced back to the genesis unit or not. Transitive systems includes IOTA \cite{popov2016tangle}, Graphchain \cite{boyen2018graphchain}, \textit{etc.} Otherwise, a user can only trace to a certain transaction in the graph. This frequently appears in the pairwise-ordering systems, such as Nano \cite{lemahieu2018nano}, Spectre \cite{sompolinsky2016spectre} and DLattice \cite{zhou2019dlattice}. The topological ordering limits the systems only suited to support cryptocurrency networks where the global state transition is not a necessity.

\smallskip
\noindent\textit{Relation with consistency.} If the units in a system are organized in a linear order, this system can achieve strict consistency. The reason is straightforward: every unit in the formed chain can be uniquely identified and recognized. The decision (including the history of previous states) of a specific position is shared by all the nodes without conflicts. However, in topological sorting, we cannot derive similar results because the properties in each type of typology are significantly varied. We only discuss the systems in the \textit{convergence} topology.  \textit{Convergence} states the idea where the generated units are expected to be close to a certain distribution. Units, being organized in linear sequence or tending to be organized in a linear sequence, could be denoted as \textit{convergence} systems. Thus, in our classification, the systems in \textit{Type V/VI} can reach strict consistency since all units will be eventually organized in a linear sequence.

%-----------------------Finality    
%-----------------------Finality
\smallskip
\noindent\textbf{Finality.}
Consensus finality (\textit{a.k.a.} ``forward security'' \cite{decker2016bitcoin}) represents the property where a confirmed unit cannot be removed from the blockchain once successfully appended to his parent unit at some point in time. Formally, as defined in \cite{vukolic2015quest}, \textit{if a correct node $i$ appends a unit $b$ to its copy of the blockchain before appending the unit $b'$, then no correct node $j$ can append the block $b'$ before $b$ to its copy of the blockchain. }  A block will be eventually either fully abandoned or fully adopted. For traditional BA-style consensus, such decisions are instantly made through two/three rounds of negotiation between the leader and replicas. In contrast, a block in NC protocols is appended in the chain with the risk of being reversed. The possibility is decreased along with the increased depth that this block is being buried. We abstract two types of finality when adapting to the DAG-based systems.

\begin{itemize}
    \item[-] \textbf{\textit{Probabilistic.}} The units appended to the system are always accompanied by the risk of being reversed. The probability is inversely proportional to its cumulative confidence (including the forms of depth/score/weight \textit{etc}). The confidence is obtained from the contributions of subsequent units.
    
    \item[-] \textbf{\textit{Deterministic.}} The units, once attached to the system at some point in time, will instantly and permanently get confirmed so that they can never be removed from the system.
\end{itemize}

\noindent\textit{Comparison with liveness.} Liveness, as we discussed in the previous content, ensures that if a unit is given as input to all honest players, it will eventually be inserted into the chain.  The property of liveness focuses on the problem: \textit{whether the systems can continuously run without crash or faults}. Network failure and malicious attacks are the two major threats against liveness. In contrast, finality guarantees that the unit inserted in the chain will be reversed with a negligible probability. This is equal to saying the units are buried deep enough. The property of finality emphasizes the problem: \textit{whether attached units are permanently valid in systems. } 

\smallskip
\noindent\textit{Approaches to achieving deterministic finality.} Achieving instant and deterministic finality relies on trusted roles (TRs) in current DAG-based systems. TRs can be instantiated as the forms of leader, validator, coordinator, \textit{etc.}  The way to elect TRs and the principles to be followed distinguish different systems. We give several examples. \textit{Byteball} \cite{churyumov2016byteball} relies on a set of reputable and honest witnesses to determine the main chain with finality. Nano \cite{lemahieu2018nano} provides quicker and deterministic transactions by a group of voted representatives via a balance-weighted vote on conflicting transactions. Vite \cite{liu2018vite} makes assigned TR create a snapshot chain to record the transactions from common nodes. The finality in Prism \cite{bagaria2019prism} is determined by its leading proposal blocks.

%=================================================
\section{Security Analysis}
\label{sec-Security}
%=================================================
This section provides security analyses (cf. \underline{Table.\ref{tab-security}}) consisting of the scopes of \textit{attack-defending}, \textit{system challenge}, \textit{theoretical analysis} and \textit{privacy-preserving}.

%============================
\subsection{Attack Defending} 
We first conclude several mainstream attacks with their assumptions and scopes. Accordingly, we list a few of the leading systems mentioned in the literature as instances to show how to mitigate these attacks. A summary is presented in Table.\ref{tab-attack}.

\smallskip
\noindent\textbf{Parasite chain attack.} The attack \cite{popov2016tangle} shares a similar mechanism as selfish mining in Bitcoin \cite{eyal2014majority}. The attack attempts to replace an honest subgraph with a prepared subgraph (\textit{a.k.a.}, parasite chain) for more profits. To launch the attack, an adversary secretly generates a subgraph offline but occasionally references the main graph to obtain a high score (equally weights/confidence/votes \textit{etc.}). Then, the adversary sends a pair of conflicting transactions separately to the main graph and his private subgraph. He continues to work for a while ensuring that his subgraph collects competitive scores for games. At this time, the conflicting transaction in the main graph may get confirmed (money spent) by several approvals from honest tips. The adversary at the same time publishes his prepared subgraph to invalidate the main graph and wins the competition with a high probability. Thus, a coin has been spent twice. 

This attack relies on the assumption that an adversary has the sufficient computing power to efficiently generate the units. Based on that, the attack targets the protocols without instant finality by powerful leaders. We show how IOTA overcomes this attack.

\smallskip
\noindent\textit{Instances.} The tip selection mechanism plays a critical role in the security of IOTA \cite{popov2016tangle}. IOTA proposes a weighted random walk (MCMC) to adjust the transition probability by a parameter $\alpha$.  Theoretically, if $\alpha$ is high enough, the system is secure under the assumption (malicious power $<51\%$). However, an excessively high $\alpha$ makes the main graph become a chain, with lots of transactions being orphaned. The study of \cite{cullen2019distributed} proposes a modified tip selection mechanism to resist the parasite chain attack. The modification focuses on the explicit formulation of the MCMC algorithm, where they employ the \textit{derivative} of original MCMC equations, called First Order MCMC. They conduct a set of experiments to simulate the effect on the likelihood of an MCMC walk terminating on tips. The derivative of MCMC formulations makes the cumulative weight on the main graph grow at a different rate. When the particle ($x$-axis) is far away from the tips, the algorithm has significant effects, whereas the particle approaches the tips, the algorithm performance starts to deteriorate. With the algorithm,  the parasite chain grows linearly with the rate of attackers' computing power, whilst the main graph grows at the rate of the rest power of the network. The parasite chain, thus, gets heavily penalized in the modified mechanism as the power of attackers is less than that of honest participants. Additionally, a model-based detection mechanism against this attack in IOTA is introduced by \cite{psdetection}.

Spectre \cite{Sompolinsky2017SPECTRES} describes a type of attack which refers to the malicious behaviors of the miners. We classify this attack as the parasite chain attack in our model due to their similar approaches, although Spectre named it the censorship attack. In the normal case, miners have to honestly verify recent blocks and immediately publish their blocks. But in this attack, dishonest miners deliberately \textit{ignore} certain blocks and transactions, preventing the blocks from being accepted due to the lack of enough votes. The merchants may consider the transactions contained in such blocks unsafe. This potentially delays the acceptance rate of transactions or even invalidates the main subgraphs. However, a successful attack requires an instant \textit{Poisson bursts} in block creation to make the blocks generated by attackers outpace others. The honest node can extend the waiting period of accepting transactions to minimize the probability of such bursts. The correct votes will eventually exceed the malicious votes since honest nodes occupy the majority. Each round of iteration strengthens the decision of correct votes that are consistent with the majority of past blocks. Spectre provides the experimental results, showing that the acceptance rate may decrease to some degree when compared to situations without attack, but still at a high-speed level.

\smallskip
\noindent\textbf{Balance attack.} The attack \cite{natoli2016balance}, also know as \textit{liveness attack}, aims to profitably keep several chains/subgraphs growing in the same pace.  The attack partitions the network into multiple subgraphs with balanced computational power.  An adversary leverages a dynamic strategy to wander across those balanced subgraphs and selects one favorable fork to maximize his/her profit.  To launch the attack, an adversary imposes a (time) delay to the subgraphs with equivalent mining power and selects one subgraph to add his issued transactions.  The miner in this subgraph performs as normal to collect transactions for block production and offer irrevocable benefits (like services, merchants, goods, \textit{etc.}) to the adversary.  In the meanwhile, the adversary accepts the benefits while issuing conflicting transactions in other subgraphs with additional computing power, to guarantee that he can arbitrarily let the second subgraph outweigh the original one. The miner dynamically maintains the balance between two (or more) subgraphs until the miner learns the fork. Even though, the adversary is still able to invalidate the miner's subgraph and conduct the double-spending.

This attack heavily relies on the assumption of a powerful adversary who: a) can split/partition the network and add the time delay; b) can temporarily and arbitrarily wander across different subgraphs without being known by honest miners; c) has enough mining power to extend one subgraph outweighing others. Thus, the balance attack targets PoW-based protocols with high block generation rates. These protocols rely on miners for block/transaction production, which is vulnerable to powerful and strategic attackers. DAG-based systems confront similar threats as it does in Bitcoin, especially for protocols that stem from the Bitcoin protocol like GHOST, Conflux, OHIE, Prism, \textit{etc.} We look through how these protocols mitigate attacks.

\smallskip
\noindent\textit{Instances.} Balance attack strategically delays the confirmation process of new blocks, instead of reverse past blocks as it is in parasite attack. We denote the block generation rate as $\lambda$ and the delay from an adversary is time $d$. As claimed in \cite{yu2020ohie}, an adversary with little computation power stalls the normal procedure of consensus when $ \lambda d > 1$.\footnote{The equation is obtained by the following assumption: the propagation delay $\lambda$ is larger than the block generation interval $\frac{1}{\lambda}$, where $d>\frac{1}{\lambda}$. } To solve the issue, reducing the rate of block generation is a reasonable way, because a higher block generation interval makes fewer blocks attached to the system. GHOST can tolerate the attack to some degree since the interval between two consequent blocks requires a long time (10 minutes in Bitcoin). The waiting period limits the behavior of adversaries who rely on delay time $d$ to launch the attack. Thus, low mining rates of GHOST \cite{kiffer2018better}\cite{kiayias2017trees} mitigate the effectiveness of the balance attack.

In contrast, this attack is effective in the early version of Conflux \cite{li2018scaling}. The reason \cite{yu2020ohie} is due to its high throughput where massive blocks are generated in each round. If the production of one block requires 1 second while the propagation needs around 10 seconds, around 10 blocks are attached to the system at the approximate same time. Such heavy pressure is imposed on subgraphs where an attacker may take advantage of the chaos. In its updated version \cite{li2020decentralized}, Conflux proposes its improved consensus GHAST \cite{li2020ghast} by providing two strategies to resist the balance attack: an optimistic strategy as GHOST and a conservative strategy.  The switch of these strategies is inherently based on an adaptive weight mechanism. When detecting a divergence of computing power, GHAST slows down the block generation rate by adaptively switching to the conservative strategy. The gap between these two strategies is that only a small fraction of blocks in the conservative strategy, instead of all, are granted weights for further branch selection. Other blocks are deemed valid but set to zero weight. If we only consider the weighted blocks, the block generation rate is sufficiently low to defend the balance attack. Another critical point is how to detect the balance attack in the network. GHAST relies on its recursive algorithm to see whether its best child has a dominant advantage compared to t the sibling blocks. If no advantages, the child may have conflicting views and GHAST thinks the attack exists.

The attack is ineffective in OHIE \cite{yu2020ohie} and Prism \cite{bagaria2019prism} due to their distinguished designs. OHIE achieves a high throughput based on collaborative chains. However, for individual chains, the improvement is not significant since the confirmation of each chain still follows the Nakamoto consensus. A miner has to solve a mathematical puzzle to generate new blocks. Thus, individual chains have low block generation rates, which can resist the balance attack as discussed in GHOST. Prism runs parallel chains with a distinguishable architecture. Generally, the balance attack impedes the confirmation of a block. But in Prism, this block is disassembled into three types of blocks according to its functionalities. These three types of blocks are independent of each other: the proposer blocks, which can pack the transaction blocks, and not necessarily wait for voters blocks to become irreversible. On the one side, a delayed transaction block will not retard the confirmation of leader blocks. An adversary cannot succeed in adding a delay to impede the procedure of other types of blocks. Neither, on the other side, the adversary can hardly partition the network that contains unevenly distributed blocks of all types. Further results on the simulation of balance attacks are provided in  \cite{bagaria2019prism}.

%============================Table
%============================Table
%============================Table
%============================Table
%============================Table

%\begin{landscape}
\begin{table*}[!htb]
%\begin{sidewaystable*}[!]

\newcommand{\tabincell}[2]{\begin{tabular}{@{}#1@{}}#2\end{tabular}}
\caption{Attacks Analysis}
\label{tab-attack}

\begin{center}
\begin{threeparttable}

\resizebox{\textwidth}{!}{

\begin{tabular}{l p{50mm}p{60mm}p{37mm}p{15mm}}
\toprule
 &   \multicolumn{1}{l}{\textbf{Behavior}} 
 &    \multicolumn{1}{l}{\textbf{Assumption}}
  &    \multicolumn{1}{l}{\textbf{Scope}} 
  &   \multicolumn{1}{l}{\textbf{Instance}} 
  
% \multicolumn{2}{c}{\textbf{Network Configuration}} 
 \\
\midrule %********************************

\multirow{1}{*}{\textit{Parasite Chain Attack}}    
& Adversaries secretly generate a subgraph to replace the honest subgraph with his prepared ones for more profits. 
& \begin{itemize}    
  \item[-] Enough power to generate a parasite chain outpacing than peers
  \item[-] No instant finality
  \end{itemize}  
& Probabilistic consensus  & \tabincell{c}{IOTA  \\ Spectre}    \\

\cmidrule{3-5}

\multirow{1}{*}{\textit{Balance attack}} 
&   Adversaries partition the network into balanced subgraphs and start mining on one subgraph to invalidate another that has a deal with merchants.
& \begin{itemize}    
\item[-] Split/partition the network
\item[-] Secretly wander across subgraphs 
\item[-] Enough power to extend subgraphs
\end{itemize}  
&  PoW-based protocols  & GHOST, Conflux, OHIE, Prism  \\

\cmidrule{3-5}

\multirow{1}{*}{\textit{Splitting attack}}   
&   Adversaries insert a conflicting transaction into two balanced branches to double spend the coins.
&  \begin{itemize}    
   \item[-] Enough power for balanced subgraphs
   \item[-] No instant finality
   \end{itemize}  
& Probabilistic consensus & IOTA  \\

\cmidrule{3-5}

\multirow{2}{*}{\textit{Large Weight Attack}} 
&  Adversaries creates a conflicting transaction with high confidence to invalidate a recently confirmed transaction.
& \begin{itemize}    
  \item[-] Units heavier than others
  \item[-] No instant finality
  \end{itemize} 
& Probabilistic consensus &\tabincell{c}{IOTA  \\ Conflux} \\

\cmidrule{3-5}

\multirow{1}{*}{\textit{Censorship Attack}} 
&  Adversaries collude enough committee members who execute the consensus to take over the system and earn profits by preventing certain transactions from being packed into blocks.
&  \begin{itemize}    
   \item[-] A closed committee
   \item[-] Collude the majorities 
   \item[-] Static committee selection mechanisms
   \end{itemize}  
& Permissioned systems &  Prism  \\

\cmidrule{3-5}

\multirow{1}{*}{\textit{Replay attack}}  
&  Adversaries reuse the same address of the victims to steal users' coins by replaying transactions multiple times.
&  \begin{itemize}    
   \item[-] Balance has to be transited
   \item[-] Reuse addresses
   \end{itemize} 
& UTXO-based systems & IOTA   \\

\cmidrule{3-5}

\multirow{1}{*}{\textit{Sybil Attack}} 
& Adversaries create massive pseudonymous identities to participate in the membership selection (to become fake miners) to conduct malicious behaviors.  
&  \begin{itemize}
   \item[-] Generate identities without any cost 
   \item[-] Enough power to create identities
   \end{itemize} 
& The committee selection and block production are separate  & Blockclique, Nano  \\

\bottomrule %************************************
\end{tabular} }
\end{threeparttable}
\end{center}

%\end{sidewaystable*}
\end{table*}
%\end{landscape}

%---------------------
\smallskip
\noindent\textbf{Splitting attack.} In \textit{splitting attack} \cite{popov2016tangle}, an adversary can launch double spending between two branches with a high probability. The adversary traverses the network to find two branches/subgraphs whose total cumulative weights are approximative. At this time, s/he attaches a pair of conflicting transactions in different branches to double spend the coins. The conflicting status will maintain for the long term if the adversary continuously sends meaningless units to keep these branches growing evenly. The adversary, thus, can conduct malicious behaviors or gain more profits during this interval. The attack follows a similar principle to the balance attack but still has differences. We observe that this attack is proved to be effective even without any explicit roles of miners. This greatly simplifies the procedures when launching attacks. An adversary only needs to send massive transactions to targeted branches, without any further actions to avoid detection from miners.

This attack relies on a simple assumption: an adversary has enough computing power to extend the targeted branches which significantly outpace peers. Thus, the splitting attack targets the naturally expanding systems without any explicitly powerful roles in deciding the consensus. Matched systems are mainly \textit{Type I} systems. We take the matured project, IOTA, as an example.

\smallskip
\noindent\textit{Instances.} Tip selection rules determine the properties of Tangle in IOTA \cite{popov2016tangle}. As aforementioned, when the parameter of $\alpha$ approaches $0$, the selection rule is uniformly random, while approaches $1$, the rule adopts the weighted random walk. A uniformly random selection rule is insecure since the adversary can easily control the branches. Therefore, to avoid the attack, a weighted random walk with the high $\alpha$ value is necessary. Some of the transactions are marked with weights, and attackers can hardly keep the balance between the two branches due to the unpredictable distribution of weights. However, a higher $\alpha$ value causes more left-behind tips, losing the confirmation rate of transactions. This design limits IOTA in resisting the splitting attack.

A temporary solution has been proposed by their foundation, introducing a central coordinator \cite{coordnator}. The coordinator acts as an authority to periodically (every two minutes) issue milestones for finality. The milestone is a special transaction that confirms the ancestor transactions with a 100$\%$ confirmation confidence. Therefore, launching the attack is impractical for adversaries. However, this approach deviates from the design principle of IOTA, making it centralized. G-IOTA \cite{bu2019g} moves the focus back to the tip selection rules. Increased left-behind tips slow down the confirmation rate since tips stay in low confidences where they are rarely approved by others. G-IOTA  accordingly adds one more verification edge for each tip, pointing to the left-behind tips. This design helps to increase the confidence of honest transactions fast, while efficiently detecting fake transactions with low confidence for a long time. It remains a higher $\alpha$ value while simultaneously keeping the confirmation rate. Further, E-IOTA \cite{bu2019metamorphic} proposes a parameterized algorithm to adjust the distribution among different types of mechanisms (0 $\alpha$, low  $\alpha$ or high  $\alpha$). This helps to achieve the balance between security and performance. Thus, the splitting attack against IOTA is mitigated by making it possible to choose a moderated $\alpha$.

%---------------------
\smallskip
\noindent\textbf{Large weight attack.} 
The attack \cite{bu2019metamorphic} occurs when a \textit{heavy} conflicting transaction invalidates a \textit{recently} confirmed transaction. The term \textit{heavy} represents the values of measurements (score/confidence/weights \textit{etc.}) The \textit{recently} means a transaction is approved by several subsequent transactions, but actually not buried in deep. This is equal to say, the (cumulative) weights of a recent transaction are not too much heavier than the weights of a newly attached transaction. In a large-weight attack, an adversary targets a recent transaction and immediately generates a conflicting transaction. The adversary improves the weights of the conflicting transactions by methods such as making it approved by lots of meaningless transactions, and appending it to the main graph until it becomes heavier than the targeted one. These two transactions belong to different paths. When an honest transaction arrives, it cannot effectively distinguish which path is correct. The transaction can only select the path with more weights. Since the coins in the original path have been irreversibly spent, double spending in the new path happens when the adversary reuses the same coins.

This attack relies on assumptions in which: a) an adversary has enough computing power to make the conflict transactions heavier than honest ones; b) the consensus is probabilistic without instant finality. Thus, the large-weight attack targets the systems with probabilistic-based consensus mechanisms. We provide examples of IOTA and Conflux.

\smallskip
\noindent\textit{Instances.} This attack is effective in the original design of IOTA \cite{popov2016tangle}. Since the tips generated by honest nodes are uniformly distributed in the network, an attacker may need a computing power rate far less than $50\%$ \cite{qin2020security}. The temporary solutions based on coordinator \cite{coordnator} shift the IOTA into a completely centralized network, but it is indeed able to resist the attack. The improved approach \cite{cullen2019distributed} can also prevent large weight attacks to some degree since the effectiveness of MCMC sharply deteriorate at the margin of the tangle. This increases the randomness of the selection procedure and reduces the probability of unluckily selecting a heavier transaction. Meanwhile, the effectiveness of the tip selection algorithm is adjustable in E-IOTA \cite{bu2019metamorphic}, which provides a similar resistance against this type of attack.

The attack is ineffective to Conflux \cite{li2020ghast}\cite{li2020decentralized}. The clever point is that Conflux even gets inspired by the concept of weight, proposing its adaptive weight algorithm to make two consensus strategies switch between each other. If an attacker generates and broadcasts meaningless blocks at a high speed to increase one's weight, the system switch to a conservative strategy. This strategy only grants a small portion of blocks with valid weights, whereas others are marked with 0 weights. The attacker cannot control the probability of being selected. Much more power is required if the attacker attempts to have a stable advantage.  Conflux also utilizes the adaptive weight mechanism to prevent the balance attack as discussed above.

%---------------------
\smallskip
\noindent\textbf{Censorship attack.}
The attack \cite{vitalik2015censorship} happens when the adversaries collude enough committee members who execute the consensus. These members may cooperatively frame certain transactions, preventing them from being packed into blocks. Malicious collusion breaks the basic security of a system since colluded members have taken over the system and earned profits according to their willingness. The attack relies on the incentive mechanism. If the members can obtain extra profits through censoring specific blocks or transactions, more members would join the game; if they obtain the profits via honestly mining, these members would keep honest. A positive incentive mechanism, like relating the reward proportional to the honest members \cite{vitalik2015censorship}, will help to maintain a healthy network. Specifically, one can earn a 100$\%$ reward if 100$\%$ of members are honest; whereas he can only obtain an 80$\%$ reward if 20$\%$ of members have colluded. Another factor is the attack must collude all targeted members within the time window. If the members leave the committee under the rotation rules, the collusion is useless and costly. These examples prove that the censorship attack is closely related to the behaviors of committee members.

The assumptions of this attack are a) the consensus is completed by a closed committee; b) an adversary has sufficient power/stake to collude a certain proportion of committee members, who may further attract other incomers; c) the committee relies on a relatively static mechanism without dramatically dynamic membership rotation. Thus, the censorship attack mainly targets the permissioned systems whose committees are fixed. Prism is an example.

\smallskip
\noindent\textit{Instances.} Prism \cite{bagaria2019prism} provides simulation results against the censorship attack with a tolerable threshold of $\beta = 0.25$, where $\beta$ indicates the fraction of hash power used for the attack. Prism simplifies the behaviors of attackers to slow down the confirmation by producing empty blocks (proposer block and voter block). Empty proposer blocks may slow down the ordering procedure of transactions since the generation of proposer blocks is delayed. Empty voter blocks may retard the voting procedure which indirectly affects the creation of proposer blocks. Based on simulation results, Prism performs better than the longest-chain protocols \textit{w.r.t} the censorship attack. The confirmation delay of Prism is greatly smaller than delays in longest-chain protocols since the actual delays are merely caused by the insertion of empty blocks. This is the key point to resist the attack, where each decoupled block would not hinder other types of blocks. In contrast, delays in the longest-chain protocols would impede all the procedures.

%---------------------
\smallskip
\noindent\textbf{Replay attack.}
The attack \cite{Josephd018replay} refers to stealing users' coins by replaying transactions multiple times. The key to performing the attack requires an adversary to reuse the address of the victims. Reusing the same address makes certain transactions repeatedly confirmed and all the addresses following this vulnerable address become vulnerable. Adversaries may drain the funds from all transactions associated with this vulnerable address. In case of the funds in this address is insufficient, the adversary first tops up the address, and then replays the transaction to steal more funds until using them up. The replay attack is mainly used for the systems based on the UTXO model, because the remaining values have to be instantly transferred from the current address to another new address, rather than directly being removed as it is in the account model. Besides, two types of variants are identified. The first one is \textit{brute force}, where an adversary tries to guess the seed. But it cost massive computing power which may not be worth it. Another type is to analyze the past transaction to find a potential input address with remaining values. All these variants still rely on reusing addresses.

The assumptions of this attack are: a) the remaining values of one address has to be transited to another address; b) the address can be reused multiple time through some technique.  Based on this, the replay attack targets the systems which adopt the UTXO model as their data structure. We take the IOTA as an example.

\smallskip
\noindent\textit{Instances.} IOTA \cite{popov2016tangle} relies on \textit{bundle} \cite{bundle} to finish the token transferring. The bundle is a virtually top-level construction used to simulate accounts for users. Each bundle links related transactions to complete the operations through a set of addresses generated by the same seed from the user. Specifically, the remaining values of transactions in the bundle need to be transferred from the current address to a new address. These two addresses are linked by the seed. The user who holds the seed can control transactions that relate to him.  The replaying attack is feasible in IOTA since the transaction bundle is inherently based on the UTXO model. Joseph \cite{Josephd018replay} shows the possibility of the attack, while Roode \textit{et al.} \cite{de2018break} provides a practical method to conduct the attack. The method modifies the functions (mainly \texttt{addRemainder}) of IOTA's API, enabling input address reused. The modification removes the procedure of generating remaining funds such that the funds can stay on the input address. Fortunately, this attack can be mitigated. As suggested by \cite{Josephd018replay}, each signed transaction bundle should be tracked via its unique hash, and each subtangle only permits one bundle. As a counterpart, this method increases the overhead.

%---------------------
\smallskip
\noindent\textbf{Sybil attack.}
The attack \cite{douceur2002sybil} is a common attack in the P2P network which is also adaptable to the blockchain network. An adversary may generate multiple pseudonymous identities to conduct malicious behaviors, such as disconnecting the channel between nodes or even taking control of the whole network. Specifically, an adversary may add a large number of pseudonymous identities in the blockchain network to participate in the membership selection. If one of them is luckily selected to be the committee member, s/he can arbitrarily generate blocks that benefit himself. The fake members can efficiently forward the block generated by adversaries whilst abandoning the genuine blocks from users. As a result, the adversary who controls the network finally reduces the overall throughput and obtains extra rewards. The more fake identities are created,  the larger chance they are selected. The key to successfully launching this attack is to generate a large number of identities with low/zero costs. Then, a malicious node will use these identities to gain disproportionately significant influence.

The attack relies on the assumptions: a) generate identities without any cost; b) the nodes have enough power to efficiently create identities. Sybil attack affects the network by its generated massive identities. Current blockchains that adopt PoS/PoW consensus effectively avoid this vulnerability since becoming a block packager (membership selection) must simultaneously solve a lot of puzzles (block production).  Therefore, this attack mainly targets the systems whose committee selection and block production are separate. We take the Blockclique as an example to show how to avoid this attack in DAG-based blockchain systems. 

\smallskip
\noindent\textit{Instances.} Blockclique \cite{forestier2018blockclique} indicates that its protocol relies on a Sybil-resistant selection mechanism. Achieving this mechanism needs to meet two requirements: a) a valid membership of nodes must be obtained with a sufficient cost, and b) the membership in the next round cannot be guessed in advance. The key of the first requirement is to explicitly select a node based on the \textit{proof of ownership of a resource}: nodes have to spend some resources and add certain delays each time participate in the node selection. The more costs or mortgages are spent, the more reputation a node will obtain. This increases the confidence of a node, decreasing friction when communicating with others. The core of the second requirement is to guarantee randomness without leakage. Blockclique utilizes an oracle to describe the random selection of nodes. All nodes need to consult this oracle to confirm their membership in a certain slot. To meets these requirements, Blockclique gives two adaptable mechanisms: the PoW-based membership selection as in OHIE \cite{yu2020ohie} or ELASTICO \cite{luu2016secure}, and the PoS-based selection as in Tezos \cite{tezos}. OHIE utilizes the residue of hash to decide which thread it switches to. ELASTICO employs PoW to generate identities and then select the membership randomly from those identities. Nodes in Tezos are randomly selected with a probability proportional to their stakes. All these selections prevent the adversaries from maliciously spawning massive nodes.

%============================
\subsection{System Challenges} 
This part focuses on design considerations from the view of system level and development feasibility.

%This subsection provides a handful of system issues that should be concerned in current DAG-based blockchain systems. 

\smallskip
\noindent\textbf{Absent Details of Committee Configuration.}
Maintaining liveness and security in committee configuration is an open and neglected issue. We observe that current committees are formed either statically or dynamically. Specifically, static committee configuration lacks liveness, making systems vulnerable to attacks like the DDoS attack, Sybil attack, and censorship attack. Dynamic committee configuration improves security by raising the bar for attacks but brings new challenges on how to maintain liveness. It includes the frequency and fraction of committee membership (epoch, dynamism). The property directly relates to the committee's security. However, the systems which seriously discuss or comprehensively analyze the issues only make up a minority.

\smallskip
\noindent\textbf{Absence of System Setup.} Setup configuration refers to the information that can be available at the onset of the protocol to each participant. Three types of setup configurations are introduced in \cite{garay2020sok}, namely \textit{no setup}, \textit{public-state setup} and \textit{private-state setup}. Most NC protocols (Bitcoin, \textit{etc.}) rely on a global-viewed genesis block as the initial state, while a number of systems (like \cite{garay2018bootstrapping}) need pre-existing configurations (such as a common reference string [CRS], or a public-key infrastructure [PKI]). In current DAG-based blockchains, some of the systems rely on the genesis block, mainly  \textit{TypeI/II/V/VI}, whereas other systems simultaneously initialize several chains in parallel, mainly \textit{Type III/IV}. It is unclear how these (parallel chain) blockchains bootstrap the systems.

\smallskip
\noindent\textbf{Absence of Incentive Mechanism.}
Three major goals of incentive mechanisms are a) letting more nodes get involved in a committee to participate in consensus; b) encouraging honest behaviors of committee members; and c) directly affecting consensus mechanisms, such as Graphchain \cite{boyen2018graphchain}. There has been little investigation into how to build incentives for increased participation. The absence of incentives makes it hard for committee members to maintain the system stable since no laws guarantee they never get compromised. Although several DAG systems claim that \textit{feeless} is one of the most outstanding advantages of DAG, they still utilize additional techniques as complementary mechanisms. For example, IOTA \cite{popov2016tangle} introduces an authority to maintain stability.

\smallskip
\noindent\textbf{Unknown Compatibility.}
Although this paper provides insights into physical components (structure, consensus) and featured techniques (see \underline{Section \ref{sec-consensus}}), there are still a large number of related components and techniques that have a significant influence on their designs/operations. Current studies lack the corresponding description, and we provide some instances as follows.

\begin{itemize}
    \item[-] \textbf{\textit{Smart contract}} is an essential component to achieve state transition in classic blockchain systems. However, none of the state-of-the-art DAG systems have implemented the mature smart contract. This is mainly because a complete state transition highly relies on the total linear order of units, ensuring the correctness and integrity of states. The systems without linear order can only encompass auxiliary components. For example, IOTA provides a decoupled upper layer solution \cite{qubic20} as the external component.

    \item[-] \textbf{\textit{Sharding}} \cite{bano2019sok} splits transactions into separated zones. Several DAG systems  (Blockchique \cite{forestier2018blockclique}, Eunomia \cite{niu2019eunomia}, Dexon \cite{chen2018dexon}) get inspired from the sharding technique to allocate the transactions into different chains in advance, preventing potential conflicts during the consensus. However, DAG systems only capture the simplest and most straightforwards concept from sharding. The issues such as how to fairly and randomly allocate the transaction, and how to maintain the separate chains relatively balances, are still left as gaps.
    
    \item[-] \textbf{\textit{Layer2 technique}} \cite{gudgeon2020sok} is an upper layer solution for linear-based blockchains. As discussed in \underline{Section \ref{subsec-difference}}, the layer2 technique attempting to solve performance issue share lots of similarities with the DAG approach in the perspective of system design. It is interesting to explore whether the layer2 technique can support more applications for DAG systems, especially the scenarios having to manage both off-chain and on-chain transactions.
    
    \item[-] \textbf{\textit{Cross-chain}} \cite{zamyatin2019sok} enables the chains that are heterogeneously structured able to communicate with each other. The usage of this technique requires powerful tools such as the global clock. Traditional blockchain systems achieve the goal of either relying on a third party as an external clock or depending on chain-dependent time definition, such as the block generation rate. The tools are relatively easy to implement on their (unique) main chains. In contrast, DAG systems can hardly achieve the regiments for cross-chain due to their inherent complex situations: they create many forks (\textit{Type V/VI}), coexist multiple chains (\textit{Type III/IV}) or even without the main chain  (\textit{Type I/II}). The difficulty to synchronize the information, both internally and externally, hinders the application across chains in DAG systems.
    
\end{itemize}

\smallskip
\noindent\textbf{Deployment Difficulty.}
Deploying proposed DAG-based systems is hard for developers. Most implemented systems, like \cite{zhou2019dlattice}\cite{forestier2018blockclique}, only present the designed protocols with incomplete evaluations, rather than releasing their source codes. A developer can hardly learn the full logic by conditional access to the repositories. Many systems \cite{li2020decentralized}\cite{liu2018vite}, with accessible repositories, do not provide proper documents on guidelines or deployment. Several systems (\textit{e.g.}, \cite{baird2016swirlds}) developed by official teams are protected by patents. We denote them as \textit{n/a}: the system can be deployed but lacks evidence to show difficulties. The current status indicates that the widespread usage of applications built on DAG-based blockchain systems is still far away from reality.

\smallskip
\noindent\textbf{Trade-off.}
Blockchain systems cannot improve the performance, scalability, security, decentralization, and strict consistency at the same time. For example, we observe that some systems, like IOTA \cite{popov2016tangle}, Graphchain \cite{boyen2018graphchain}, improves the performance and scalability at the expense of security and consistency.  While some systems, such as Prism \cite{bagaria2019prism}, OHIE \cite{yu2020ohie} ensures strict consistency, but the scalability and performance are sacrificed as a balance. In traditional distributed systems, FLP Impossibility theorem \cite{FischerLP85} and CAP \cite{GilbertL02} theorem help us understand the limitation of some factors. However, in DAG-based systems, or even blockchain systems, there is no theorem or experience to clarify the boundaries of each factor, the way of reaching a balance between multiple factors, or even the range of factors that are involved.

%============================
\subsection{Theoretical Analysis} 
Formal treatment is important for system security. However, only a minority of systems present analysis under specific models.

\smallskip
\noindent\textbf{Informal Model.} Many systems are informally presented and lack formal models either on security or properties. This leads to confused proofs or discussions when they claimed to be secure. Several systems are presented in a formal manner, formalizing models consistent with their properties. Although their models vary from system to system, we still list them as typical samples (cf. \underline{Table.\ref{tab-formalModel}}). Besides, several studies provide analysis with formal analyses, including topics on properties \cite{birmpas2019fairness} or performance (of parallel-chain systems) \cite{fitzi2018parallel}.  We encourage future work to conform to this format, enabling them to be thoroughly discussed and critically proved.

\begin{table}[!hbt]
%\begin{table}[H]
 \caption{DAG Systems with Formal Models} 
 \label{tab-formalModel}
  \centering
 \resizebox{0.95\linewidth}{!}{ \begin{tabular}{lc}
    \toprule
      & \textbf{Adopted Model}  \\
    \midrule
    IOTA \cite{ferraro2018iota}\cite{popov2019equilibria}\cite{kusmierz2018extracting}  &  Markov Process  \\
    Avalanche \cite{rocket2019scalable}  & Continuous-Time Markov Process  \\
    GHOST/Inclusive \cite{sompolinsky2015secure}\cite{lewenberg2015inclusive}  &  Nakamoto's   \\ 
    Spectre/Phantom \cite{sompolinsky2020phantom}   &  Self-defined \\
    Prism \cite{bagaria2019prism}  &  Nakamoto's (Backbone \cite{garay2015bitcoin}) \\ 
    OHIE \cite{yu2020ohie} &  Nakamoto's (Backbone \cite{garay2015bitcoin})  \\ 
    Conflux \cite{li2020ghast} &  Nakamoto's   \\ 
    Haootia \cite{tang2020haootia} & Crypto-style proof \\
     
    \cmidrule{1-2}
     \multirow{2}{*}{\textit{Informal}}  & \cite{boyen2018graphchain}\cite{sompolinsky2016spectre}\cite{lemahieu2018nano}\cite{bentov2017tortoise}\cite{he2019consensus}\cite{baird2016swirlds}\cite{xiang2019jointgraph}\cite{will2019chainweb}\cite{gkagol2019aleph}    \\ 
         &  \cite{choi2018fantom}\cite{liu2018vite}\cite{zhou2019dlattice}\cite{amiri2019caper}\cite{danezis2018blockmania}\cite{gupta2019cdag}\cite{yin2019streamnet}  \\ 
    \bottomrule
  \end{tabular}}
  \label{server}
%\end{table}
\end{table}

Informal models lead to non-unified properties in different systems. We mention several types of properties that defined in concurrent literature, such as \textit{effectiveness}, \textit{atomicity} and \textit{timeliness} in \cite{zamyatin2019sok}; \textit{efficiency} in \cite{birmpas2019fairness}; \textit{liveness} and \textit{correntness} in \cite{bonneau2015sok}; \textit{persistence} and \textit{liveness} in \cite{garay2015bitcoin}; \textit{etc}.

\smallskip
\noindent\textbf{Whitepaper Information.}
Whitepapers were the earliest documents to express ideas and proposals from developers and technicians. These documents encompass brief models, related components, and, more importantly, the developing plan, which targets future investors. In this paper, not all the discussed DAG systems are written in the form of academic papers. For those non-academic documents, we only select and discuss systems with formal documents, which present design details and technical models consistent with the literature. Besides selected systems, a series of ever-existing projects confront the situation of non-updating. Most of them follow similar models and we omit them here.

%============================
\subsection{Privacy-Preserving Properties} 
Privacy in classical blockchain systems \cite{conti2018survey} is a complicated issue. Generally, the privacy of blockchain systems includes two categories: \textit{anonymity of identities} and \textit{privacy of metadata}. Firstly, Bitcoin allows users to generate multiple addresses. Adversaries can obtain the link between these addresses and user entities by collecting massive historical transactions and analyzing the relations to pursue profits \cite{miller2017empirical}. Most DAG systems following this concept will confront the same threats. Secondly, the metadata on the ledger is shown in plaintext as default. Adversaries can trace any interested transactions (\textit{e.g.}, with large balance) and conduct harmful activities by locating their associated accounts \cite{wang2020preserving}. Unfortunately, we find that none of the existing DAG-based blockchain systems (at least till the completion of this paper) empower their system with privacy-related properties. No (effective) privacy-preserving solutions have been applied to DAG systems so far. Therefore, ways to properly enhance privacy are a promising area.

\begin{table*}[!ht]
\newcommand{\tabincell}[2]{\begin{tabular}{@{}#1@{}}#2\end{tabular}}
\caption{Attack Defense, System Challenge, Model Availability and Privacy\tnote{1} }
\label{tab-security}
\begin{center}
\begin{threeparttable}

\resizebox{\textwidth}{!}{
%\begin{tabular}{l p{1mm}p{1mm}p{1mm}p{1mm}p{1mm}p{1mm}p{1mm}p{1mm}p{1mm}p{1mm}p{1mm}p{1mm}p{1mm}p{1mm}p{1mm}p{1mm}p{1mm}p{1mm}p{1mm}p{1mm}p{1mm}p{1mm}p{1mm}p{1mm}p{1mm}p{1mm}p{1mm}p{1mm}p{1mm}p{1mm}p{1mm}p{1mm}p{1mm}p{1mm}p{1mm}p{1mm}p{1mm}p{1mm}}
\begin{tabular}{l lllll lllll lllll lllll lll lllll lllll lllll }
\toprule
& \multicolumn{4}{c}{\textbf{Type I}} 
& \multicolumn{3}{c}{\textbf{Type II}}
& \multicolumn{9}{c}{\textbf{Type III}} 
& \multicolumn{8}{c}{\textbf{Type IV}} 
& \multicolumn{3}{c}{\textbf{Type V}} 
& \multicolumn{5}{c}{\textbf{Type VI}}
\\

\cmidrule(lr){2-5} 
\cmidrule(lr){6-8}
\cmidrule(lr){9-17}
\cmidrule(lr){18-25}
\cmidrule(lr){26-28}
\cmidrule(lr){29-33}

& 
\rotatebox{90}{\textbf{IOTA}} & 
\rotatebox{90}{\textbf{IOTA}$^{+}$ \cite{cullen2019distributed} } & 
\rotatebox{90}{Graphchain} & 
\rotatebox{90}{Avalanche} & 
\rotatebox{90}{\textbf{Spectre}} &
\rotatebox{90}{Phantom} & 
\rotatebox{90}{Meshcash} & 
\rotatebox{90}{\textbf{Nano}} &
\rotatebox{90}{Hashgraph}  &
\rotatebox{90}{DLattice} & 
\rotatebox{90}{Jointgraph} & 
\rotatebox{90}{Chainweb}  & 
\rotatebox{90}{Aleph} &
\rotatebox{90}{Vite} &
\rotatebox{90}{Caper}  &
\rotatebox{90}{$\mathcal{L}$-series }  &
\rotatebox{90}{\textbf{Prism} } & 
\rotatebox{90}{\textbf{OHIE}} & 
\rotatebox{90}{Sphinx }  & 
\rotatebox{90}{Blockmania } &
\rotatebox{90}{\textbf{Blockclique} } &
\rotatebox{90}{Eunomia}  &
\rotatebox{90}{Dexon} & 
\rotatebox{90}{PARSEC} & 
\rotatebox{90}{Byteball}  & 
\rotatebox{90}{Haootia} &
\rotatebox{90}{JHdag}  &
\rotatebox{90}{\textbf{GHOST}} & 
\rotatebox{90}{Inclusive} & 
\rotatebox{90}{CDAG}  & 
\rotatebox{90}{\textbf{Conflux}} &
\rotatebox{90}{StreamNet}  
\\   
\midrule %********************************  

Parasite chain at. & & \checkmark & & & \checkmark  \\

Balance at. &&&&&&&&&&&&&&&&& \checkmark &  \checkmark &&&&&&&&&& \checkmark &&& \checkmark    \\

Large Weight at.  &\checkmark & \checkmark   &&&&&&&&&&&&&&&&&&&&&&&&&&&&& \checkmark\\

Censorship at.  &&&&&&&&&&&&&&&&& \checkmark  \\

Sybil at. &&&&&&&& \checkmark &&&&&&&&&&&&& \checkmark \\

\midrule

Committee Conf.  & - &- & - & - & - & - & - &  dy  &  dy & dy & st & dy & st  & st & st & dy  &  dy & st & st & dy & dy & dy & st & st & dy & - & - & - & - & - & - & -\\ %\cellcolor{gray}

System Init.  &  \multicolumn{7}{c}{genesis unit}  &  \multicolumn{17}{c}{CRS / PKI / genesis unit}  &  \multicolumn{8}{c}{genesis unit}  \\ 

Incentive &&& \checkmark &   &  \multicolumn{24}{c}{\XSolidBrush} \\ 

Compatibility (SC)  &&&&&& \checkmark  &&&&&& & \checkmark & \checkmark & \checkmark & \checkmark  & & \checkmark &&&& \checkmark & \checkmark & \checkmark   && \checkmark & \checkmark & \checkmark &  \checkmark & \checkmark & \checkmark & \checkmark  \\ 

Deployment  & \checkmark & - &  \multicolumn{3}{c}{n/a} & - & - & \checkmark  & \multicolumn{3}{c}{n/a} & - & - & - & n/a & -& \multicolumn{5}{c}{n/a}  &- &- & - & \multicolumn{2}{c}{n/a}  & - &  \checkmark &  \checkmark &  \multicolumn{2}{c}{n/a} & -     \\ 

\midrule

Formal Model  &  \checkmark &  &  &  \checkmark & \checkmark & \checkmark &&&&&&&&&&& \checkmark & \checkmark &&&&&&&& \checkmark && \checkmark  & \checkmark  & & \checkmark  \\ 

Document  & $\dag\ddag$  & $\dag$ & $\dag\ddag$  & $\dag\ddag$  & $\dag\ddag$ & $\dag$ & $\dag$ & $\dag\ddag$  & $\ddag\flat$ & $\dag\ddag$ & $\dag\ddag$ & $\dag$  & $\dag$ & $\ddag$ & $\dag\ddag$ & $\dag$ & $\dag\ddag$ & $\dag\ddag$ & $\dag\ddag$ & $\dag\ddag$ & $\dag\ddag$ & $\dag$ & $\dag$ & $\dag$  & $\dag\ddag$ & $\dag\ddag$ & $\dag$  & $\dag\ddag$ & $\dag\ddag$ & $\dag\ddag$ & $\dag\ddag$ & $\dag$  \\  

\midrule

Privacy & \multicolumn{32}{c}{\XSolidBrush} \\ 

\bottomrule %************************************
\end{tabular}

}

\begin{tablenotes}
       \footnotesize
       \item[] \scriptsize \checkmark = prevent attacks / with the property / easy to use; default (blank) = not mentioned or N/A;  - = do not apply; \\ \XSolidBrush =  without consideration.  CRS = common reference string; SC= smart contract; 
        at. = attack; conf.= configuration; init. = initiation; \\ dy = dynamic setup; st = static setup; 
        $\dag$ = Academic documents available; $\ddag$ = Implementation available;  \\ $\flat$ = Patent protection;  
        default =  Whitepaper/Concept;  n/a = can be deployed but lack of evidence to show difficulties.
       
     \end{tablenotes}
   \end{threeparttable}
\end{center}
\end{table*}
 
%---------------------
%---------------------

\subsection{Summary} 
\textit{\textbf{Firstly}}, \textit{removing the leader/miner makes DAG systems vulnerable towards attacks with intensive-computing power.} The systems with natural topology (\textit{Type I/II/V/VI}) require auxiliary techniques to maintain stability and security, \textit{e.g.}, the coordinator in IOTA \cite{coordnator} and the mainchain in \textit{Type V/VI} systems. \textit{\textbf{Secondly}}, \textit{building smart contracts or state transition applications rely on linear order.} The transition of states differs from the operation of a token transfer. State transition requires multiple steps where each state can be traced back to the initial state. In contrast, a token transfer finishes once both the sender and receiver agree to it. \textit{\textbf{Thirdly}}, \textit{very few systems consider the incentive mechanisms.} The systems that follows the Nakamoto's model (like GHOST \cite{sompolinsky2015secure}, Inclusive \cite{lewenberg2015inclusive}, Conflux \cite{li2020decentralized}) inherit the reward system of Bitcoin. This is feasible because the system still relies on a mainchain. Other types of systems do not provide solutions on how to inspire more participants. \textit{\textbf{Next}}, \textit{none of DAG blockchains equips privacy-related techniques or crypto primitives.} We think the reason comes from their goals. Improving the performance and scalability is the key target of DAG systems. Additional privacy protocols will greatly increase the complexity and slow down the confirmation time. \textit{\textbf{Last}}, \textit{a universal theoretical analysis model will not appear (at least in a short time) in DAG systems.} We have identified six types of models, but none of them can share the same model. A suitable analysis closely depends on its design model.

%=================================================
\section{Performance}
\label{sec-performance}
%=================================================

In terms of performance, we consider \textit{throughput} (\textit{i.e.}, the maximum rate at which values can be agreed upon by the consensus protocol), \textit{scalability} (\textit{i.e.}, the system's ability to achieve greater throughput when consensus involves a larger number of nodes) and \textit{latency} (\textit{i.e.}, the time it takes from when a value is proposed, until when consensus has been reached on it). Here are the details.

\begin{itemize}

    \item[-] \textbf{\textit{Throughput}} ($\lambda$) is the maximum rate of confirmed units over the network. It is affected by various factors, including the limitations of bandwidth, the design of consensus protocols, and the fraction $\beta$ of (hashing power of) the adversaries. The throughput in DAG systems calculates the confirmed total units during a specific time frame, no matter they are collected from one single chain or multiple chains. We follow the classic measures called \textit{TPS} (Transaction Per Seconds). 

    \item[-] \textbf{\textit{Latency}} ($\tau$) mainly consists of two parts, the unit \textit{propagation time} and \textit{confirmation time}. The propagation time is the length of time that a unit reaches its destination. It is closely related to the parameter \textit{coverage}, used to describe how many nodes received these broadcasted units. The confirmation time represents the length of time that a unit is deemed as confirmed. For instant confirmation protocols (\textit{e.g.}, BFT-style), the consumed time is static on average. For delayed confirmation protocols (\textit{e.g.}, NC-based), the time is unpredictable. The probability $\varepsilon$ of a unit is removed by its child will drop as the chain/graph grow. When the unit is buried deep enough, an approximate time is obtained.     

    \item[-] \textbf{\textit{Scalability}} ($\phi$) shows the ability of a system to handle a growing amount of units when adding a large number of nodes. The assumption of consistency is the leading factor that influences the property. The strict consistency (which requires total linear order of the units) will significantly retard the performance since the complexity of consensus increase with the enlarged committee. In contrast, a weak consistency can greatly scale the blockchain due to its tolerance of forks. Other factors including the design of consensus protocols, the fraction of malicious nodes, \textit{etc}.

\end{itemize}

%============================
\subsection{General Analysis}
Performance is the bottleneck of Bitcoin due to its security limitation. Simply increasing the mining rate can improve the performance but with the expense of decreased security. Suppose the block creation rate of an attacker is greater than the growth rate of the main chain, the malicious chain will eventually outpace the honest chain. Thus, increasing the mining rate while maintaining security is critical for improving the throughput and latency of protocols. Current (leading) systems avoid the aforementioned limitations in three independent ways as follows.

The \textbf{\textit{first}} line of work  (GHOST, Conflux) provides the solutions by \textit{maximizing the mining rate until reaching the security-limitation}. These solutions follow the main architecture of Bitcoin, but with the modifications of adopting more sophisticated extension rules to solve forks. The systems mainly belong to \textit{Type V/VI}. These systems have to extend graphs and sort units in the main chain. Such designs confront a similar bottleneck with classic blockchain systems. Improving performance relies on decreasing the confirmation time of the main chain. The time cost from conflict solving and unit sorting limits the upper bound of throughput.

The \textbf{\textit{second}} line of work (Prism, OHIE) aims to \textit{horizontally scale the blockchain by enabling multiple chains processed in parallel}. The system adopting this type of method generally has a fixed security threshold.  Within this threshold, the throughput can maximally approach the communication limitation. Conversely, once exceeding the threshold, the system is limited by its security bound. Improving scalability without undermining safety is the key point in this line of studies. This design pattern mainly appears in \textit{Type III} and \textit{Type IV}.  The systems are based on structured DAGs and each individual chain speeds up simultaneously. 

\textbf{\textit{Another}} line of work (IOTA, Avalanche, Spectre, Phantom) \textit{reconstructs systems to reach the (physical) communication-limitation}. The approach greatly modifies the original designs,  decreasing the confirmation time as well as improving scalability. However, as a sacrifice, systems inevitably either weaken the security promise or increase the system's complexity. Such limitations arise from excessive random forks due to unstructured networks when the mining rate is increased. These systems usually appear in \textit{Type I/II}.

\begin{figure}[!hbt]
\centering
\includegraphics[width=0.42\textwidth]{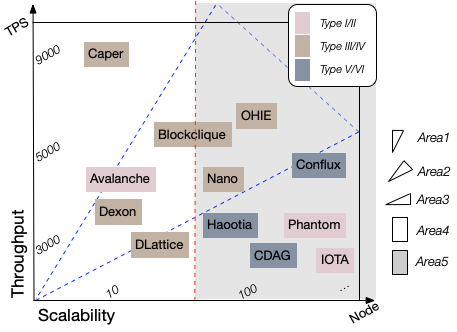}
\caption{Throughput and Scalability}
\label{fig-performance}
\end{figure}

\medskip
\noindent\textbf{Observations.} We further capture several experimental testing results from the corresponding literature and list them in \underline{Fig.\ref{fig-performance}}. The performance of a DAG-based blockchain system can be measured in its \textit{total TPS}, equivalently, the multiplication of \textit{individual chain throughput} (in \textbf{y} axis), and participated nodes (\textit{scalability} in \textbf{x} axis). The entire figure can be made up by two ways: \textit{Area1}, \textit{Area2} and \textit{Area3} separated by blue lines, or \textit{Area4} and \textit{Area5} by red lines. Based on results, we observe several facts:

\begin{itemize}
\item[$\diamond$] Most DAG blockchains locate in \textit{Area2} and \textit{Area3}. This indicates the performance improvements of DAG solutions mainly come from the contribution of scalability. Systems rely on parallel chains and side chains to carry more transactions in each short period.

\item[$\diamond$] \textit{Type I/II} and \textit{Types V/VI} systems locate in \textit{Area3} with a high probability while \textit{Type III/IV} in \textit{Area3}. This reflects DAG blockchains with natural-graph topology (mainchain-based designs included) perform better at scalability, but are poor in individual throughput. The is because frequent forks in the network retard the confirmation time of transactions. In contrast, parallel-chain based protocols sitting in the middle of the figure balance these two factors. The more nodes joining the networks, the more witnesses are required in each round of voting, requiring much time for negotiation.

\item[$\diamond$] Several systems locate in \textit{Area4} while others in \textit{Area5} without seemly associations. We review their decoupled components and find some commons. \textit{Area4} systems mainly adopt BFT-style consensus mechanisms or rely on TR-based protocols with instant decisions. This would facilitate confirmation procedures but cannot scale chains very well. On the contrary, \textit{Area5} systems leverage NC/NC-variants protocols with permissionless settings. Such designs provide satisfactory scalability but are poor in throughput.

\item[$\diamond$] As discussed, we identify two main factors that have significant impacts on the system performance, namely \textit{structure topology} and \textit{consensus mechanism}. A graph-based topology can greatly improve scalability while BFT-style consensus brings benefits in chain throughput (fast confirmation). Besides, other techniques also play critical roles. We discuss them separately in instance analyses.

\item[$\diamond$] We observe that existing DAG blockchain systems cannot achieve both individual throughput and scalability at the same time. Frankly speaking, it is not clear what the optimal solution is for the sweet spot when considering a trade-off between scalability and throughput. The bottleneck might be limited by security considerations or physical boundaries.
\end{itemize}

%=========================
\subsection{Instance Analysis}
We provide detailed analyses of mentioned instances. Due to the differences between experimental settings and hardware devices, we focus on qualitative reasons that help to gain high performance or cause potential sacrifices, instead of quantitative outcomes.

GHOST \cite{sompolinsky2015secure} follows the model of Bitcoin, but with modifications on chain selection rules. The protocol extends the chain by selecting the \textit{weightest} subtree, instead of the \textit{longest} subchain. It greatly changes the way to organize blocks. Unlike Bitcoin's linear structure, GHOST organizes blocks in a tree structure which improves the utilization of mining power under high contention. This design enables GHOST to significantly improve the block generation rate, and decrease the confirmation time of blocks. A high generation rate leads to unsolvable forks in Bitcoin, whereas a large proportion of the forks are tolerable in GHOST. As proved in \cite{sompolinsky2015secure}, when the rate moronically increases, the security threshold of Bitcoin sharply decreases, whilst the GHOST protocol maintains stability. Forks frequently appear in a high block generation rate, but GHOST is more resilient and scalable than Bitcoin.

Conflux \cite{li2020decentralized} inherits the design of GHOST to achieve high performance without undermining security. Standard NC-based protocols reach a total linear order of transactions while confirming blocks. The key insight of Conflux is to decouple the confirmation of blocks against transactions. Differing from Bitcoin and GHOST, Conflux defers the steps of the transaction sorting, making it processed after the block confirmation, rather than at the same time. The order of transactions can be deterministically derived from the agreed order of blocks. The dependency between these two components is thus changed. Each time a block only needs to concentrate on a gross-grain sorting of blocks, without having to simultaneously consider the fine-grain sorting of transactions. Besides, Conflux enables each block to reference multiple edges at each time, which allows concurrent transactions and blocks. Conflux also stresses its adaptive weight mechanisms to switch the strategies between a fast model and a slow model. This design smoothly balances security as well as performance.

Spectre \cite{sompolinsky2016spectre} and Phantom \cite{sompolinsky2020phantom} embrace the DAG structure to support faster block generation and larger block volume. The improved performance of Spectre comes from two aspects: a) The system structures blocks to form a topological network. Transactions can concurrently attend to the network, making the system scalable. b) Increasing the block generation rate contributes to performance. This is due to the fact that Spectre only requires a pairwise ordering between two blocks, avoiding the barrier of conflicting states among many blocks. Similarly to Spectre, blocks are created and attached concurrently in Phantom. The main difference is that Phantom requires a total linear order of blocks and transactions. This makes Phantom have to maintain a trade-off between security and performance. Simply increasing the block generation rate will unpredictably affect several associated factors. Thus, the confirmation time of transactions is much slower than that in Spectre but still faster than any NC-based blockchain system under the single linear chain structure.

IOTA \cite{popov2016tangle}, adopting the third method, both decrease the confirmation time and improve the system scalability. It relies on a simple and intuitive tip selection rule to extend the graph without any further limitations. The fast block creation rate directly improves the confirmation rate for pending transactions. A peak rate ($>400$ TPS) has been observed with a confirmation coverage of over 99\%. It is sufficient for most application cases. Further, we capture two features with respect to the performance of IOTA. a) Theoretically, the performance is merely limited by the physical hardware, including bandwidth and propagation delay. The Tangle can grow without hindrance and forks are tolerable.  b) The confirmation time of transactions significantly varies according to their confidence. The higher confidence a transaction initially obtains, the faster it can be confirmed. Uneven distribution of confirmation time is formed since most of the transactions follow biased incentives. As a sacrifice, the design of IOTA inherently weakens the strong guarantee of consistency and security. IOTA has to (temporarily) rely on a centralized Coordinator to periodically take snapshots of the history to balance stability and performance.

Avalance \cite{rocket2019scalable} redesigns the consensus protocols by leveraging network sampling. Similar to IOTA, Avalanche achieves high throughput and scalability with the expense of undermining strict consistency. The high performance derives from three aspects: a) Leaderless BFT protocols in Avalanche avoid the congestion and vulnerability of the centralized authority; b) The system only maintains correct pairwise orders rather than a total linear sequence of transactions. c) The complexity of handling a message is fixed ($O(k)$), free from the expansion of the network. Besides these three intuitive reasons, an additional factor is the fraction of Byzantine nodes in the network. The probabilistic guarantee of its safety will significantly degrade when Byzantine participants exceed $f$ due to the exponential effect of the repeated sampling. Keeping the fraction at a low level enables the subsampled voting mechanism to reach the consensus more safely and efficiently.

Prism \cite{bagaria2019prism} deconstructs the Nakamoto consensus into basic functionalities and reconstructs its new system by scaling up these functionalities. The bottleneck is limited by physical conditions, rather than security requirements. The key point is that all functionalities are decoupled and independent. This design greatly improves the performance in the following aspects: a) The transaction blocks are instantly confirmed by the powerful leader blocks without delay. The confirmation does not rely on any intensive resources like computing power or stakes, which enables fast execution. b) Th leader block confirms a list of possible ledgers instead of a unique ledger, making non-conflicted transactions simultaneously processed. c) The confirmation of a leader block will not be hindered by the states of voter blocks. Once the leader block receives a majority of votes from voter blocks, it becomes permanent without having to worry about situations of voter blocks. Prism can be deemed as the first NC-based protocol (\textit{w.r.t} the leader chain) that breaks the confirmation barrier of transactions.

OHIE \cite{yu2020ohie} scales NC-based protocols by processing chains in parallel. The system achieves high performance as well as maintains a linear order of transactions thanks to the design that speeds up almost every procedure. Specifically, OHIE makes improvements in the following aspects: a) A pre-allocation of transactions into individual chains avoids duplicated confirmation and potential conflicts. Assume a total of $k$ transactions and $n$ nodes, each chain just needs to take $1/n$ workload to process $k/n$ transactions. b) The system applies NC-based extension rules to every single chain, rather than the whole system. This is essential to overcome the bottleneck because much of the time is wasted under the high contention of forks. Instead, OHIE sorts the blocks by comparing their intrinsic parameters (positions) at a high speed. 3) OHIE decouples the functions of confirmation and sorting where sorting blocks happen before the confirmation of blocks. The sorting relies on the location of a block, while the confirmation is based on its proposed \textit{confirm bar}. This design provides an additional benefit of scalability. Unlike BFT-style protocols, newly joint nodes (with legal permission) will not increase the burden of existing ones and the system can still reach peak performance. The limitation of OHIE comes from its sorting algorithm.

\subsection{Summary}
Based on previous discussions, we conclude with some insights. Since the design of each DAG system varies from one to another, we capture several common features as follows.

\begin{itemize}
  \item[$\diamond$] \textit{We can hardly improve the scalability and throughput at the same time under the assumption of strict consistency}. The more nodes are involved, the more time will be spent on decisions. Even though each type of system is radically different, they still have to reach the trade-off between security and performance to a certain degree. 
  
  \item[$\diamond$] \textit{To achieve strict consistency, the sorting algorithm might become the bottleneck in DAG-based systems}. The computing complexity in sorting algorithms exponentially increase with the participating members. One particular case is that the committee member only has one member (might dynamically change) -- a leader who can decide the sequence, such as Vite \cite{liu2018vite}, Jointgraph \cite{xiang2019jointgraph}. This design maximally decreases the confirmation time but results in centralization. 

  \item[$\diamond$] \textit{Weakening the strict requirements of consistency enables high scalability and fast confirmation}. Systems like IOTA \cite{popov2016tangle}, Nano \cite{lemahieu2018nano}, Spectre \cite{Sompolinsky2017SPECTRES}, only guarantee the partial/pairwise order of the transactions. This is enough for asset transferring between two/multiple parties, but cannot support strict state transitions like smart contract components.
  
  \item[$\diamond$] For the NC/NC-variant based DAG systems, we observe that \textit{the \textit{depth} $d$ only has impacts on the confirmation latency, rather than the throughput and the propagation latency}. \textit{Depth} is used to measure how deep a block is buried in the chain.
  
  \item[$\diamond$] \textit{Theoretical analysis of performance usually adopts Nakamoto's model} \cite{garay2015bitcoin}, such as the analysis in GHOST \cite{sompolinsky2015secure}, Prism \cite{bagaria2019prism}, OHIE \cite{yu2020ohie} Conflux \cite{li2020ghast}. Unfortunately, only a handful of them provides experimental evaluations.

\end{itemize}

%=================================================
\section{Further Discussion}
\label{sec-discussion}
%=================================================

In this section, we discuss a) ways to integrate each component (structure, consensus, properties) and their impact on each other; b) differences between the DAG technique and existing scaling approaches; c) related components with their impact on DAG systems; d) challenges surrounding DAG-based blockchains.

%---------------------------------------Components
%---------------------------------------Components
\smallskip
\subsection{Intra-Component}
This subsection explores the mechanisms, features, and differences of each sub-component inside each component. We discuss two physical components \textit{structure} and \textit{consensus mechanism} and one virtual component \textit{property}.

\textbf{\textit{Structure}} defines the way units are recorded in DAG systems. It consists of two aspects, namely \textit{data structure} and \textit{graph topology}. Traditional blockchain systems neglect the second factor since they rely on the linear-chain topology as default. In contrast, graph-based topology in DAG systems is the most distinguishable feature. Here, we provide more discussion.

\textit{Data structure.} The metadata stored on the ledgers can be structured in either transactions or blocks. A major difference between these ``containers'' is their permission to operate the ledgers. This includes the rights of \textit{read} and \textit{write}. \textit{Read} means a node can access the states in the ledger and obtain the messages to generate his new transaction without changing any previous states. \textit{Write} means the node can change the states of the ledger by adding new content. In the transaction-container systems, all participants can both \textit{read} and \textit{write} the data into ledgers. The nodes who receive legal transactions will directly append them into their ledgers. However, in block-container systems, the committee members (miner/validator/\textit{etc.}) have the rights of \textit{read} and \textit{write}, whereas the ordinary participant can only \textit{read}. Transactions sent by the ordinary nodes require to be packaged by the committee nodes and then written to the ledger after the consensus. Ordinary participants cannot directly affect the status of the account. Additionally, the difference also lies in its role of \textit{filter}. The committee nodes can filter malicious transactions to some extent. A double-spending transaction will be abandoned directly, rather than being appended to the ledger. This effectively resists the flooding attack and replay attack.

\textit{Topology.} Topology refers to the graph formed by the units and their relationships. It also describes the way to organize each unit in the network. Three design patterns are identified in \underline{Section \ref{sec-modeling}}, namely \textit{divergence} ($\widehat{D}$), \textit{parallel} ($\widehat{P}$), and \textit{convergence} ($\widehat{C}$). The topology cannot determine the final properties. It only shows the visualized graph made by edges and vertexes. Additionally, the topology can indirectly reflect the complexity of the protocols.  $\widehat{D}$ type contains the most flexible systems without strict/uniform consensus mechanisms. Transactions and blocks can arbitrarily be dispersed in the network. Only a simple rule (TSA/RTA/GA/Sampling) is applied to each system. $\widehat{P}$ type includes the systems with a group of nodes who separately maintain their chains/accounts. The complexity depends on their aimed properties, such as the linear order. An additional sorting algorithm is introduced to the system for the final sequence. $\widehat{C}$ type generally consists of two main steps, from the main chain and determine the sequence. The main chain both records the history as a trusted snapshot chain and solves the conflicts among blocks and transactions.

\textbf{\textit{Consensus mechanism}} defines the way to reach the agreement. Traditional linear-based blockchain systems decouple the consensus into two main questions: \textit{who} conducts the consensus and \textit{how to} operate the consensus. We follow this category (see \underline{Section \ref{sec-consensus}}) with additional details as follows.

\textit{Openness} and \textit{membership selection} answer the question of \textit{who}. In blockchain systems, a group of nodes that participate in the consensus procedure is called the committee. These sub-components define the rules on how to select a set of trustworthy nodes to form the committee. The committee is essential for successfully running consensus algorithms since it improves the security threshold of the systems. Specifically, for an open environment (permissionless), nodes compete for winners according to their resources, like computing power, stake, and reputation. For a closed environment (permissioned), the member is generally assigned by the developer team.  Adversaries cannot become legal members if they have not invested in large amounts. Even though an attacker temporarily holds sufficient resources, the advantages cannot last for the long term. For example, IOTA \cite{popov2016tangle} only utilizes the PoW as the anti-spam tool, rather than the rules of membership selection. An attacker may easily launch the parasite attack and splitting attack (see \underline{Section \ref{sec-Security}}). Thus, applying the coordinator to the IOTA network is evidence of improving the security threshold. Similarly, assigning several authority nodes as legal members also exclude unfamiliar nodes outside the committee. Besides, the small size committee enables the systems to run optimally with lightweight payloads. This is especially critical for parallel-chain based systems (\textit{Type III, Type IV}). The more nodes are involved, the more messages are broadcast.

\textit{Unit allocation}, \textit{unit positioning}, \textit{extension rule}, \textit{conflict solving} and \textit{featured technique} answer the question of \textit{how}. Classic blockchain systems focus on the procedure of extension rule where the chain only needs to consider how to select its child. The rule is straightforwards since the metric of child selection is based on one dimensional parameters, such as length (Bitcoin \cite{nakamoto2008peer}), weight \cite{sompolinsky2015secure}, age \cite{king2012ppcoin}.  In contrast, DAG-based systems are complex. Firstly, the role of the units is uncertain. A unit might be a transaction, an event, or a block and the role of a block can even be decoupled into different functionalities (see Prism \cite{bagaria2019prism}). Secondly, locating a unit, as the pre-step for the sorting algorithm, is hard in the graph-based network. A well-structured DAG (such as \textit{Type III}, \textit{Type IV}) can position a specific unit by at least two parameters, $(i,h)$ in \underline{Table.\ref{tab-consensus}}. The sub-component \textit{Unit position} helps to mitigate this issue. However, an unstructured DAG (\textit{Type I}, \textit{Type II}) cannot precisely locate their units. Thirdly, each DAG system has its designs with customized aims and properties. Not all of the systems require strict consistency or total ordered sequence. This leads to different extension rules in multidimensional metrics. We separately use \textit{extension rule} and \textit{conflict solving} to describe their consensus approaches. The extension rule mainly refers to the principles of tie-breaking, while conflict-solving represents the way to sort the selected units into the final sequence. \textit{Technical feature} emphasizes key techniques adopted by each system. A complete consensus mechanism in DAG systems might include all sub-components or several of them.

\textbf{\textit{Properties}} are not physical components (like structure and consensus). Instead, properties describe the final goals of each system. It is not standalone since the aimed properties determine how to design the algorithms and integrate them. Three properties are identified in \underline{Section \ref{sec-modeling}}, namely \textit{consistency}, \textit{order}, and \textit{finality}.  Consistency describes whether a state is globally viewed among different ledgers; order shows how each unit is ordered, and finality represents the possibility that a ledger can be reversed. These three factors are orthogonal in most cases but might have overlaps in a few cases. For example, Vite \cite{liu2018vite} consists of multiple parallel chains and one snapshot chain. If we watch from the whole system, consistency is not achieved since the ledgers in each node have totally different views. In contrast, if we watch from the snapshot chain, consistency is achieved. Such conflicts are caused by the role of its main (snapshot) chain. In most cases, the main chain records the full history and simultaneously conducts the consensus. The decisions will be synchronized to each distributed node, and they can reach the same view of states. However, if the main chain only records the history without further consensus steps, common nodes cannot obtain any synchronized states. In this situation, we measure the consistency in the view of its main chain.

%---------------------------------------inter-component
\smallskip
\subsection{Inter-Component}
This subsection explores the connections between individual components and their impacts on each other.  Two physical components \textit{structure} and \textit{consensus mechanism} and one virtual component \textit{property} are included. Here, we provide more details.

\textbf{\textit{Structure and Consensus.}}
The structure defines the way transactions are recorded in the systems. It greatly impacts the consensus of the unit types they agreed upon. In our classification, two major types of units are identified: \textit{transaction} and \textit{block}. If the system is directly based on transactions, the consensus tends to be simpler than that based on blocks. The former only requires the rules of conflict solving and unit sorting, whereas the latter needs additional procedures such as membership selection, the configuration of committees, \textit{etc.}  Conversely,  the consensus also diversifies the types of structural design. For example, an additional snapshot block in Vite \cite{liu2018vite} is introduced into the system due to the design of consensus. We ignore the classification of block types at more specific levels, such as vote/propose/transaction blocks in Prism \cite{bagaria2019prism}. We consider they are encompassed in the consensus procedures.

\textbf{\textit{Consensus and Property.}} 
The Properties indicate the goal of consensus, and conversely, consensus algorithms are the way to achieve properties. Traditional blockchain systems do not often mention this virtual component since they attempt to achieve strict consistency, or the total linear order, as their default requirements. In contrast, DAG-based systems relax such requirements, letting systems achieve partial consistency or stay in topological order. Thus, the consensus algorithms under different properties vary from system to system. The consensus procedures and complexity to achieve the total linear order are much more complicated than staying in the topological order. But the latter sacrifices security and compatibility to some degree.

\textbf{\textit{Structure and Property.}} 
Properties indirectly impose requirements on the structure design. For example, a unit to be deemed as valid or confirmed requires high confidence to prove it is attached by sufficient child units. This makes the unit equipped with a counter to measure confidence. For systems based on transactions, the counter can only be initiated as an additional field: weight, vote, \textit{etc.} In contrast, for the systems based on blocks, the counter may consider more options like the block volume or the block size, which are used to show the collected/verified transactions in every single block. Also, some powerful blocks may instantly have decisions on contained transactions, like the proposed block in Prism \cite{bagaria2019prism}.

%---------------------------------------Differences
%---------------------------------------Differences
\smallskip
\subsection{Differences with Parallel Approaches}\label{subsec-difference}

The performance and scalability can be optimized by many techniques. We conclude with the most prevailing approaches and highlight their differences compared with DAG.

\textbf{\textit{Sharding technique}} \cite{wang2019sok} splits the transactions into disjoint shards to enable them processed in parallel. DAG approach differs from the sharding technique in the following aspects: a) The systems adopting the sharding approach rely on strict consistency. In contrast, strict consistency is not necessarily for DAG systems. b) Sharding-based systems still utilize the linear-chain structure to maintain their ledgers for final states. The separated zones have to reach an agreement in each epoch across multiple shards, which makes c) the multiple-committee consensus becomes essential in the sharding technique. The key idea of such consensus protocols is to assign the nodes to committees in a non-deterministic way, preventing the adversary from threatening specific members. Conversely, the DAG systems are generally based on single-committee consensus. Some studies \cite{bano2019sok} claim that single-committee consensus is not scalable where newly participated nodes decrease the throughput. This is only one side of the coin since the assumption of consistency is neglected. If the system only requires partial consistency, like Nano \cite{lemahieu2018nano}, a single committee will not become the limitation for scalability.

\textbf{\textit{Layer2 technique}} \cite{gudgeon2020sok}\cite{jourenko2019sok}, refers to the solutions that process certain transactions outside of the main chain, but the consensus of these transactions still relies on a parent chain. The parent chain is a traditional linear-based blockchain, selectively recording the important transactions (\textit{e.g.} balance, final state). Compared with DAG, this technique is inherently an upper-layer protocol without changing the structure or undermining the properties of its parent chain. Layer2 technique cannot maximally improve the performance since the consensus is still limited by its linear-based parent chain. The bottleneck is significantly affected by its main chain. Another critical problem is to reliably guarantee the consistency between off-chain states and on-chain states. Relying on authority nodes for endorsement or charging the deposit in advance might increase extra costs either computationally or economically.

\textbf{\textit{Sidechain technique}} \cite{back2014enabling} deviates from the layer2 technique in whether they have own consensus algorithms. The sidechain is a separate blockchain that processes transactions individually. It interacts with the main chain and tokens can be transferred between them. Several DAG systems (such as Vite/Parsec) get inspired by this design to some degree, which introduces a hybrid architecture with one main chain to sort the units and multiple ordinary (side) chains to process transactions in parallel. However, a major difference is the coupling between the main chain and side chains. In DAG-based systems, the main chain and side chains are closely integrated. The states (metadata) are recorded on all involved side chains and the consensus across these chains (ledgers) is then finished with the help of the main chain. In contrast, the main chain and side chain are completely decoupled in this technique.  The side chain is merely an auxiliary chain, providing a very small proportion of information to the complete system.

\textbf{\textit{Heterogeneous structure}} represents a vertically structural shift. The approach changes originally homogeneous blocks into heterogeneous blocks by adding new block types. It assigns the blocks with different functionalities, generally, including two types: the \textit{keyblocks} used to conduct consensus and the \textit{microblocks} to vote for leaders and carry transactions. This line of studies include Bitcoin-NG \cite{eyal2016bitcoin}, Fruitchain \cite{pass2017fruitchains}, ByzCoin \cite{kogias2016enhancing}, and ComChain \cite{vizier2018comchain}. All these systems improve the throughput of performance without terribly undermining the security promise. The most significant difference is that: the DAG approach inherently is the horizontally scaling solution that aims to enable transactions processed in parallel. Several DAG-based systems directly adopt the concepts of this technique, such as Prism \cite{bagaria2019prism}, to decouple the functionality of blocks; while some systems indirectly utilized the concept, such as Conflux \cite{li2020decentralized}/Vite\cite{liu2018vite}, to organize the key blocks in a separate (main) chain for consensus.

\textbf{\textit{Hybrid consensus solution}} refers to the systems which combine two or more consensus protocols. These systems attempt to abstract the advantages of each system and integrate them into one protocol. Thunderella \cite{pass2018thunderella} replaces the complex process of view-change in PBFT with the Nakamoto consensus. This makes the system smoothly switch between the optimistic conditions and the worst-case conditions. In optimistic conditions, Thunderella provides a high throughput since the adversary is weak, whereas, in the worst-case conditions, it falls back to the NC protocol for the conservative guarantee of security. Other systems includes Tangaroa (Raft+BFT) \cite{copeland2016tangaroa}, Tendermint (PoS+PBFT) \cite{buchman2016tendermint}, \textit{etc}. However, these systems only optimize the performance at the consensus level, which cannot overcome the bottleneck from the linear-based structure or topology. DAG-based systems also adopt the concept of this approach to construct their integrated components, especially how to combine the procedure of committee formation and agreement decision.

\textbf{\textit{Other techniques}} \cite{hafid2020scaling} include on-chain solutions that propose the modifications towards blockchain protocols. Increasing the size or improving the volume of blocks is the key feature. BCH \cite{bch17} modifies the hardcoded parameter, adjusting the block size from 1M to 8M. Larger blocks increase the total throughput and reduce transaction fees. SigWit \cite{sigwit} splits a classic block (1M) into the associated blocks: a base transaction block with a size of 1 MB and an extended block with 3 MB. These blocks individually take different functionalities. However, these systems typically require a hard/soft fork from the original blockchain, which costs much time for the negotiation with their communities. Compared to DAG systems, this approach merely changes the components of blockchains, rather than any structure or topology. The performance bottleneck still depends on its base blockchain.

%=================================================
\section{Concluding Remarks}
\label{sec-conclu}
%=================================================

We provide a handful of open problems come from developers and communities and finally conclude this paper.

%---------------------------------------Myth

\noindent\textbf{Questions from Communities.} 
We also collect several highly mentioned problems from communities. At the end of the paper, we provide our answers as follows.

\smallskip
\noindent\hangindent 1em   \textit{Do we need blocks?} 
    We do not always need blocks. The systems in \textit{Type I/III/V} show us a series of examples. 
    
\smallskip
\noindent\hangindent 1em   \textit{What's the main difference between DAG-based and classic blockchains? }
    The main difference is the organization of units that transits from the linear-based model to the graph-based model. Another point that is often neglected is their properties. Classic blockchain systems set strict consistency as their default requirements, whereas DAG systems may greatly weaken such requirements.
    
\smallskip
\noindent\hangindent 1em   \textit{Does DAG indeed help to improve the performance?}
    DAG solutions do improve the performance of blockchain, in three aspects: scalability, throughput, and confirmation time. A system may improve one or two of them, which depends on its design of the protocol and requirements of the properties. 
    
\smallskip
\noindent\hangindent 1em   \textit{Can smart contract be supported?}
    Ensuring correct state transitions is the most fundamental task of smart contracts. This requires units sorted in  a linear sequence for a consistent view across different ledgers. Thus, only the system designed with a linear sorting algorithm has the chance to establish upper-layer components such as the smart contract.
  
\smallskip
\noindent\hangindent 1em   \textit{What types of applications can be applied? }
    Existing systems can only support basic functionalities like asset transferring. Upper layer components like the smart contract have not been implemented so far (except for external components such as \cite{iotacoordicide}). The applications, heavily relying on state transitions, cannot be directly applied to current DAG systems.
    
\smallskip
\noindent\hangindent 1em   \textit{Is there a blockchain design that simultaneously scales throughput, storage efficiency, and security \cite{wang2019sok}? } 
    Currently, no protocols in the context of DAG blockchains exist and meet all these requirements at the same time. Moreover, the metrics mentioned in the question are even controversial among peer researchers.  
    
\smallskip
\noindent\hangindent 1em   \textit{What types of paradigm can we learn? } 
    We have identified six types of DAG-based blockchain systems under two-dimensional metrics. No concurrent work can outpace ours with respect to the refinement of classification. Fitzi \textit{et al.} introduces the parallel chain model, which in fact contains \textit{Type III/IV}. Other studies \cite{sompolinsky2020phantom}, mentioning the blockDAG or TxDAG,  cannot clearly and comprehensively present the features and differences among current DAG systems, either.

\smallskip
\noindent\textbf{Conclusion.}
Constrained by limitations on performance and scalability, the revolution of classic blockchains is required. DAG-based systems provide innovative models whose underlying structures enable high throughput and large scalability. However, the ﬁeld has grown increasingly complex with different designs and patterns, making newcomers confused. In this work, we provide the first structured analysis of DAG-based blockchains. We provide the overview by reviewing and analyzing ever-existing and ongoing studies. Then, we abstract a general model to capture the features of DAG and identify six types of design patterns. We analyze the collected systems by respectively evaluating their structure, consensus mechanism, property, security, and performance, followed by discussions on their impacts, comparisons, and challenges. To the best of our knowledge, this paper provides the first comprehensive and insightful analysis and summary of DAG-based blockchain systems, making a timely contribution to the prolific and vibrant area of this field. Optimistically, we believe new mechanisms will be progressively proposed to improve the current systems. Future developments, especially structure shifts, will impact performance, scalability, or security in a variety of ways.

%=================================================
\bibliographystyle{unsrt}
\bibliography{bib}
%=================================================

%-------------------------------------------------
\section*{Appendix A: Mathematical Model Explanation}
%-------------------------------------------------

We give more detailed explanation of the property \textit{acyclic} in our DAG-based model. The definition in \underline{Section \ref{sec-modeling}} is stated as:

\begin{equation*}
\begin{array}{ll}
\begin{aligned}
  & \ddag: \, \textit{\textbf{Assume that}} \,\, u_i \in \{  u_1,u_2,...,u_l \} \subseteq \mathcal{V}: \,\,\, \textcolor{blue}{\forall \{i,j,k\} \in [1,l]},\, \\
  & \quad where \,\, k>j>i,\,\, \textcolor{blue}{j\subset\{\varnothing, ...,\{i+1,...,k-1\}\} }, \,\,\,  \\
   &  \quad \textbf{if}\,\, u_i \gets u_j, u_j \gets u_k,\,\, \textbf{then,}\,\,  u_k \gets u_i \,\,\textit{does not exist}.\\
\end{aligned}
\end{array}
\end{equation*}

We emphasize two critical points in our mathematical model (highlighted in \textcolor{blue}{blue}).
The first first key point of avoiding the loop graph is the settings of dynamical boundary of the range $[i, k]$. The parameter $i$ and $k$ are set to be variable, acting as a sliding window. The second key point lies in the settings of parameter set $j$. Here, $j$ can be either a single element or a subset that contains multiple elements in the range of $(i,k)$.

\begin{table*}[!hbt]
  \centering
    \resizebox{\linewidth}{!}{  
    \begin{tabular}[t]{p{3cm} p{8cm}p{4.5cm}}
    \toprule
    \textbf{Elements}  & \textbf{Vertex Sets }  & \textbf{Boundary Settings} \\   \midrule
    $\textrm{size}(j)=0$ & $\{i,i+1\},...,\{j,j+1\},...,\{k-1,k\}$   & $i\in[1,l-1], k\in[2,l]$\\ 
    $\textrm{size}(j)=1$ & $\{i,i+1,i+2\},...,\{j-1,j,j+1\},...,\{k-2,k-1,k\}$  & $i\in[1,l-2], k\in[3,l]$\\
    $\textrm{size}(j)=2$ & $\{i,i+1,i+2,i+3\},...,\{k-3,k-2,k-1,k\}$ & $i\in[1,l-3], k\in[4,l]$\\
    $\textrm{size}(j)=n$ & ... & $i\in[1,l-n-1], k\in[1+n,l]$\\
    $\textrm{size}(j)=l-2$ & $\{i,i+1,,...,k-1,k\}$ & $i=1, k=l$  \\
    \bottomrule
    \end{tabular}
    }
\end{table*}

We give concrete examples as follows. 

\begin{itemize}
    \item[-] \textit{Case I}: Assume that $j=\varnothing$, the definition deteriorates into the first property of \textit{unidirectional} where $ u_i \gets u_k $ and $ u_k \gets u_i $ cannot co-exist.
    \item[-] \textit{Case II}: Assume that $j$ is a single element which belongs to $(i,k)$. It ensures when $u_i \gets u_j $ and $ u_j \gets u_k$, no edge of $ u_k \gets u_i$ exists. This avoid all loops made up by three vertices . 
     \item[-] \textit{Case III}: Assume that $j$ a subset that contains several elements. For simplicity, we suppose totally $5$ vertices in the graph, where $u_i \in \{  u_1,u_2,...,u_5 \}$ and $l=5$. In this case, as $1 \leq \textrm{size}(j) \leq 3 $ (here, $3=l-2$), we list all potential vertex sets covered by our model.
     \begin{itemize}
        % \item[$\circ$] $\textrm{size}(j)=0$: \{1,2\},\{1,3\},\{1,4\},\{1,5\},\{2,3\},\{2,4\},\{2,5\},\{3,4\},\{3,5\},\{4,5\}
         \item[$\circ$] $\textrm{size}(j)=1$: \{1,2,3\}, \{1,2,4\}, \{1,2,5\}, 
          \{2,3,4\}, \{2,3,5\}, \{3,4,5\}. In this situation, our definition avoids loops that are consisted of $3$ vertices.
         \item[$\circ$] $\textrm{size}(j)=2$: \{1,2,3,4\}, \{2,3,4,5\}. In this case, our model avoids loops made by $4$ vertices.
         \item[$\circ$] $\textrm{size}(j)=3$: \{1,2,3,4,5\}. In this case, our model avoids loops containing $5$ vertices.
     \end{itemize}

\end{itemize}

\section{Appendix B. General Description of Our Types}

We analyze consensus mechanisms and corresponding features of existing DAG systems. In this part, we merely provide a very rough description for quick access. \textit{Detailed discussions for each DAG blockchain system are presented in \text{\underline{Section.\ref{sec-consensus}}}. The following contents are aligned with our summaries in  \text{\underline{Table.\ref{tab-consensus}}}.}

\textit{Type I} systems are blockless. The topology of transactions is a naturally expanding graph. The systems in this type follow the simplest design when compared to others -- a transaction can directly verify the ancestors. The only restriction comes from their rules of graph extension. IOTA \cite{popov2016tangle} and Graphain \cite{boyen2018graphchain} adopt a similar tip selection rule (TSA), in which one newly attached transaction can verify multiple ancestor transactions. Avalanche \cite{tanana2019avalanche} utilizes a sampling method to randomly select the child set of transactions in each round. \textit{Type II} systems deviate from \textit{Type I} by introducing the block to package transactions. These blocks are structured as a naturally expanding graph. Spectre \cite{Sompolinsky2017SPECTRES} employs the recursive traverse algorithm to collect transactions and determines a pairwise order between two blocks by voting. Phantom \cite{srivastava2018phantom} extends the graph via a greedy algorithm to gradually include valid blocks. Meshcash \cite{bentov2017tortoise} combines slow PoW-based protocols with a fast consensus protocol. Systems of these types obtain scalability (processing) and flexibility (design).

\textit{Type III} systems are based on the blockless data structure while \textit{Type IV} systems are block-based structures. The units of systems are maintained by individual nodes and finally form multiple parallel chains. Transactions in Nano \cite{lemahieu2018nano} are organized by pairs. The system employs a pairwise voting algorithm to determine the priority between two blocks.  Hashgraph \cite{baird2016swirlds}\cite{baird2018hedera}, DLattice \cite{zhou2019dlattice}, Jointgraph \cite{xiang2019jointgraph}, Aleph \cite{gkagol2019aleph}, Caper \cite{amiri2019caper} and PARSEC \cite{chevalier2019protocol} utilize the asynchronous Byzantine agreement protocol to achieve consensus. This protocol differs from traditional BFT-style consensus in the ways of their (committee) membership selection and conflict solving. Traditional BFT protocols have fixed members who are assigned in advance by certain rules. The case is different in DAG systems because all the nodes that maintain chains have the possibility to become committee members or even conduct the task of the leader. Blockmania \cite{danezis2018blockmania} and Dexon \cite{chen2018dexon} also slightly modify the BFT-protocols. Sphinx \cite{wang2021weak} modifies the PBFT protocol to a leaderless protocol by weakening its strict consistency. Besides, Chainweb \cite{will2019chainweb} extends individual chains with the cross-reference link between them. Vite \cite{liu2018vite}, instead, introduces a powerful snapshot chain to record all the individual chains. Lachesis-class \cite{Choi2018OPERARA}\cite{choi2018fantom}\cite{Nguyen2019ONLAYOL}\cite{Nguyen2019StakeDagSC}\cite{Nguyen2019StairDagCV} contains a family of protocols. They structure the transactions in DAG and then pack them into blocks in an on-top main chain. In contrast, Prism \cite{bagaria2019prism},  OHIE \cite{yu2020ohie}, Blockclique \cite{forestier2018blockclique} and Eunomia \cite{niu2019eunomia} apply the extended Nakamoto consensus to systems, and further utilize sorting algorithms to order blocks into a linear sequence. 

\textit{Type V} systems are blockless while \textit{Type VI} systems are based on blocks. The appended transactions gradually converge into the main chain. Byteball \cite{churyumov2016byteball} introduce a group of witness nodes. These witnesses form a man chain to determine the graph-based network. Haootia \cite{tang2020haootia} establishes a two-layer framework of consensus. The first layer embraces generated transactions and organizes them in a naive graph topology. The second layer is a PoW-based backbone chain with key blocks to decide the total order of transactions. \textit{JHdag} \cite{he2019consensus} applies Nakamoto consensus into the systems. But each block only contains one transaction to limit the block size and simplify the cryptographic puzzles of PoW. GHOST \cite{sompolinsky2015secure}, Inclusive \cite{lewenberg2015inclusive}, Conflux \cite{li2020decentralized}, and CDAG \cite{gupta2019cdag} all employ the modified Nakamoto consensus as their extension rule by replacing the longest-chain rule with weightiest-subtree rule. This design reduces resource waste among competitive chains and also involves more transactions during the same time interval. StreamNet \cite{yin2019streamnet} adopts the concept from Conflux \cite{li2018scaling} and IOTA \cite{popov2016tangle} to form a permissionless network with a main chain.

\end{document}